\documentclass{LMCS}

\usepackage{amssymb}
\usepackage{stmaryrd}
\usepackage{bussproofs}
\usepackage{float}
\usepackage{pst-tree}

\usepackage{enumerate}
\usepackage{hyperref}
\usepackage{url}

\makeatletter
\def\paragraph{\@startsection{paragraph}{4}%
  \z@\z@{-\fontdimen2\font}%
  \itshape}
\makeatother

\newcommand{\online}[1]{Available at \url{#1}}

\newcommand{\lcalc}{\ensuremath{\lambda}-calculus}
\newcommand{\lmcalc}{\ensuremath{\lambda\mu}-calculus}
\newcommand{\lterm}{\ensuremath{\lambda}-term}
\newcommand{\lmterm}{\ensuremath{\lambda\mu}-term}
\newcommand{\logsys}[1]{\textnormal{\textsf{#1}}}
\newcommand{\NJ}{\logsys{NJ}}
\newcommand{\NK}{\logsys{NK}}
\newcommand{\PCF}{\logsys{PCF}}
\newcommand{\ie}{\emph{i.e.}}
\newcommand{\eps}{\varepsilon}
\newcommand{\Nat}{\ensuremath{\mathbb{N}}}
\newcommand{\imp}{\rightarrow}
\newcommand{\reduct}{\rightarrow}
\newcommand{\pair}[2]{{\langle #1,#2\rangle}}
\newcommand{\sem}[1]{\llbracket #1\rrbracket}
\newcommand{\subst}[3]{#1[{^{#2}/_{#3}}]}
\newcommand{\A}{\Gamma}
\newcommand{\B}{\Delta}
\newcommand{\AIC}[1]{\AxiomC{\ensuremath{#1}}}
\newcommand{\UIC}[1]{\UnaryInfC{\ensuremath{#1}}}
\newcommand{\BIC}[1]{\BinaryInfC{\ensuremath{#1}}}
\newcommand{\ZIC}[1]{\AIC{}\UIC{#1}}
\newcommand{\RL}[1]{\RightLabel{\ensuremath{#1}}}
\newcommand{\DP}{\DisplayProof}
\newcommand{\PPP}{\noLine\UIC{\vdots}\noLine}

\newcounter{cmove}
\newcommand{\Garcangle}{30}
\newcommand{\Garcanglem}{-30}
\newcommand{\Gnodesep}{0.1}
\newcommand{\ocolname}{red}
\newcommand{\pcolname}{blue}
\newcommand{\mucolname}{black}
\newcommand{\ocol}{\csname \ocolname\endcsname}
\newcommand{\pcol}{\csname \pcolname\endcsname}
\newcommand{\move}[1]{\textnormal{\texttt{#1}}}
\newcommand{\defmove}{\ensuremath{\bullet}}
\newcommand{\playmove}[2][1]{\stepcounter{cmove}\rnode{play\arabic{cmove}}{\move{#2}}\ncarc[arcangle=\Garcangle,nodesep=\Gnodesep]{-}{play\arabic{cmove}}{play#1}}
\newcommand{\playmovem}[3][1]{\stepcounter{cmove}\rnode{play\arabic{cmove}}{\move{#3}}\ncarc[arcangle=\Garcangle,nodesep=\Gnodesep]{-}{play\arabic{cmove}}{play#1}\ncarc[linecolor=\mucolname,linestyle=dashed,arcangle=\Garcanglem,nodesep=\Gnodesep]{-}{play\arabic{cmove}}{play#2}}
\newcommand{\colmove}[3][1]{\playmove[#1]{#2#3}}
\newcommand{\colmovem}[4][1]{\playmovem[#1]{#2}{#3#4}}
\newcommand{\pmove}[2][1]{\colmove[#1]{\pcol}{#2}}
\newcommand{\pmovem}[3][1]{\colmovem[#1]{#2}{\pcol}{#3}}
\newcommand{\omove}[2][1]{\colmove[#1]{\ocol}{#2}}
\newcommand{\omovem}[3][1]{\colmovem[#1]{#2}{\ocol}{#3}}
\newcommand{\ob}[1][1]{\omove[#1]{\defmove}}
\newcommand{\pb}[1][1]{\pmove[#1]{\defmove}}
\newcommand{\pbm}[2][1]{\pmovem[#1]{#2}{\defmove}}
\newcommand{\oq}[1][1]{\omove[#1]{q}}
\newcommand{\pq}[1][1]{\pmove[#1]{q}}
\newcommand{\oa}[1][1]{\omove[#1]{a}}
\newcommand{\pa}[1][1]{\pmove[#1]{a}}
\newcommand{\newplay}{\renewcommand{\Garcangle}{30}\renewcommand{\Garcanglem}{-30}\renewcommand{\Gnodesep}{0.1}\setcounter{cmove}{0}}
\newcommand{\newlinplay}{\renewcommand{\Garcangle}{-50}\renewcommand{\Garcanglem}{50}\renewcommand{\Gnodesep}{0}\setcounter{cmove}{0}}
\newenvironment{playenv}[2]{%
  \newplay%
  \begin{array}[t]{#1}#2\\%
}{\end{array}}
\newcommand{\nextplay}{\hline\newplay}

\floatstyle{boxed}\restylefloat{table}
\psset{levelsep=6mm,radius=1.2mm}
\newenvironment{treeenv}{\begin{center}\psset{nodesep=0,framesize=6pt}}{\end{center}}
\newcommand{\nod}[3][]{\TC*[#1]~[tnpos=l]{\ensuremath{#2}}~[tnpos=r]{\ensuremath{#3}}}
\newcommand{\onod}[3][]{\nod[linecolor=\ocolname#1]{#2}{#3}}
\newcommand{\pnod}[3][]{\nod[linecolor=\pcolname#1]{#2}{#3}}
\newcommand{\sqnod}[3][]{\Tf*[#1]~[tnpos=l]{\ensuremath{#2}}~[tnpos=r]{\ensuremath{#3}}}
\newcommand{\osqnod}[3][]{\sqnod[linecolor=\ocolname#1]{#2}{#3}}
\newcommand{\psqnod}[3][]{\sqnod[linecolor=\pcolname#1]{#2}{#3}}
\newcommand{\os}[1][1]{\omove[#1]{\Tf*[linecolor=\ocolname,framesize=4pt]}}
\newcommand{\ps}[1][1]{\pmove[#1]{\Tf*[linecolor=\pcolname,framesize=4pt]}}

\renewcommand{\L}{\ensuremath{\mathcal{L}}}
\newcommand{\var}{\ensuremath{\mathcal{V}}}
\newcommand{\avar}{\ensuremath{\mathcal{A}}}
\newcommand{\ovar}{\ensuremath{\mathcal{O}}}
\newcommand{\pvar}{\ensuremath{\mathcal{P}}}
\newcommand{\oenum}{o}
\newcommand{\oren}{\varsigma}
\newcommand{\osubst}{\vartheta}
\newcommand{\asubst}{\theta}
\newcommand{\ren}[2]{#1 #2}
\newcommand{\projl}[1]{\pi_1 #1}
\newcommand{\projr}[1]{\pi_2 #1}
\newcommand{\proji}[1]{\pi_i #1}
\newcommand{\cst}{\star}
\newcommand{\foapp}[1]{\{#1\}}
\newcommand{\expand}{\leftarrow}

\newcommand{\opol}{\ensuremath{O}}
\newcommand{\ppol}{\ensuremath{P}}
\newcommand{\node}[1]{\textnormal{\texttt{#1}}}
\newcommand{\unode}[1]{\textnormal{\texttt{#1}}}
\newcommand{\play}[1]{\textnormal{\texttt{#1}}}
\newcommand{\enabl}[1]{\vdash_{#1}}
\newcommand{\proj}[2]{{{#1}\restriction_{#2}}}
\newcommand{\prodproj}[2]{{{#1}\lbag_{#2}}}
\newcommand{\id}[1]{\textnormal{\texttt{id}}_{#1}}
\newcommand{\fol}{\mathbin{;}}
\newcommand{\diag}[1]{\Delta_{#1}}
\newcommand{\weak}[1]{\textnormal{\texttt{wk}}_{#1}}
\newcommand{\contr}[1]{\textnormal{\texttt{ctr}}_{#1}}
\renewcommand{\int}[1]{\textnormal{\texttt{int}}(#1)}
\newcommand{\plays}[1]{\ensuremath{\mathcal{P}_{#1}}}
\newcommand{\pplays}[1]{\ensuremath{\mathcal{P}^P_{#1}}}
\newcommand{\pref}{\leq}
\newcommand{\ppref}{\leq^P}
\newcommand{\ppreview}[1]{\ensuremath{\underline{\ulcorner#1\urcorner}}}
\newcommand{\pview}[1]{\ensuremath{\ulcorner#1\urcorner}}
\newcommand{\vc}[1]{\textnormal{\texttt{VC}}(#1)}
\newcommand{\inst}[1]{\play{inst}_{#1}}
\newcommand{\umove}{\ensuremath{\textnormal{\texttt{*}}}}

\newcommand{\Qn}{\textnormal{\texttt{Q}}}
\newcommand{\Ar}{\textnormal{\texttt{A}}}

\newcommand{\unfold}[1]{\ensuremath{\widehat{#1}}}
\newcommand{\fold}[1]{\ensuremath{\underline{#1}}}

\newcommand{\Scat}{\mathcal{S}}
\newcommand{\CScat}{\mathcal{S}^c}
\newcommand{\Gcat}{\mathcal{G}}
\newcommand{\Gcatgr}{\mathcal{G}_{00}}
\newcommand{\Gcattf}{\mathcal{G}^{\texttt{tf}}}
\newcommand{\Gcata}{\mathcal{G}_1}
\newcommand{\Gcatatf}{\mathcal{G}_1^{\texttt{tf}}}
\newcommand{\grfunc}{\textnormal{GR}}

\newcommand{\trad}[1]{#1^\star}
\newcommand{\bigtrad}[1]{\trad{(#1)}}
\newcommand{\tradb}[1]{#1^\bullet}

\newcommand{\reduce}{\succ}
\newcommand{\psep}{\bowtie}
\newcommand{\U}{\ensuremath{\mathcal{U}}}
\newcommand{\V}{\ensuremath{\mathcal{V}}}
\newcommand{\At}{\ensuremath{\mathcal{A}}}

\def\doi{6 (4:3) 2010}
\lmcsheading%
{\doi}
{1--50}
{}
{}
{Nov.~\phantom07, 2008}
{Oct.~20, 2010}
{}

\begin{document}

\title[Game semantics for first-order logic]{Game semantics for first-order logic}
\author[Olivier LAURENT]{Olivier LAURENT}
\address{Preuves Programmes Syst\`emes, CNRS -- Universit\'e Paris VII, UMR 7126 -- Case 7014, 75205 Paris Cedex 13 -- FRANCE}
\email{Olivier.Laurent@ens-lyon.fr}

\keywords{ game semantics; first-order logic; classical logic; 
intuitionistic logic; lambda-calculus; lambda-mu-calculus; full 
completeness; type isomorphisms; realizability; control category}
\subjclass{F.3.2, F.4.1, F.3.3}

\begin{abstract}
We refine HO/N game semantics with an additional notion of pointer (mu-pointers) and extend it to first-order classical logic with completeness results. We use a Church style extension of Parigot's lambda-mu-calculus to represent proofs of first-order classical logic.

We present some relations with Krivine's classical realizability and applications to type isomorphisms.
\end{abstract}

\maketitle

\section*{Introduction}

Game interpretations of logic and programming languages have been initially developed on the logic side (for example~\cite{lorenzengames} for intuitionistic logic).
From the beginning of the 90s, most of the attention has been turned to programming languages with the introduction of \emph{game semantics}~\cite{ajm,ho,gamesn,phdmccusker,phdharmer}.

Our goal is to develop a game model of first-order classical logic based on the HO/N model (more precisely its ``classical'' version presented in~\cite{controlgames} which relaxes the bracketing condition).
We have to work between the model of the \lcalc{} presented in~\cite{gamesgoi} (which is fully complete for the \lcalc{} and thus not general enough to allow for the interpretation of classical features) and the model of~\cite{controlgames} where the use of answers might allow for too general behaviours (from the logical point of view). The difference between these two models can be figured out by looking at the interpretation of atoms: \cite{gamesgoi} uses a \emph{one-move} game, while \cite{controlgames} uses a \emph{two-moves} game.
The key ingredient will be the introduction of additional \emph{$\mu$-pointers} in the model of~\cite{gamesgoi} (together with the usual justification pointers, or $\lambda$-pointers). The extension of the one-move model allows for the interpretation of classical logic (and not only the \lcalc).

\bigskip

Game models provide accurate interpretations of logical systems and programming languages as given by \emph{full completeness} results (any element of the model acting on the interpretation of a formula/type is the interpretation of a proof/term). A natural companion property is \emph{faithful completeness} (two different syntactic objects have different interpretations). When both are satisfied, the distinction between syntax and semantics becomes almost irrelevant as suggested by Girard~\cite{meaning1}.

To be slightly more precise, we say that a denotational model is \emph{equivalence complete} if it is fully and faithfully complete, and if each object of the model is isomorphic to the interpretation of a formula/type. That is, if the interpretation functor into the model defines an \emph{equivalence of categories} between the syntactic category and the model.

We will show how the appropriate notion of canonical form for a Church style first-order extension of Parigot's \lmcalc~\cite{lambdamu} together with our first-order game model give an equivalence completeness result. This is the main theorem of this paper.

Starting from the first-order case, we derive a few (known and new) models for different sub-systems. We also study the relation between our one-move model with $\mu$-pointers and the two-moves model (in the setting of propositional logic).
Here is a summary of the main game models considered in the paper:
\begin{center}
\begin{tabular}{|c||ccc@{\;}c|}
\hline
atomic & propositional & propositional & propositional & first-order \\
games & (simply typed)  & \NJ/\lcalc & \NK/\lmcalc & logic\\
& \lcalc{} over 1 atom & & & \\
\hline
1 move & \cite{gamesgoi}  & Section~\ref{secmodellcalc} & Section~\ref{secunfolding} & Section~\ref{secgamemodel} \\
2 moves & \cite{ho,gamesn} & Section~\ref{secunfolding} & \cite{controlgames} & end of Section~\ref{secunfolding} \\
\hline
\hline
structures & $\lambda$-pointers & atomic labels & $\mu$-pointers & first-order labels \\
used here & & & & instantiations \\
\hline
\end{tabular}
\end{center}
There are equivalence completeness results underlying all these models.

Following the method developed in~\cite{classisos}, we apply our game model to the purely syntactic problem of characterizing the type isomorphisms of call-by-name first-order classical logic. This is a new result in the topic of type isomorphisms.

\bigskip

We end the paper with the presentation of a close relation between game semantics and Krivine's classical realizability \cite{realgeocalslides}.
This is important for two reasons. First, the idea of introducing $\mu$-pointers in the one-move game models came from an analysis of the interpretation of proofs through Krivine's realizability.
Second, game semantics and realizability are two of the most important tools developed along the Curry-Howard correspondence to relate logic and computer science and to derive computational interpretations of proofs. Being able to conciliate these two approaches is a very pleasant thing.

In this paper we focus on the logical aspects of game semantics. However games are also a crucial tool in the study of the semantics of programming languages. The \lmcalc{} appears as a natural bridge since it is known to provide both a term syntax for proofs in classical logic and a foundation for functional programming languages with control operators. The link between games and realizability, which is presented here, offers another bridge between games and the theory of programming languages.

\paragraph{Related works.}

The game setting developed in~\cite{lorenzengames,felschergames} is quite similar to our proposal concerning the notion of play (and view). However it is done in an intuitionistic setting and without any particular interest for the composition of strategies which is at the core of HO/N games. In this line of work, Coquand~\cite{coquandgames} has explicitly worked on composition and in a classical setting but in relation with a quite different syntactic system: Novikoff's calculus. Finally Herbelin~\cite{gamescpcf} (following Coquand) and Laird~\cite{controlgames} (following HO/N) arrived to a meeting point by giving a fully complete game model for a classical extension (\emph{\`a la $\lambda\mu$-calculus}) of PCF, that is without propositional variables or quantification.

A key ingredient which is new with respect to those works is the notion of $\mu$-pointer. It happens that they appear to be a particular case of the contingency pointers introduced by Laird for local exceptions~\cite{excepgames}.

A more algebraic approach (by means of generators and relations) to game semantics for first-order quantification is developed in~\cite{algfogames}. The underlying logic is very basic: linear and without propositional connectives.

\tableofcontents

\section{Notations and used languages}

\subsection{First-order logic}

In the whole paper, we consider a fixed first-order language \L, that is a countable set of function symbols (with given arities), denoted by $f$, $g$,~... and a countable set of relation symbols (with given arities), denoted by $X$, $Y$,~... (arities are natural numbers). We assume given a countable set of first-order variables $\var$, denoted $x$, $y$,~...

To clarify the different uses we will have of first-order variables and of first-order terms, we consider the set of variables $\var$ as the disjoint union of three countable sets of variables: \avar-variables, \ovar-variables and \pvar-variables. And we assume given an enumeration $(\oenum_i)_{i\in\Nat}$ of \ovar-variables.

First-order terms are defined as:
\begin{eqnarray*}
t & ::= & x \mid f \vec{t}
\end{eqnarray*}
where function application respects the arity of symbols.

As sub-classes, we will use \avar\pvar-terms (first-order terms built from \avar-variables and \pvar-variables only) and \ovar\pvar-terms (first-order terms built from \ovar-variables and \pvar-variables only). This will be done in the spirit of Barendregt's convention~\cite[2.1.13 page~26]{barendregt}: different names are used for different purposes (see in particular Section~\ref{secgamemodel}). \avar-variables will be used for bound occurrences in types/formulas and arenas, \ovar-variables for bound occurrences in \lmterm s and strategies and \pvar-variables for free occurrences.

Formulas are defined as:
\begin{eqnarray*}
A & ::= & \top \mid \bot \mid X \vec{t} \mid A\imp A \mid A\wedge A \mid \forall x A
\end{eqnarray*}
where relation application respects the arity of symbols, $x$ is an \avar-variable and $\vec{t}$ are \avar\pvar-terms.

An \emph{atomic formula} is  $X t_1 \dots t_k$, $\top$ or $\bot$, denoted $R$, $S$,~... If it is neither $\top$ nor $\bot$ it is a \emph{non-constant} atomic formula.

The (now quite popular~\cite{realzf,controlcat,polgamesapal}) restriction of the set of connectives to the so-called ``negative'' ones is what makes the framework much easier to manage. Note that the other connectives are easy to define from their negative dual by means of negation (for example $\exists x A\equiv (\forall x(A\imp\bot))\imp\bot$).

\subsection{Church style \texorpdfstring{\lmcalc}{lambda-mu-calculus} for first-order logic}

In order to describe proofs in first-order classical logic, we use (according to the Curry-Howard correspondence) a Church style extension of Parigot's \lmcalc~\cite{lambdamu} with abstraction and application for first-order universal quantification. First-order formulas are used as types.

Given two disjoint countable sets of variables ($\lambda$-variables, denoted $a$, $b$,~... and $\mu$-variables, denoted $\alpha$, $\beta$,~...), the corresponding \lmterm s are:
\begin{eqnarray*}
M & ::= & a \mid \lambda a.M \mid (M)M \mid \pair{M}{M} \mid \projl{M} \mid \projr{M} \mid \cst \mid [\alpha]M \mid \mu\alpha.M \mid \Lambda x.M \mid M\foapp{t}
\end{eqnarray*}
where $x$ is an \ovar-variable and $t$ is an \ovar\pvar-term. We use the simplified notation $\mu\alpha[\beta]M$ instead of $\mu\alpha.[\beta]M$ when these two constructions come together.

\lmterm s are considered up to $\alpha$-equivalence for $\lambda$-variables bound by $\lambda$, $\mu$-variables bound by $\mu$ and \ovar-variables bound by $\Lambda$. We consider only \lmterm s without free \ovar-variables (Barendregt's convention). A \lmterm{} is \emph{closed} if it contains neither free $\lambda$-variables nor free $\mu$-variables (it may contain free \pvar-variables).

Typing judgments are of the shape $\A\vdash M:A\mid\B$ where $\A$ is a set of typing declarations for distinct $\lambda$-variables (\ie{} pairs $a:A$) and $\B$ is a set of typing declarations for distinct $\mu$-variables (\ie{} pairs $\alpha:A$).
The derivation rules for this system are given in Table~\ref{tabtyprules}.

\begin{table}
\begin{gather*}
\ZIC{\A,a:A\vdash a:A\mid\B}
\DP
\quad\;
\AIC{\A,a:A\vdash M:B\mid\B}
\UIC{\A\vdash \lambda a.M:A\imp B\mid\B}
\DP
\quad\;
\AIC{\A\vdash M:A\imp B\mid\B}
\AIC{\A\vdash N:A\mid\B}
\BIC{\A\vdash (M)N:B\mid\B}
\DP
\\[2ex]
\ZIC{\A\vdash \cst:\top\mid\B}
\DP
\quad\qquad
\AIC{\A\vdash M:A\mid\B}
\AIC{\A\vdash N:B\mid\B}
\BIC{\A\vdash \pair{M}{N}:A\wedge B\mid\B}
\DP
\quad\qquad
\AIC{\A\vdash M:A_1\wedge A_2\mid\B}
\UIC{\A\vdash \proji{M}:A_i\mid\B}
\DP
\\[2ex]
\AIC{\A\vdash M:A\mid\B,\alpha:A}
\UIC{\A\vdash [\alpha]M:\bot\mid\B,\alpha:A}
\DP
\quad\qquad
\AIC{\A\vdash M:\bot\mid\B,\alpha:A}
\UIC{\A\vdash \mu\alpha.M:A\mid\B}
\DP
\\[2ex]
\AIC{\A\vdash \subst{M}{y}{z}:\subst{A}{y}{x}\mid\B}
\RL{y\notin \A,M,A,\B}
\UIC{\A\vdash \Lambda z.M:\forall x A\mid\B}
\DP
\quad\qquad
\AIC{\A\vdash M:\forall x A\mid\B}
\UIC{\A\vdash M\foapp{t}:\subst{A}{t}{x}\mid\B}
\DP
\\
\text{$x$ is an \avar-variable, $y$ is a \pvar-variable and $z$ is an \ovar-variable.}
\end{gather*}
\caption{Typing rules for the first-order \lmcalc}\label{tabtyprules}
\end{table}

Through the Curry-Howard correspondence, type inhabitance corresponds to provability.

\begin{prop}[Provability]
The formula $A$ is provable in first-order classical logic if and only if there exists a closed \lmterm{} $M$ such that $\vdash M:A\mid$ is derivable.
\qed
\end{prop}

The equality between proofs is the congruence generated by the equational theory $\beta\eta\mu\rho\theta$ on \emph{typed} \lmterm s given in Table~\ref{tabeqlm}.

\begin{table}
\begin{equation*}
\begin{array}{rlllcl}
(\lambda a.M)N              & =_\beta  & \subst{M}{N}{a} & :A \\
\lambda a.(M)a              & =_\eta   & M               & :A\imp B & \qquad & a \notin M \\
\projl{\pair{M}{N}}         & =_\beta  & M               & :A \\
\projr{\pair{M}{N}}         & =_\beta  & N               & :A \\
\pair{\projl{M}}{\projr{M}} & =_\eta   & M               & :A\wedge B \\
\cst                        & =_\eta   & M               & :\top \\
(\Lambda x.M)\foapp{t}      & =_\beta  & \subst{M}{t}{x} & :A \\
\Lambda x.M\foapp{x}        & =_\eta   & M               & :\forall x A & & x \notin M \\
(\mu\alpha.M)N              & =_\mu    & \mu\alpha.\subst{M}{[\alpha](L)N}{[\alpha]L} & :A \\
\projl{\mu\alpha.M}         & =_\mu    & \mu\alpha.\subst{M}{[\alpha]\projl{L}}{[\alpha]L} & :A \\
\projr{\mu\alpha.M}         & =_\mu    & \mu\alpha.\subst{M}{[\alpha]\projr{L}}{[\alpha]L} & :A \\
(\mu\alpha.M)\foapp{t}      & =_\mu    & \mu\alpha.\subst{M}{[\alpha]L\foapp{t}}{[\alpha]L} & :A \\
{}[\beta]\mu\alpha.M        & =_\rho   & \subst{M}{\beta}{\alpha} & :\bot \\
\mu\alpha[\alpha]M          & =_\theta & M               & :A & & \alpha\notin M \\
{}[\alpha]M                 & =_\rho   & M               & :\bot
\end{array}
\end{equation*}
\begin{tabular}{p{0.95\textwidth}}
where $\subst{M}{\mathcal{C}[L]}{[\alpha]L}$ is obtained by substituting any sub-term of $M$ of the shape $[\alpha]L$ by $\mathcal{C}[L]$.
\end{tabular}
\caption{Equalities between \lmterm s}\label{tabeqlm}
\end{table}

\subsection{The syntactic category}
\label{secsyntcat}

The \emph{syntactic category} $\Scat$ has objects given by types and morphisms  from $A$ to $B$ obtained by quotienting the set of closed \lmterm s of type $A\imp B$ by the congruence generated by $\beta\eta\mu\rho\theta$. The identity morphism is the equivalence class of the \lmterm{} $\lambda a.a$ of type $A\imp A$. The composition of two equivalence classes containing $M:A\imp B$ and $N:B\imp C$ is the class of $\lambda a.(M)(N)a:A\imp C$ ($a\notin M$, $a\notin N$).

In order to simplify our work in the rest of the paper, we are going to move from the syntactic category to an equivalent one.

Concerning formulas, we first define $\imp$-canonical forms (non-terminal $\textsf{Q}$ in Table~\ref{tabgramcanform}):\label{defimpcanform}
\begin{equation*}
\forall\vec{x}(Q_1\imp\cdots\imp Q_k\imp R)
\end{equation*}
with $R$ atomic but different from $\top$ (called the \emph{final atom} of the formula) and the $Q_j$s in $\imp$-canonical form.
Then \emph{canonical forms} are:
\begin{equation*}
\bigwedge_{1\leq i\leq n} \forall\vec{x}(Q^i_1\imp\cdots\imp Q^i_{k_i}\imp R^i)
\end{equation*}
with $n\geq 0$ (where $\bigwedge_{1\leq i\leq 0}Q_i=\top$, $\bigwedge_{1\leq i\leq 1}Q_i=Q_1$ and $\bigwedge_{1\leq i\leq n+1}Q_i=(\bigwedge_{1\leq i\leq n}Q_i)\wedge Q_{n+1}$), with the $\forall\vec{x}(Q^i_1\imp\cdots\imp Q^i_{k_i}\imp R^i)$s in $\imp$-canonical form. This corresponds to the non-terminal $\textsf{C}$ in the grammar of Table~\ref{tabgramcanform}.

\begin{table}
\begin{align*}
A\wedge(B\wedge C) &= (A\wedge B)\wedge C \\
A\wedge\top &= A \\
\top\wedge A &= A \\[2ex]
(A\wedge B)\imp C &= A\imp(B\imp C) \\
\top\imp A &= A \\[2ex]
A\imp(B\wedge C) &= (A\imp B)\wedge(A\imp C) \\
A\imp\top &= \top \\[2ex]
\forall x (A\wedge B) &= \forall x A \wedge \forall x B \\
\forall x \top &= \top \\
A\imp\forall x B &= \forall x (A\imp B) & x\notin A \\[2ex]
A\wedge B &= B\wedge A \\
\forall x\forall y A &= \forall y\forall x A
\end{align*}
\caption{Type isomorphisms}\label{tabtypisos}
\end{table}

\begin{table}
  \begin{align*}
    \textsf{R} &::= X \vec{t} \mid \bot \\
    \textsf{A} &::= \textsf{R} \mid \textsf{Q}\imp\textsf{A} \\
    \textsf{Q} &::= \textsf{A}\mid \forall x\textsf{Q} \\
    \textsf{B} &::= \textsf{Q}\mid \textsf{B}\wedge\textsf{Q} \\
    \textsf{C} &::= \textsf{B} \mid \top
  \end{align*}
\caption{Canonical forms for formulas}\label{tabgramcanform}
\end{table}

\begin{prop}[Canonical forms for formulas]
If we consider formulas up to the equations of Table~\ref{tabtypisos}\footnote{These equations are validated by syntactic isomorphisms, see Proposition~\ref{propsynisos} page~\pageref{propsynisos}.} (except the last two), any formula can be written in canonical form.
\end{prop}

\proof
We consider the equations of Table~\ref{tabtypisos} (except the last two) as rewriting rules from left to right.

We define the two functions $\phi$ and $\psi$ from formulas to integers greater or equal to $2$:
\begin{align*}
  \phi(\top)=\phi(\bot)=\phi(R)&=\psi(\top)=\psi(\bot)=\psi(R) = 2 \\
  \phi(A\wedge B) &= 2(\phi(A)+1)\phi(B) \\
  \psi(A\wedge B) &= 2(\psi(A)+1)\psi(B) \\
  \phi(A\imp B) &= \phi(B)^{\phi(A)} \\
  \psi(A\imp B) &= \psi(B)^{\psi(A)} \\
  \phi(\forall x A) &= \phi(A)^2 \\
  \psi(\forall x A) &= 2\psi(A)
\end{align*}
We can easily check that for each rewriting rule $A\mapsto B$, $(\phi(A),\psi(A))>(\phi(B),\psi(B))$ (with respect to the lexicographic order).
Finally, if $A$ is a formula such that no rewriting rule applies to it, then $A$ is in canonical form.
\qed

Up to the $\beta\eta\mu\rho\theta$ equational theory, any closed \lmterm{} whose type is a canonical form can be written as a \emph{canonical normal form} which is either $\cst$ or a tuple of terms of the shape: 
\begin{equation*}
\Lambda\vec{x}.\lambda\vec{a}.\underline{\mu[\,]}((b)\foapp{\vec{t}})\vec{M}
\end{equation*}
where $\underline{\mu[\,]}$ is of the shape $\mu\alpha[\beta]$ except that $[\beta]$ disappears if $((b)\foapp{\vec{t}})\vec{M}$ has type $\bot$ and that $\mu\alpha$ disappears if $\underline{\mu[\,]}((b)\foapp{\vec{t}})\vec{M}$ has type $\bot$ (see Appendix~\ref{appcanformterm} for a proof of this result).

The \emph{syntactic category} $\CScat$ is the category in which objects are types in canonical form and morphisms are closed \lmterm s in canonical normal form quotiented by $\beta\eta\mu\rho\theta$\footnote{We will see in fact in Corollary~\ref{corcanform} that there is no quotient involved here since two different canonical normal forms cannot be equalized through $\beta\eta\mu\rho\theta$.}.
According to the previous remarks, this is a category equivalent to $\Scat$.

\section{A game model of first-order logic}
\label{secgamemodel}

\subsection{Arenas}
\label{secarena}

The notions of \emph{forest} and \emph{tree} will occur at different places in this work. Sometimes enriched with some additional structure (such as labels or pointers) and sometimes not. Here we consider forests and trees as finite objects defined by mutual induction:
\begin{enumerate}[$\bullet$]
\item a finite list of trees is a forest,
\item a node together with a forest is a tree, the node is called a \emph{root} and the roots of the trees of the forest are the \emph{sons} of this root and the trees of the forest are the \emph{immediate sons} of this tree.
\end{enumerate}
This definition is well founded by using the case of an empty list of trees as a forest. Notice that a tree is never empty while a forest could perfectly be empty.

The root of a tree is considered as the \emph{top element} of the tree, so that we can speak about a node \emph{above} or \emph{below} another in a tree/forest.
The \emph{polarity} of a node is the parity of the length of the path from a root to this node (in particular the polarity of roots is even).

If $\mathcal{F}$ is a forest and $\mathcal{T}$ is a tree, the \emph{graft} of $\mathcal{F}$ on $\mathcal{T}$ is the tree which has the same root as $\mathcal{T}$ and with immediate sons obtained by concatenating $\mathcal{F}$ (on the left) to the list of the immediate sons of $\mathcal{T}$. If $\mathcal{F}'$ is a forest, the graft of $\mathcal{F}$ on $\mathcal{F}'$ is obtained by grafting $\mathcal{F}$ on each tree of $\mathcal{F}'$ (this may entail duplications of $\mathcal{F}$, and if $\mathcal{F}$ comes with some additional structure, this structure is also duplicated).

If $\mathcal{T}_1$ and $\mathcal{T}_2$ are two trees, the \emph{merging} of $\mathcal{T}_1$ and $\mathcal{T}_2$ is the tree obtained by grafting the list of immediate sons of $\mathcal{T}_1$ (which is a forest) on $\mathcal{T}_2$. This means that the two roots are identified and the two lists of immediate sons are concatenated.
If $\mathcal{F}_1=[\mathcal{T}_1,\dots,\mathcal{T}_p]$ and $\mathcal{F}_2=[\mathcal{T}'_1,\dots,\mathcal{T}'_q]$ are two forests, the \emph{merging} of $\mathcal{F}_1$ and $\mathcal{F}_2$ is the forest $[\mathcal{T}_1^1,\dots,\mathcal{T}_1^q,\mathcal{T}_2^1,\dots,\mathcal{T}_2^q,\dots,\mathcal{T}_p^1,\dots,\mathcal{T}_p^q]$ where the tree $\mathcal{T}_i^j$ is the merging of $\mathcal{T}_i$ and $\mathcal{T}'_j$.

\begin{exa}
If we consider the forest $\mathcal{F}$ and the two trees $\mathcal{T}_1$ and $\mathcal{T}_2$:
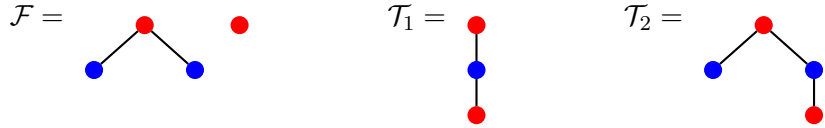
\begin{center}
$\mathcal{F}=\pstree{\onod{}{}}{\pnod{}{} \pnod{}{}} \pstree{\onod{}{}}{}$
\qquad\qquad
$\mathcal{T}_1=\pstree{\onod{}{}}{\pstree{\pnod{}{}}{\onod{}{}}}$
\qquad\qquad
$\mathcal{T}_2=\pstree{\onod{}{}}{\pnod{}{} \pstree{\pnod{}{}}{\onod{}{}}}$
\end{center}

The graft of $\mathcal{F}$ on $\mathcal{T}_2$ is the following tree:
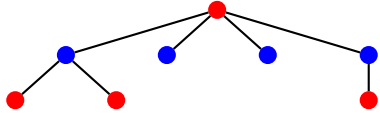
\begin{treeenv}
\pstree{\onod{}{}}{\pstree{\pnod{}{}}{\onod{}{} \onod{}{}} \pnod{}{} \pnod{}{} \pstree{\pnod{}{}}{\onod{}{}}}
\end{treeenv}

The merging of $\mathcal{T}_1$ and $\mathcal{T}_2$ is:
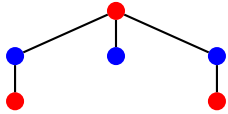
\begin{treeenv}
\pstree{\onod{}{}}{\pstree{\pnod{}{}}{\onod{}{}} \pnod{}{} \pstree{\pnod{}{}}{\onod{}{}}}
\end{treeenv}
\end{exa}

We define a notion of arena adapted to the presence of first-order quantification (in the spirit of polymorphic arenas~\cite{gamesF} developed for second-order quantification).

\begin{defi}[Arena]
An \emph{arena} is a forest with nodes labelled with:
\begin{enumerate}[$\bullet$]
\item a list of first-order \avar-variables, called the \emph{first-order label} of the node;
\item a list of non-constant atomic formulas (using only \avar\pvar-terms), called the \emph{atomic label} of the node (in such a way that \avar-variables appearing in an \avar\pvar-term already appear in the first-order label of the node or of a node above it).
\end{enumerate}
The nodes of the forest are called \emph{moves}. Concerning the \emph{polarity}, we also use \opol{} for even and \ppol{} for odd.

If the move \node{m} is the son of the move \node{n} in the arena $A$, we say that \node{n} \emph{enables} \node{m} (denoted by $\node{n}\enabl{A}\node{m}$). Roots are also called \emph{initial moves} denoted by $\enabl{A}\node{m}$.
\end{defi}

In this paper, we have to deal with a bunch of binding structures. For each of them we can use binding through names and $\alpha$-renaming, de~Bruijn indexes, pointers,~... We decide to use explicit names for first-order variables in arenas. If an \avar-variable $x$ appears in an \avar\pvar-term of the atomic label of a move \node{m} and also in the first-order label of a move \node{n} above \node{m} (or of \move{m} itself), $x$ has to be considered as bound in the arena. We will not explicitly deal with arenas up to $\alpha$-conversion of these bound $\avar$-variables. However we will assume all the elements of the first-order labels of an arena to be different. This could require implicit renaming in the arena constructions.

\begin{exa}\label{exarena}
If we represent first-order labels on the left-hand side and atomic labels on the right-hand side of each move, here is an arena:
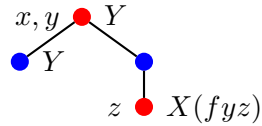
\begin{treeenv}
  \pstree{\onod{x,y}{Y}}{\pnod{}{Y} \pstree{\pnod{}{}}{\onod{z}{X (f y z)}}}
\end{treeenv}
where $X$ has arity $1$, $Y$ has arity $0$, and $f$ has arity $2$.

Remember that, in the general case, the atomic label of a move may contain more than one element.

For the following examples, we name the root as $\move{a}_0$, its sons as $\move{a}_1$ and $\move{a}_2$ and the son of $\move{a}_2$ is named $\move{a}_3$.
\end{exa}

\begin{defi}[Arrow arena]
Let $A$ and $B$ be two arenas, the \emph{arrow arena} $A\imp B$ is the graft of $A$ on $B$.
\end{defi}

\subsection{Sequences of moves}
\label{secseqmoves}

Game semantics usually deals with sequences of moves which are equipped with some additional structure. Due to the presence of first-order quantification, we will use some even richer structure.

We first introduce pointers on sequences of moves. Two kinds of pointers are required in our setting.

\begin{defi}[Justified sequence]
A \emph{justified sequence} \play{s} on the arena $A$ is a sequence of moves of $A$ together with:
\begin{enumerate}[$\bullet$]
\item for each occurrence of a non-initial move \move{m}, we give a \emph{justification pointer} (or \emph{$\lambda$-pointer}) to an earlier occurrence of move in \play{s} (that corresponds to giving an integer smaller than the index of \move{m} in \play{s}) which enables \move{m} in $A$;
\item for each occurrence of a move \move{m} with atomic label $l$ in $A$, we give, for each element of $l$, at most one \emph{$\mu$-pointer} to an element of the atomic label of an earlier occurrence of move \move{n} of opposite polarity (this can be represented as, for each element of $l$, an integer corresponding to the index of \move{n} and then an integer giving the chosen element in the atomic label $l'$ of \move{n}).
\end{enumerate}
\end{defi}

\begin{exa}
Here is a justified sequence on the arena $A$ of Example~\ref{exarena}:
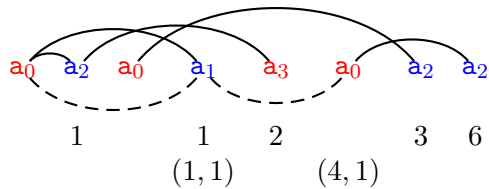
\vspace{3ex}
\begin{equation*}
\newlinplay
\begin{array}{cccccccc}
\omove[0]{a$_0$} & \pmove{a$_2$} & \omove[0]{a$_0$} & \pmovem{1}{a$_1$} & \omove[2]{a$_3$} & \omovem[0]{4}{a$_0$} & \pmove[3]{a$_2$} & \pmove[6]{a$_2$} \\[3ex]
 & 1 & & 1 & 2 & & 3 & 6 \\
 & & & (1,1) & & (4,1) &
\end{array}
\end{equation*}
$\lambda$-pointers are represented as plain lines and $\mu$-pointers as dashed lines.

The second line gives $\lambda$-pointers as integers.
The third line gives $\mu$-pointers as pairs of integers (the second index is always $1$ since the length of the atomic labels of $A$ is at most $1$, and there is at most one pair for each move for the same reason).
\end{exa}

We now introduce first-order instantiations on sequences of moves (independently of the pointer structure).
An \emph{\ovar-instantiation} of a move \node{m} of an arena $A$, which has a first-order label of length $n$, is a list of $n$ \ovar-variables. A \emph{\pvar-instantiation} (which is \emph{not} dual to \ovar-instantiation) of a move \node{m} of an arena $A$, which has a first-order label of length $n$, is a list of $n$ \ovar\pvar-terms.

\begin{defi}[Instantiated sequence]
An \emph{instantiated sequence} \play{s} on the arena $A$ is a sequence of moves of $A$ together with:
\begin{enumerate}[$\bullet$]
\item an \ovar-instantiation for each \opol-move
\item a \pvar-instantiation for each \ppol-move
\end{enumerate}
such that all the \ovar-variables appearing in the \ovar-instantiations are different.
\end{defi}

We consider the possibility of modifying the \ovar-variables: an \emph{\ovar-renaming} is an injection from the set of \ovar-variables to itself. If \play{s} is an instantiated sequence and if $\oren$ is an \ovar-renaming, $\ren{\play{s}}{\oren}$ is the instantiated sequence obtained by substituting $\oenum$ by $\oren(\oenum)$ in any instantiation of \play{s}.

More generally, an \emph{\ovar-substitution} is a function from \ovar-variables to \ovar\pvar-terms. If \play{s} is an instantiated sequence and if $\osubst$ is an \ovar-substitution, $\ren{\play{s}}{\osubst}$ is obtained by substituting $\oenum$ by $\osubst(\oenum)$ in any \pvar-instantiation of \play{s}.

The combination of pointers and instantiations is required for interaction sequences and plays.

\begin{defi}[Interaction sequence]
Let $A$, $B$ and $C$ be three arenas, an \emph{interaction sequence} $\play{u}$ on $A$, $B$ and $C$ is a justified sequence on $(A\imp B)\imp C$ (without any $\mu$-pointer between a move of $A$ and a move of $C$) together with:
\begin{enumerate}[$\bullet$]
\item for each \opol-move played in $C$ and for each \ppol-move played in $A$, an \ovar-instantiation;
\item for each \ppol-move played in $C$ and for each \opol-move played in $A$, a \pvar-instantiation;
\item for each move played in $B$, a pair of an \ovar-instantiation and of a \pvar-instantiation;
\end{enumerate}
such that all the \ovar-variables appearing in the \ovar-instantiations are different.
This turns $\play{u}$ into an instantiated sequence on $A\imp(B\imp C)$ by forgetting: pointers, \ovar-instantiations for \opol-moves played in $B$ and \pvar-instantiations for \ppol-moves played in $B$.

The set of all interaction sequences on $A$, $B$ and $C$ is noted $\int{A,B,C}$.
\end{defi}

An instantiated justified sequence $\play{s}$ on an arena $A$ generates substitutions of the \avar-variables appearing in the first-order labels of $A$ by the \ovar\pvar-terms appearing in the instantiations. Let \move{m} be an occurrence of move in \play{s} with instantiation $[t_1,\dots,t_k]$ and let $[x_1,\dots,x_k]$ be the first-order label of $\unode{m}$ in $A$, we define the substitution $\asubst_{\move{m}}$ as $\{x_1\mapsto t_1,\dots,x_k\mapsto t_k\}$ if $\unode{m}$ is an initial move, and $\asubst_{\move{n}}\cup\{x_1\mapsto t_1,\dots,x_k\mapsto t_k\}$ where \move{n} is the occurrence of move justifying \move{m} in \play{s} otherwise.

\begin{defi}[Play]
A \emph{play} on the arena $A$ is an instantiated justified sequence on $A$ such that:
\begin{enumerate}[$\bullet$]
\item polarities of moves are alternating;
\item there are no $\mu$-pointers from Opponent moves;
\item there is exactly one $\mu$-pointer for each element of the atomic label of each Player move;
\item for each $\mu$-pointer going from a formula $X t_1\dots t_k$ labelling (in $A$) an occurrence of move \move{m} to a formula $Y u_1\dots u_p$ labelling (in $A$) an occurrence of move \move{n}, we have $X=Y$, $k=p$, $t_1\asubst_{\move{m}}=u_1\asubst_{\move{n}}$,~...,~$t_k\asubst_{\move{m}}=u_k\asubst_{\move{n}}$;
\item all the \ovar-variables appearing in a \pvar-instantiation have appeared in a previous \ovar-instantiation.
\end{enumerate}
The set of all plays on $A$ is noted $\plays{A}$. The set of even length plays on $A$ is noted $\pplays{A}$. The prefix order on plays is noted $\pref$ and we use the notation $\play{s}\ppref\play{t}$ ($\play{s}$ is \emph{\ppol-prefix} of $\play{t}$) for $\play{s}\pref\play{t}\wedge\play{s}\in\pplays{A}$.
\end{defi}

We can summarize the structure put on moves of a play:
\begin{enumerate}[$\bullet$]
\item an Opponent move is equipped with a justification pointer and with an \ovar-instanti\-ation;
\item a Player move is equipped with a justification pointer, with a list of $\mu$-pointers and with a \pvar-instantiation.
\end{enumerate}
\ovar-variables are introduced by Opponent and then used by Player.

\begin{exa}
On the arena:
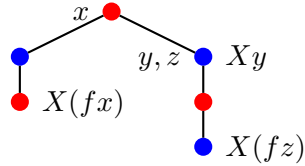
\begin{treeenv}
  \pstree{\onod{x}{}}{\pstree{\pnod{}{}}{\onod{}{X (f x)}} \pstree{\pnod{y,z}{X y}}{\pstree{\onod{}{}}{\pnod{}{X (f z)}}}}
\end{treeenv}
if we name the root as $\move{b}_0$, then we name the moves along the first branch $\move{b}_1$ and $\move{b}_2$ and along the second branch $\move{b}_3$, $\move{b}_4$ and $\move{b}_5$, we have the following play (above the line):
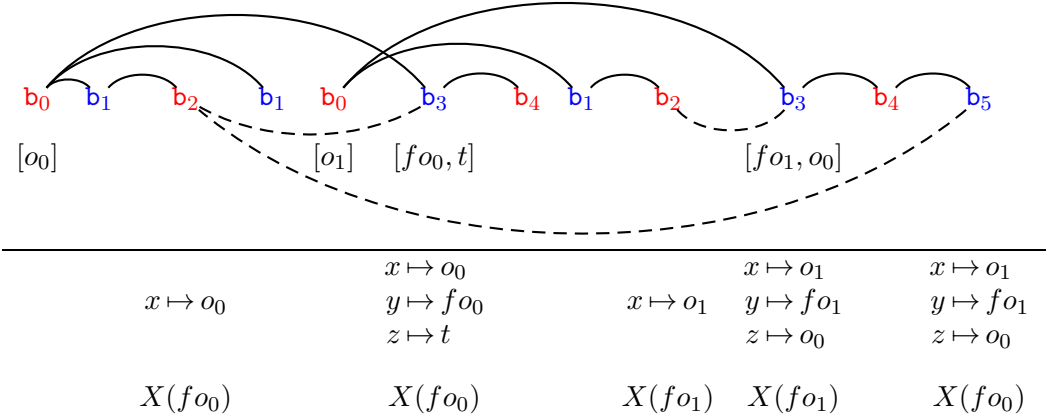
\vspace{6ex}
\begin{equation*}
\newlinplay
\begin{array}{cccccccccccc}
\omove[0]{b$_0$} & \pmove{b$_1$} & \omove[2]{b$_2$} & \pmove{b$_1$} & \omove[0]{b$_0$} & \renewcommand{\Garcanglem}{30}\pmovem{3}{b$_3$} & \omove[6]{b$_4$} & \pmove[5]{b$_1$} & \omove[8]{b$_2$} & \pmovem[5]{9}{b$_3$} & \omove[10]{b$_4$} & \renewcommand{\Garcanglem}{40}\pmovem[11]{3}{b$_5$} \\[2ex]
[\oenum_0] & & & & [\oenum_1] & [f\oenum_0,t] & & & & [f\oenum_1,\oenum_0] & & \\[6ex]
\hline
& & \setlength{\arraycolsep}{1pt}\begin{array}[c]{rcl}x &\mapsto & \oenum_0\end{array} & & & \setlength{\arraycolsep}{1pt}\begin{array}[c]{rcl}x &\mapsto & \oenum_0\\ y &\mapsto & f\oenum_0\\ z &\mapsto & t\end{array} & & & \setlength{\arraycolsep}{1pt}\begin{array}[c]{rcl}x &\mapsto & \oenum_1\end{array} & \setlength{\arraycolsep}{1pt}\begin{array}[c]{rcl}x &\mapsto & \oenum_1\\ y &\mapsto & f\oenum_1\\ z &\mapsto & \oenum_0\end{array} & & \setlength{\arraycolsep}{1pt}\begin{array}[c]{rcl}x &\mapsto & \oenum_1\\ y &\mapsto & f\oenum_1\\ z &\mapsto & \oenum_0\end{array} \\[5ex]
& & X (f\oenum_0) & & & X (f\oenum_0) & & & X (f\oenum_1) & X (f\oenum_1) & & X (f\oenum_0) \\
\end{array}
\end{equation*}
For each occurrence of move $\move{m}$ with a non-empty atomic label $[R]$ (there is no atomic label of greater length in the considered arena), we have indicated (below the line) the corresponding substitution $\asubst_{\move{m}}$ and the associated formula $R\asubst_{\move{m}}$.

It is thus easy to check that $\mu$-pointers validate the condition on atomic labels given in the definition of play.
The formula $R\asubst_{\move{m}}$ gives some dynamic content of the move which depends on the position in the play (thus in a proof on the syntactic side). A comment on the logical meaning of this is given in the beginning of Section~\ref{secconcl}.
\end{exa}

We define various notions of projections of sequences of moves.

If \play{s} is an instantiated justified sequence on $A\imp B$, $\proj{\play{s}}{A}$ (resp.\ $\proj{\play{s}}{B}$) is the subsequence (with some pointers and some instantiations) of \play{s} containing the moves belonging to $A$ (resp.\ $B$), with their justification pointers (except for initial moves of $A$ which do not have justification pointers anymore), with their $\mu$-pointers going to moves in $A$ (resp.\ $B$) (the others disappear), and with their instantiations.
It is an instantiated justified sequence.

If \play{u} is an interaction sequence on $A$, $B$ and $C$, we define the following sequences (with some pointers and some instantiations):
\begin{enumerate}[$\bullet$]
\item $\proj{\play{u}}{A\imp B}$ is the subsequence of \play{u} containing moves in $A$ and moves in $B$ with their pointers (if they arrive to a move in $A$ or $B$ and are not $\mu$-pointers starting from a Player move of $B$ in \play{u}) and with their instantiation for moves in $A$, and their \ovar-instantiation for \ppol-moves in $\play{u}$ played in $B$ and their \pvar-instantiation for \opol-moves in $\play{u}$ played in $B$.
\item $\proj{\play{u}}{B\imp C}$ is the subsequence of \play{u} containing moves in $B$ and moves in $C$ with their pointers (if they arrive to a move in $B$ or $C$ and are not $\mu$-pointers starting from an Opponent move of $B$ in \play{u}) and with their instantiation for moves in $C$, their \ovar-instantiation for \opol-moves in $\play{u}$ played in $B$ and their \pvar-instantiation for \ppol-moves in $\play{u}$ played in $B$.
\item $\proj{\play{u}}{A\imp C}$ is the subsequence of \play{u} containing moves in $A$ and moves in $C$ with their justification pointer if it arrives to a move in $A$ or $C$.

For any initial move \move{m} in $A$, whose justifier must be an initial move $\move{m}'$ in $B$ itself justified by an initial move $\move{m}''$ in $C$, we put $\move{m}''$ as justifier of \move{m}.

The $\mu$-pointers of this justified sequence are given by: we put a $\mu$-pointer from the formula $R$ associated with the occurrence of move \move{m} to the formula $S$ associated with the occurrence of move \move{n}, if there exists a sequence of $\mu$-pointers $p_1,\dots,p_n$ ($n>0$) in $\play{u}$ such that:
  \begin{enumerate}[$-$]
  \item the source of $p_1$ is $R$ associated with the occurrence of move \move{m}
  \item the target of $p_n$ is $S$ associated with the occurrence of move \move{n}
  \item the source of $p_i$ is the target of $p_{i-1}$ ($2\leq i\leq n$)
  \item the target of $p_i$ is in $B$ ($1\leq i\leq n-1$)
  \end{enumerate}
This means that we find a path of $\mu$-pointers from $R$ to $S$ going only through labels of moves in $B$ (if the path contains only one edge, it has not to go through $B$).

Since with any move of \play{u} in $B$ are associated both an \ovar-instantiation and a \pvar-instantiation (of the same length), we can define an \ovar-substitution $\osubst$: the \ovar-variable $x$ is substituted by $t$ if $x$ appears in $k$th position in the \ovar-instantiation of an occurrence of move \move{m} of \play{u} in $B$ and $t$ is the $k$th element of the \pvar-instantiation of \move{m}.
The instantiations in $\proj{\play{u}}{A\imp C}$ are obtained from the instantiations in $\play{u}$ by applying $\osubst$.
\end{enumerate}
The objective of these projections of interaction sequences is to extract candidate plays for $A\imp B$, $B\imp C$ and $A\imp C$, as given in the definition of the composition of strategies below.

\begin{defi}[Strategy]
A \emph{strategy} $\sigma$ on the arena $A$, denoted $\sigma:A$, is a non-empty set of even length plays which is closed under even length prefixes, and:
\begin{enumerate}[$\bullet$]
\item \emph{deterministic}: if $\play{s}\move{m}\in\sigma$ and $\play{s}\move{n}\in\sigma$ then $\play{s}\move{m}=\play{s}\move{n}$;
\item \emph{uniform}: if $\play{s}\in\sigma$ and $\oren$ is an \ovar-renaming then $\ren{\play{s}}{\oren}\in\sigma$.
\end{enumerate}
\end{defi}

\noindent
A particular kind of strategy playing $\mu$-pointers and instantiations in a very constrained way is useful. A play is \emph{$\mu$-rigid} if:\label{defmurigid}
\begin{enumerate}[$\bullet$]
  \item the atomic label of a Player move always has the same length as the atomic label of the previous move,
  \item a $\mu$-pointer is always going to the corresponding element of the atomic label of the previous move,
  \item the instantiation of a Player move is always the same as the instantiation of the previous move.
\end{enumerate}
A strategy is \emph{$\mu$-rigid} if all its plays are.

In order to define a category, we consider the following identities and composition.

If $A$ is an arena, the \emph{identity} $\id{A}$ on $A\imp A$ is given by:
\begin{multline*}
\id{A} = \{\play{s}\in\pplays{A_1\imp A_2} \mid \forall \play{t}\ppref\play{s}, \proj{\play{t}}{A_1}=\proj{\play{t}}{A_2}\\
\wedge \text{$\mu$-pointers are going to the corresponding element of the previous move}\}
\end{multline*}
It contains only $\mu$-rigid plays.

If $\sigma:A\imp B$ and $\tau:B\imp C$ are two strategies, the \emph{composition} of $\sigma$ and $\tau$ is given on $A\imp C$ by:
\begin{equation*}
\sigma\fol\tau = \{\proj{\play{u}}{A\imp C}\in\pplays{A\imp C} \mid \play{u}\in\int{A,B,C} \wedge \proj{\play{u}}{A\imp B}\in\sigma \wedge \proj{\play{u}}{B\imp C}\in\tau\}
\end{equation*}

These two constructions give rise to strategies and we obtain a category of arenas and strategies (see Appendix~\ref{appcat}).

\subsection{Innocence}

In order to restrict the set of strategies to those corresponding to proofs in first-order logic, we introduce the notions of view and innocence.

\begin{defi}[View]
A \emph{view} on the arena $A$ is a play \play{s} on $A$ such that:
\begin{enumerate}[$\bullet$]
\item Opponent moves in \play{s} are all $\lambda$-justified by the preceding move;
\item the list of \ovar-variables played by Opponent (obtained by concatenating the \ovar-instantiations in \play{s} according to the order in which they appear in \play{s}) is a prefix of the enumeration $(\oenum_i)_{i\in\Nat}$.
\end{enumerate}
\end{defi}

\noindent
The condition on \ovar-variables is related with the notion of \emph{skeleton} in~\cite{felschergames}.

If $\play{s}$ is an instantiated justified sequence, the \emph{pre-view} $\ppreview{\play{s}}$ of $\play{s}$ is defined by:
$\ppreview{\eps}=\eps$,
$\ppreview{\play{s}\move{m}}=\ppreview{\play{s}}\move{m}$ if $\move{m}$ is a Player move,
$\ppreview{\play{s}\move{m}}=\move{m}$ if $\move{m}$ is an initial Opponent move, 
$\ppreview{\play{s}\move{m}\play{t}\move{n}}=\ppreview{\play{s}\move{m}}\move{n}$ if $\move{n}$ is an Opponent move justified by $\move{m}$.

The \emph{view} $\pview{\play{s}}$ of \play{s} is obtained from its pre-view by applying the \ovar-renaming required to respect the naming condition of views. If $[x_0,x_1,\dots,x_n]$ is the list of \ovar-variables played by Opponent in $\ppreview{\play{s}}$ (obtained by concatenating the \ovar-instantiations in $\ppreview{\play{s}}$ according to the order in which they appear), we consider an \ovar-renaming $\oren$ satisfying $\oren(x_i)=\oenum_i$ for $0\leq i\leq n$ (we call it a \emph{canonical renaming} induced by $\ppreview{\play{s}}$) and we define $\pview{\play{s}}=\ren{\ppreview{\play{s}}}{\oren}$ (the value of $\oren$ outside $\{x_0,\dots,x_n\}$ has no impact).

If $\play{s}$ is a play then $\pview{\play{s}}$ is a view and if $\play{s}$ is a view then $\pview{\play{s}}=\play{s}$.

We choose the presentation of \emph{innocent strategies} based on their underlying view functions.

\begin{defi}[View function]
A \emph{view function} on the arena $A$ is a non-empty set of even length views on $A$ which is closed under even length prefixes and deterministic: if $\play{s}\move{m}\in\sigma$ and $\play{s}\move{n}\in\sigma$ then $\play{s}\move{m}=\play{s}\move{n}$.
\end{defi}

Note that a view function is not a strategy since it violates the uniformity condition.

Let $\sigma$ be a view function on $A$, its \emph{view closure} $\vc{\sigma}$ is given by $\eps\in\vc{\sigma}$, and if $\play{s}\in\vc{\sigma}$, $\play{s}\move{m}\move{n}\in\plays{A}$ and $\pview{\play{s}\move{m}\move{n}}\in\sigma$ then $\play{s}\move{m}\move{n}\in\vc{\sigma}$.
\begin{lem}[View closure]
If $\sigma$ is a view function then 
$\vc{\sigma}$ is a strategy.
\end{lem}

\proof
By definition, $\vc{\sigma}$ is a non-empty \ppol-prefix closed set of even-length plays. If $\play{s}\move{m}\in\vc{\sigma}$ and $\play{s}\move{n}\in\vc{\sigma}$ then $\pview{\play{s}\move{m}}=\pview{\play{s}}\move{m}_0\in\sigma$ and $\pview{\play{s}\move{n}}=\pview{\play{s}}\move{n}_0\in\sigma$ (where $\move{m}_0$ and $\move{n}_0$ are obtained from $\move{m}$ and $\move{n}$ by applying a canonical \ovar-renaming induced by $\pview{\play{s}}$) thus $\pview{\play{s}\move{m}}=\pview{\play{s}\move{n}}$ by determinism of $\sigma$ and finally $\play{s}\move{m}=\play{s}\move{n}$.

By induction on the length of $\play{s}$, we can show that $\play{s}\in\vc{\sigma}$ implies $\ren{\play{s}}{\oren}\in\vc{\sigma}$ for any \ovar-renaming $\oren$. This is an easy consequence of $\pview{\play{s}}=\pview{\ren{\play{s}}{\oren}}$.
\qed

Composition of view functions is given by:
$\sigma\fol\tau = \{\pview{\play{s}}\mid \play{s}\in\vc{\sigma}\fol\vc{\tau}\}$, and the identity view function is $\pview{\id{}}$.

An \emph{innocent strategy} is a strategy obtained as the view closure of a view function.
We will now consider only innocent strategies and just say ``strategy''. Moreover we will mainly say ``strategy'' for the underlying view function.

\begin{prop}[Category of innocent games]\label{propcat}
Arenas and view functions give a category $\Gcat$.
\end{prop}

\proof
All the technical results on strategies corresponding to the categorical structure are developed in Appendix~\ref{appcat}.
\qed

\subsection{Constructions}

The notion of arrow arena was already required to define morphisms between arenas. We now turn to other constructions on arenas and strategies to describe the richer structure of the category of games: a control category~\cite{controlcat}. We first start with the propositional constructions.

\paragraph{Arena constructions.}
Let $A$ and $B$ be two arenas:
  \begin{description}
  \item[Empty] The \emph{empty arena} $\top$ is the empty forest.
  \item[Unit] The \emph{unit arena} $\bot$ is the forest with only one tree with only one node \umove{} (empty labels).
  \item[Atom] If $R$ is a non-constant atomic formula, the corresponding \emph{atomic arena} $R$ is the unit arena with $[R]$ as atomic label for its unique node (empty first-order label).
  \item[Sum] The \emph{sum} $A+B$ of $A$ and $B$ is the concatenation of $A$ and $B$.
  \item[Product] The \emph{product} $A\times B$ of $A$ and $B$ is the merging of $A$ and $B$. The labels of roots are obtained by concatenation from the labels of the corresponding roots in $A$ and $B$ (the first-order labels of $A$ and $B$ are supposed to be disjoint). A move in $A\times B$ is represented as a pair of moves $(\move{m},\move{n})$ of $A$ and $B$ such that at least one is initial.
  \end{description}

\begin{exa}\label{exprodarena}
Starting from the arena $A$ of Example~\ref{exarena}, the product $A\times A$ (where we put ``primes'' on the second copy) is the arena:
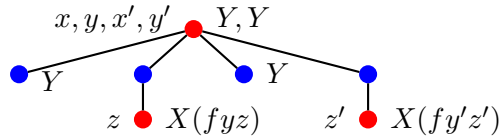
\begin{treeenv}
  \pstree{\onod{\raisebox{4pt}{$x,y,x',y'$}}{\raisebox{4pt}{$Y,Y$}}}{\pnod{}{\raisebox{-3pt}{$Y$}} \pstree{\pnod{}{}}{\onod{z}{X (f y z)}} \pnod{}{Y} \pstree{\pnod{}{}}{\onod{z'}{X (f y' z')}}}
\end{treeenv}
with root named $(\move{a}_0,\move{a}'_0)$, its sons named $(\move{a}_1,\move{a}'_0)$, $(\move{a}_2,\move{a}'_0)$, $(\move{a}_0,\move{a}'_1)$ and $(\move{a}_0,\move{a}'_2)$, the son of $(\move{a}_2,\move{a}'_0)$ is $(\move{a}_3,\move{a}'_0)$ and the son of $(\move{a}_0,\move{a}'_2)$ is $(\move{a}_0,\move{a}'_3)$.
\end{exa}

\paragraph{Strategy constructions.}

\begin{defi}[Linear strategy]
Let $\sigma$ be a view function on $A\imp B$, $\sigma$ is \emph{linear} if:
\begin{enumerate}[$\bullet$]
\item for each initial move \move{m} in $B$, there is a play $\move{m}\move{n}$ in $\sigma$ with \move{n} in $A$
\item for each view $\move{m}\move{n}\play{s}$ in $\sigma$, \move{n} is the unique move in $A$ justified by \move{m}
\end{enumerate}
\end{defi}

\noindent
Let $\sigma:A\imp C$ and $\tau:B\imp D$ be two view functions:
  \begin{description}
  \item[Sum] The view function $\sigma+\tau$ is obtained by the union of the view functions. 
Its view closure is $\{\play{s}\in\pplays{A+B\imp C+D}\mid\proj{\play{s}}{A\imp C}\in\vc{\sigma}\wedge\proj{\play{s}}{B\imp D}\in\vc{\tau}\}$.
If both $\sigma$ and $\tau$ are linear then $\sigma+\tau$ is linear.
  \item[Product] Assume $\sigma:A\imp C$ is linear. For each initial move \move{c$_0$} in $C$, there is a unique move \move{a$_0$} in $A$ such that \play{c$_0$a$_0$} belongs to $\sigma$. A view \play{s} in $A\times B\imp C\times D$ \emph{respects} $\sigma$ if, for \move{(c$_0$,d$_0$)} its initial move, any move \move{(a,b)} in \play{s} with \move{a} initial satisfies $\move{a}=\move{a$_0$}$ (with $\play{c$_0$a$_0$}\in\sigma$ and \move{a$_0$} justified by \move{c$_0$}) and any move \move{(c,d)} in \play{s} with \move{c} initial satisfies $\move{c}=\move{c$_0$}$. If $\play{s}$ respects $\sigma$, $\prodproj{\play{s}}{\tau}$ is obtained by replacing any move \move{(a$_0$,b)} by \move{b} and any move \move{(c$_0$,d)} by \move{d} (with the appropriate pointers and instantiations) and by removing the other moves. If $\play{s}$ respects $\sigma$, we consider $\play{s}_0$ obtained by replacing any move \move{(a,b$_0$)} with \move{a} non initial in $A$ and \move{b$_0$} initial in $B$ by \move{a} and any move \move{(c,d$_0$)} with \move{c} non initial in $C$ and \move{d$_0$} initial in $D$ by \move{c} (with the appropriate pointers and instantiations). We define $\prodproj{\eps}{\sigma}=\eps$ and, if $\play{s}$ is not empty, $\prodproj{\play{s}}{\sigma}$ is $\move{c$_0$}\move{a$_0$}\play{s$_0$}$ in which \move{a$_0$} is justified by \move{c$_0$} and the moves of $\play{s$_0$}$ enabled by \move{a$_0$} in $A$ are justified by \move{a$_0$}.
We define $\sigma\times\tau=\{\play{s}\in\pplays{A\times B\imp C\times D}\mid\forall \play{t}\ppref\play{s}, \text{$\play{t}$ is a view respecting $\sigma$} \wedge \prodproj{\play{t}}{\sigma}\in\sigma\wedge\prodproj{\play{t}}{\tau}\in\tau\}$.

If $\tau$ is linear (but $\sigma$ is not) we can proceed in a symmetric way for defining $\sigma\times\tau$.
If both $\sigma$ and $\tau$ are linear, the two definitions coincide and $\sigma\times\tau$ is linear.
  \item[Projections] The linear view function $\pview{\id{A}}:A\imp A$ is also a linear view function on $A+B\imp A$ and on $B+A\imp A$.
  \item[Diagonal] We can consider moves in $A\imp B$ as moves in $A\imp B+B$ by identifying the original $B$ with either the left one or the right one. In this way we can see $\pview{\id{A}}$ as a set of plays $\id{A}^1$ in $A\imp A+A$ by considering the left embedding and also as a set of plays $\id{A}^2$ in $A\imp A+A$ by considering the right embedding. The linear view function $\diag{A}$ on $A\imp A+A$ is the union of $\id{A}^1$ and $\id{A}^2$.
  \item[Weakening] The linear view function $\weak{A}$ on $\bot\imp A$ is $\{\eps\}\cup\{\move{m}\umove\mid\text{\move{m} initial in $A$}\}$ (this means that $\move{m}$ comes with the unique possible $\ovar$-instantiation leading to a view and that $\umove$ is justified by $\move{m}$).
  \item[Contraction] We consider a play \play{s} in $A\times A\imp A$ (we add indexes: $A_1\times A_2\imp A_0$ to make things clearer). An occurrence of move in $A_0$ is called a left move if the previous move in $A_1\times A_2$ was in $A_1$ (and the same with ``right move'' and $A_2$). If only moves from $A_0$ were played before, we consider it both as a left move and as a right move. We define $\play{s}_l$ as the subsequence of \play{s} containing \move{m}: if $(\move{m},\move{n})$ is an occurrence of move in $A_1\times A_2$ with \move{n} initial in $A_2$, or if \move{m} is a left move in $A_0$. $\play{s}_r$ is given in a symmetric way. $\play{s}_l$ and $\play{s}_r$ can be seen as sequences of moves in $A\imp A$.
The linear view function $\contr{A}$ on $A\times A\imp A$ is $\{\play{s}\in\pplays{A\times A\imp A}\mid\text{\play{s} view}\wedge\play{s}_l\in\id{A}\wedge\play{s}_r\in\id{A}\}$.
  \end{description}

\begin{exa}
The definition of product gives the following kind of view in $\sigma\times\tau$:
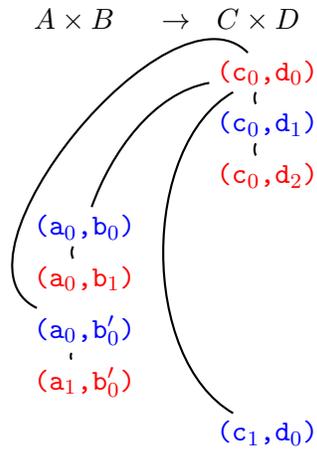
\begin{equation*}
\begin{playenv}{lcl}{A \times B & \imp & C\times D}
    & & \omove[0]{(c$_0$,d$_0$)} \\
    & & \pmove{(c$_0$,d$_1$)} \\
    & & \omove[2]{(c$_0$,d$_2$)} \\
    \pmove{(a$_0$,b$_0$)} \\
    \omove[4]{(a$_0$,b$_1$)} \\
    \renewcommand{\Garcangle}{100}\pmove{(a$_0$,b$'_0$)} \\
    \omove[6]{(a$_1$,b$'_0$)} \\
    & & \renewcommand{\Garcangle}{60}\pmove{(c$_1$,d$_0$)}
\end{playenv}
\end{equation*}
where $\sigma:A\imp C$ is linear and contains the view:
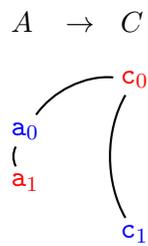
\begin{equation*}
\begin{playenv}{lcl}{A & \imp & C}
    & & \omove[0]{c$_0$} \\
    \pmove{a$_0$} \\
    \omove[2]{a$_1$} \\
    & & \pmove{c$_1$}
\end{playenv}
\end{equation*}
and $\tau:B\imp D$ contains the view:
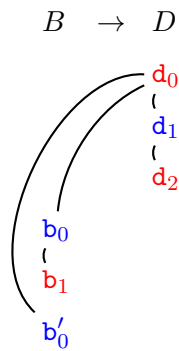
\begin{equation*}
\begin{playenv}{lcl}{B & \imp & D}
    & & \omove[0]{d$_0$} \\
    & & \pmove{d$_1$} \\
    & & \omove[2]{d$_2$} \\
    \pmove{b$_0$} \\
    \omove[4]{b$_1$} \\
    \renewcommand{\Garcangle}{70}\pmove{b$'_0$}
\end{playenv}
\end{equation*}
\end{exa}

\begin{exa}
The (quite complicated) definition of contraction gives the following (simple) views:
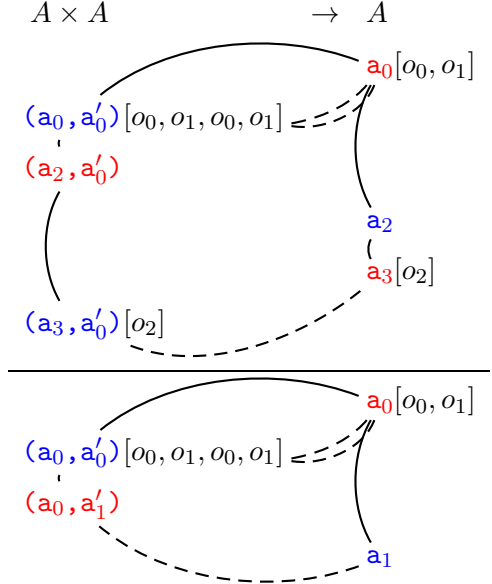
\begin{equation*}
\begin{playenv}{lcl}{\;A \times A & \imp & A}
    & & \omove[0]{a$_0$}[\oenum_0,\oenum_1] \\
    \pmove{(a$_0$,a$'_0$)}[\oenum_0,\oenum_1,\oenum_0,\oenum_1]\pnode{h}\ncarc[linecolor=\mucolname,linestyle=dashed,arcangle=-20,nodesep=\Gnodesep]{-}{h}{play1}\ncarc[linecolor=\mucolname,linestyle=dashed,arcangle=-40,nodesep=\Gnodesep]{-}{h}{play1} \\
    \omove[2]{(a$_2$,a$'_0$)} \\
    & & \pmove{a$_2$} \\
    & & \omove[4]{a$_3$}[\oenum_2] \\
    \pmovem[3]{5}{(a$_3$,a$'_0$)}[\oenum_2] \\[2ex]
\nextplay
    & & \omove[0]{a$_0$}[\oenum_0,\oenum_1] \\
    \pmove{(a$_0$,a$'_0$)}[\oenum_0,\oenum_1,\oenum_0,\oenum_1]\pnode{h}\ncarc[linecolor=\mucolname,linestyle=dashed,arcangle=-20,nodesep=\Gnodesep]{-}{h}{play1}\ncarc[linecolor=\mucolname,linestyle=dashed,arcangle=-40,nodesep=\Gnodesep]{-}{h}{play1} \\
    \omove[2]{(a$_0$,a$'_1$)} \\
    & & \renewcommand{\Garcanglem}{30}\pmovem{3}{a$_1$}
\end{playenv}
\end{equation*}
where $A$ is the arena of Example~\ref{exarena} and $A\times A$ is described in Example~\ref{exprodarena}.
\end{exa}

\begin{thm}[Control category of games]\label{thmcontrolcat}
The category $\Gcat$ of arenas and view functions is a control category.
\end{thm}

In this control category, central morphisms are linear strategies.

\proof
All the technical results on strategies corresponding to the control category structure are developed in Appendix~\ref{appcat}.
\qed

\begin{defi}[Total strategy]
A strategy $\sigma:A$ is \emph{total} if whenever $\play{s}\in\sigma$ and $\play{s}\move{m}\in\plays{A}$, there exists some $\play{s}\move{m}\move{n}$ in $\sigma$.
\end{defi}

A total strategy is \emph{maximal} for inclusion: if $\sigma$ is total and $\sigma\subseteq\tau$ then $\sigma=\tau$.

\begin{defi}[Finite strategy]
The \emph{size} of a strategy is the sum of the lengths of its views.
A strategy is \emph{finite} if its size is finite.
\end{defi}

The identity strategy is total and finite, and total and finite strategies compose (see Appendix~\ref{appcat}). This allows us to define the sub-category $\Gcattf$ of $\Gcat$ containing only total finite strategies, which is also a control category (easy to check).

We now turn to the constructions corresponding to quantification.

\paragraph{First-order constructions.}
Concerning arenas, if $A$ is an arena, $x$ is a $\pvar$-variable and $y$ is a fresh $\avar$-variable, the \emph{quantification} $\forall y \subst{A}{y}{x}$ is obtained by renaming $x$ into $y$ in $A$ and then by pushing $y$ on the first-order label of each root. We will sometimes use the notation $\forall x A$ for this arena (since the particular choice of the name $y$ is not important, see the discussion on bound variables in Section~\ref{secarena}).

\begin{exa}\label{exarenaform}
The arena of Example~\ref{exarena} is the interpretation of the formula $\forall x\forall y(Y\imp(\forall z X (f y z)\imp\bot)\imp Y)$.
\end{exa}

Let $\sigma:A\imp B$ be a view function (with a \pvar-variable $x\notin A$), $\forall x.\sigma$ is the view function on $A\imp\forall y\subst{B}{y}{x}$ given by: $\forall x.\sigma=\{\eps\}\cup\{\subst{\subst{(\move{m}[x]\play{s})}{\oenum_{i+1}}{\oenum_i}}{\oenum_0}{x}\mid\move{m}\play{s}\in\sigma\}$ where $\move{m}[x]$ is obtained from \move{m} by pushing $x$ on its instantiation. If $\sigma$ is linear then $\forall x.\sigma$ is still linear.

Let $A$ be an arena, the \emph{linear} view function $\inst{t}$ on $\forall x A\imp\subst{A}{t}{x}$ is given by: $\inst{t}=\{\play{s}\in\pplays{\forall x A\imp\subst{A}{t}{x}} \mid \text{\play{s} view}\wedge \forall \play{t}\ppref\play{s}, \proj{\play{t}}{\forall x A}[t]=\proj{\play{t}}{\subst{A}{t}{x}}\}$ where $\play{t}[t]$ is obtained from $\play{t}$ by pushing $t$ on the instantiation of its initial move.

\subsection{Interpretation of the \texorpdfstring{\lmcalc}{lambda-mu-calculus}}
\label{secinterp}

A typing derivation ending with a judgment $\A\vdash M:A\mid\B$ is interpreted as a strategy $\sem{M}$ on $\sum\A\imp A\times\prod\B$. Using Theorem~\ref{thmcontrolcat}, there is a canonical way of interpreting the usual propositional constructions of the \lmcalc{} in our model (following~\cite{controlcat}). Moreover this ensures the validation of those of the $\beta\eta\mu\rho\theta$ equalities which are not dealing with first-order constructs.

For the interpretation of the ($\forall$-introduction) rule, we transform $\sigma$ into $\forall x.\sigma$ (since the arenas $\forall x (A\times B)$ and $(\forall x A)\times B$ are isomorphic if $x\notin B$).
For the interpretation of the ($\forall$-elimination) rule, we transform $\sigma$ into $\sigma\fol\inst{t}$.

\begin{lem}[First-order correctness]
The following equalities are valid through the interpretation in games:
\begin{equation*}
\begin{array}{rlllcl}
(\Lambda x.M)\foapp{t}      & =_\beta  & \subst{M}{t}{x} & :A \\
\Lambda x.M\foapp{x}        & =_\eta   & M               & :\forall x A & & x \notin M \\
(\mu\alpha.M)\foapp{t}      & =_\mu    & \mu\alpha.\subst{M}{[\alpha]L\foapp{t}}{[\alpha]L} & :A
\end{array}
\end{equation*}
\end{lem}

\proof\hfill
  \begin{enumerate}[$\bullet$]
  \item $(\Lambda x.M)\foapp{t}$: The view function interpreting this term is $\forall x.\sigma\fol(\inst{t}\times\id{\B})$ which is $\{\subst{\play{s}}{t}{x}\mid\play{s}\in\sigma\}$. One easily checks it is also the interpretation of $\subst{M}{t}{x}$.
  \item $\Lambda x.M\foapp{x}$: An immediate computation shows the interpretation of this term to be the same as the interpretation of $M$.
  \item $(\mu\alpha.M)\foapp{t}$: this case is a consequence of the centrality of the morphism $\inst{t}$ in the control category of games (see~\cite[Chapter~7]{phddelataillade} for example).
  \qed
  \end{enumerate}

\noindent
We denote by $\Gcata$ (resp.\ $\Gcatatf$) the full sub-category of $\Gcat$ (resp.\ $\Gcattf$) containing only arenas with at most one element in the atomic labels of moves (called \emph{$1$-arenas}). According to the previous interpretation of typing derivations, these categories are expressive enough to interpret the closed terms of our \lmcalc{}\footnote{These two categories are not control categories, since the product of two $1$-arenas is not a $1$-arena in general.
The existence of surrounding control categories would allow us to extend the \lmcalc{} with a disjunction connective in types. We prefer not to do it since the calculus would become even more complex.} and we will mainly focus on them in the sequel.

\begin{exa}
The interpretation of the closed term $\lambda a.a\foapp{t}$ of type $\forall x Xx\imp Xt$ is the view function containing the empty view and the view:
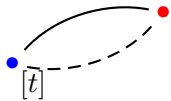
\begin{equation*}
\begin{playenv}{ccc}{\forall x Xx & \imp & Xt}
& & \ob[0] \\
\pbm{1}\raisebox{-8pt}{$[t]$}
\end{playenv}
\end{equation*}

The interpretation of the closed term:
\begin{equation*}
\lambda f.(f\foapp{x})\Lambda y.\lambda d.\mu\alpha.(f\foapp{y})\Lambda z.\lambda a.\mu\delta[\alpha]a
\end{equation*}
of type $\forall x(\forall y(Xx\imp Xy)\imp\bot)\imp\bot$ is the view function containing the following unique maximal view:
\begin{equation*}
\begin{playenv}{ccccccccccccc}{\forall x & ( & \forall y & ( & Xx & \imp & Xy & ) & \imp & \bot & ) & \imp & \bot}
& & & & & & & & & & & & \ob[0] \\
& & & & & & & & & \pb{}[x] \\
& & & & & & \ob[2][\oenum_0] \\
& & & & & & & & & \pb{}[\oenum_0] \\
& & & & & & \ob[4][\oenum_1] \\
& & & & \renewcommand{\Garcanglem}{30}\pbm[5]{3}
\end{playenv}
\end{equation*}
\end{exa}

In order to prove the completeness of the model, instead of working by induction on the size of strategies and of building incrementally the corresponding term, we will use a more geometric and global approach through $\lambda\mu$-forests (an intermediate notion between terms and strategies in the spirit of B\"ohm trees --- a similar approach is used in~\cite{gamescpcf}).
Let us start with the simple case of formulas and arenas. A formula in canonical form $\bigwedge_{1\leq i\leq n} \forall\vec{x}(A^i_1\imp\cdots\imp A^i_{k_i}\imp R^i)$ can be rewritten into $\bigwedge_{1\leq i\leq n} [R^i,\vec{x}](A^i_1,\dots,A^i_{k_i})$. By considering $[R^i,\vec{x}]$ as an operator/constructor with $k_i$ arguments and by looking at the forest given from the syntactic trees of the $[R^i,\vec{x}](A^i_1,\dots,A^i_{k_i})$, we obtain nothing but the arena associated with the original formula. This shows in particular that nodes in the arena are in bijection with occurrences of atomic formulas in a canonical form.

\begin{exa}
  The formula given in Example~\ref{exarenaform} and interpreted by the arena of Example~\ref{exarena} would be represented as $[Y,x,y]([Y],[\bot]([X (f y z),z]))$.
\end{exa}

We now develop the same kind of correspondence at the level of terms, $\lambda\mu$-forests and strategies.

\begin{defi}[$\lambda\mu$-forest]
A \emph{$\lambda\mu$-forest} is a forest with two additional disjoint finite sets of edges --- $\lambda$-edges (labelled with a natural number) and $\mu$-edges --- and with a list of \ovar\pvar-terms associated to each node and $\lambda$- or $\mu$-variables associated to some nodes, satisfying:
\begin{enumerate}[$\bullet$]
\item The nodes of even polarity have exactly one son.
\item The source of an edge is always a node of odd polarity and the target is always of even polarity. Moreover the target is above the source.
\item A node cannot be the source of two different $\lambda$-edges or of two different $\mu$-edges.
\item If a node of odd polarity is not the source of a $\lambda$-edge, it is labelled with a $\lambda$-variable. If a node of odd polarity is not the source of a $\mu$-edge, it may be labelled with a $\mu$-variable.
\item The list of terms associated with a node of even polarity is a list of \ovar-variables.
\item An \ovar-variable appearing in an \ovar\pvar-term of the list associated with a node of odd polarity must appear above in the list associated with a node of even polarity.
\item The list of \ovar-variables appearing along a branch (obtained by concatenating the lists associated with the nodes of even polarity in the branch according to the order in which they appear) is a prefix of the enumeration $(\oenum_i)_{i\in\Nat}$.
\end{enumerate}
If there is no node labelled with a $\lambda$- or $\mu$-variable,
the $\lambda\mu$-forest is \emph{closed}.
\end{defi}

With any \lmterm{} in canonical normal form is naturally associated a $\lambda\mu$-forest:
\begin{enumerate}[$\bullet$]
\item with $\cst$ is associated the empty forest
\item with a tuple of \lmterm s is associated the forest whose trees correspond to each \lmterm
\item with a \lmterm{} $\Lambda\vec{x}.\lambda\vec{a}.\underline{\mu\alpha[\beta]}(b\foapp{\vec{t}})\vec{M}$ is associated the following tree: we first consider the tree with a root \move{r} which has one son \move{n} whose sons are the trees corresponding to the $\vec{M}$s then
  \begin{enumerate}[$-$]
  \item we put the labels $b$ and $\beta$ on \move{n}
  \item we apply the substitution $[^{\oenum_{i+k}}/_{\oenum_i}\mid i\in\Nat]$ (with $\vec{x}=x_1\dots x_k$)
  \item we associate the list $[\vec{x}]$ with \move{r} and the list $[\vec{t}]$ with \move{n}
  \item we apply the substitution $[^{\oenum_0}/_{x_1},\dots,^{\oenum_{k-1}}/_{x_k}]$
  \item for each node labelled with the $\lambda$-variable $a_i$ ($\vec{a}=a_1\dots a_k$), we remove the label and we put a $\lambda$-edge with target \move{r} and label $i$
  \item for each node labelled with the $\mu$-variable $\alpha$, we remove the label and we put a $\mu$-edge with target \move{r}
  \end{enumerate}
\end{enumerate}
If the \lmterm{} is closed then the associated $\lambda\mu$-forest is closed.

\begin{exa}\label{exlmforest}
We represent $\lambda$-edges by plain edges and $\mu$-edges by dashed edges in $\lambda\mu$-forests.

\medskip

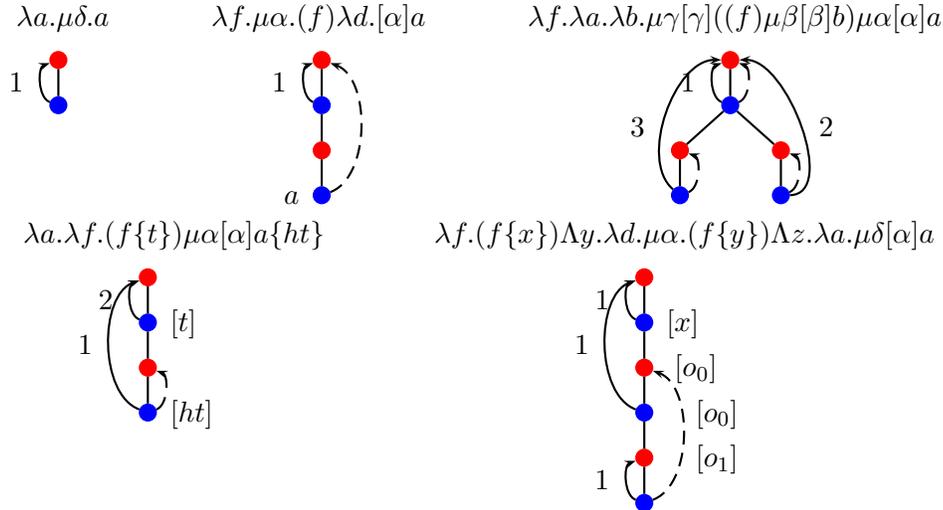
\hfil
\begin{tabular}[t]{c}
$\lambda a.\mu\delta.a$ \\
\pstree{\onod[,name=r]{}{}}{\pnod[,name=n]{}{}}
\ncarc[arcangle=70]{->}{n}{r}^1
\end{tabular}
\hfil
\begin{tabular}[t]{c}
$\lambda f.\mu\alpha.(f)\lambda d.[\alpha]a$ \\
\pstree{\onod[,name=r]{}{}}{\pstree{\pnod[,name=f]{}{}}{\pstree{\onod{}{}}{\pnod[,name=a]{a}{}}}}
\ncarc[arcangle=70]{->}{f}{r}^1
\ncarc[arcangle=-70,linestyle=dashed]{->}{a}{r}
\end{tabular}
\hfil
\begin{tabular}[t]{c}
$\lambda f.\lambda a.\lambda b.\mu\gamma[\gamma]((f)\mu\beta[\beta]b)\mu\alpha[\alpha]a$ \\
\pstree{\onod[,name=r]{}{}}{\pstree{\pnod[,name=f]{}{}}{\pstree{\onod[,name=alpha]{}{}}{\pnod[,name=a]{}{}} \pstree{\onod[,name=beta]{}{}}{\pnod[,name=b]{}{}}}}
\ncarc[arcangle=70]{->}{f}{r}^1
\ncarc[arcangle=70]{->}{a}{r}^3
\ncarc[arcangleA=-110,arcangleB=-70]{->}{b}{r}_2
\ncarc[arcangle=-70,linestyle=dashed]{->}{f}{r}
\ncarc[arcangle=-70,linestyle=dashed]{->}{a}{alpha}
\ncarc[arcangle=-70,linestyle=dashed]{->}{b}{beta}
\end{tabular}
\hfil

\medskip

\hfil
\begin{tabular}[t]{c}
$\lambda a.\lambda f.(f\foapp{t})\mu\alpha[\alpha]a\foapp{h t}$ \\
\pstree{\onod[,name=r]{}{}}{\pstree{\pnod[,name=f]{}{[t]}}{\pstree{\onod[,name=alpha]{}{}}{\pnod[,name=a]{}{[h t]}}}}
\ncarc[arcangle=70]{->}{f}{r}^2
\ncarc[arcangle=70]{->}{a}{r}^1
\ncarc[arcangle=-70,linestyle=dashed]{->}{a}{alpha}
\end{tabular}
\hfil
\begin{tabular}[t]{c}
$\lambda f.(f\foapp{x})\Lambda y.\lambda d.\mu\alpha.(f\foapp{y})\Lambda z.\lambda a.\mu\delta[\alpha]a$ \\
\pstree{\onod[,name=r]{}{}}{\pstree{\pnod[,name=f]{}{[x]}}{\pstree{\onod[,name=alpha]{}{\;[\oenum_0]}}{\pstree{\pnod[,name=fbis]{}{\quad[\oenum_0]}}{\pstree{\onod[,name=delta]{}{\quad[\oenum_1]}}{\pnod[,name=a]{}{}}}}}}
\ncarc[arcangle=70]{->}{f}{r}^1
\ncarc[arcangle=70]{->}{fbis}{r}^1
\ncarc[arcangle=70]{->}{a}{delta}^1
\ncarc[arcangle=-70,linestyle=dashed]{->}{a}{alpha}
\end{tabular}
\hfil

\noindent All these $\lambda\mu$-forests, except the second one, are closed.
\end{exa}

By translating \lmterm s as $\lambda\mu$-forests, there is a loss of information. For example $\lambda a.a$ and $\lambda a.\lambda b.a$ are both translated as:
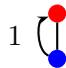
\begin{treeenv}
  \pstree{\onod[,name=r]{}{}}{\pnod[,name=n]{}{}}
  \ncarc[arcangle=70]{->}{n}{r}^1
\end{treeenv}
We have to use types to recover the missing information. A $\lambda\mu$-forest is \emph{typed} if a formula in $\imp$-canonical form (see page~\pageref{defimpcanform}) is associated with each node in such a way that:
\begin{enumerate}[$\bullet$]
\item if the node \move{n} (with formula $A$) is source of a $\lambda$-edge with label $i$ and target \move{r} (with formula $B$ and list of terms $\vec{\oenum}$), then $B=\forall\vec{x}(B_1\imp\cdots\imp B_k\imp S)$ with $A=\subst{B_i}{\vec{\oenum}}{\vec{x}}$;
\item if the node \move{n} (with formula $A$ and list of terms $\vec{t}$) is source of a $\mu$-edge with target \move{r} (with formula $B$ and list of terms $\vec{\oenum}$), then $A=\forall\vec{x}(A_1\imp\cdots\imp A_k\imp R)$ and $B=\forall\vec{z}(B_1\imp\cdots\imp B_p\imp S)$ with $\subst{R}{\vec{t}}{\vec{x}}=\subst{S}{\vec{\oenum}}{\vec{z}}$;
\item if the node \move{n} of odd polarity has formula $A=\forall\vec{x}(A_1\imp\cdots\imp A_k\imp R)$ then $R=\bot$ if and only if \move{n} is neither the source of a $\mu$-edge nor labelled with a $\mu$-variable;
\item if the node \move{r} of even polarity (with formula $A$) is the $i$th son of the node \move{n} (with formula $B$ and list of terms $\vec{t}$), then $B=\forall\vec{x}(B_1\imp\cdots\imp B_k\imp S)$ with $A=\subst{B_i}{\vec{t}}{\vec{x}}$.
\end{enumerate}
The type of the $\lambda\mu$-forest is the conjunction of the types of its roots.

We can extend the translation from \lmterm s to $\lambda\mu$-forests with types: when translating $\Lambda\vec{x}.\lambda\vec{a}.\mu\alpha[\beta](b\foapp{\vec{t}})\vec{M}$ of type $A$ with $b$ of type $B$, we associate $A$ with \move{r} and $B$ with \move{n}.

\begin{exa}
The $\lambda\mu$-forests of Example~\ref{exlmforest} can be turned into typed $\lambda\mu$-forests with the following respective types:
\begin{gather*}
  \bot\imp X \\
  ((X\imp\bot)\imp\bot)\imp X \\
  (Y\imp X\imp Z)\imp X\imp Y\imp Z \\
  \forall x X x\imp(\forall x(X (h x)\imp\bot))\imp\bot \\
  \forall x (\forall y(Xx\imp Xy)\imp\bot)\imp\bot
\end{gather*}
\end{exa}

Starting from a typed $\lambda\mu$-forest, we can build a unique typed \lmterm. We decompose the $\lambda\mu$-forest into $\lambda\mu$-trees, we compute the corresponding \lmterm s and the \lmterm{} associated with the $\lambda\mu$-forest is the induced tuple. Concerning a $\lambda\mu$-tree $T$, if the root \move{r} has formula $A=\forall\vec{x}(A_1\imp\cdots\imp A_k\imp R)$, we introduce $k$ fresh $\lambda$-variables $a_1$,~\dots,~$a_k$, and a fresh $\mu$-variable $\alpha$. To any node which is source of a $\lambda$-edge with index $i$ and target \move{r}, we add the label $a_i$. To any node which is source of a $\mu$-edge with target \move{r}, we add the label $\alpha$. Let \move{n} be the son of \move{r} and let $b$ and $\beta$ be the labels obtained for it. Let $\vec{M}$ be the \lmterm s inductively associated with the sub-trees under \move{n}. The \lmterm{} associated with $T$ is $\Lambda\vec{y}\lambda a_1\dots\lambda a_k\underline{\mu\alpha[\beta]}(b\foapp{\vec{t}})\vec{M}$ where $\vec{y}$ are the terms labelling \move{r} and $\vec{t}$ are the terms labelling \move{n}.

With any total finite view function on the arena associated with a type $A$ is associated a closed $\lambda\mu$-forest: we consider views ordered with the prefix ordering (so that moves in the views give nodes in the forest), we remove the $\lambda$-pointers of \opol-moves, the other pointers give the ($\lambda$ and $\mu$) edges of the $\lambda\mu$-forest, the instantiations give the lists of terms. Concerning the labels of $\lambda$-edges, if the node \move{n} corresponds to the move \move{m} corresponding itself to the occurrence $R$ of an atomic formula in $A$ and if $R$ appears in a sub-formula $\forall\vec{x}(B_1\imp\cdots\imp B_k\imp S)$ of $A$ as the final atom of $B_i$ then the $\lambda$-edge with source \move{n} has label $i$.

With any node of a closed typed $\lambda\mu$-forest of type $A$, we can associate a move of the arena associated with $A$:
\begin{enumerate}[$\bullet$]
\item if \move{r} is the $i$th root of the $\lambda\mu$-forest, the corresponding move is the $i$th root of the arena;
\item if the node \move{n} of even polarity has a $\lambda$-edge with label $i$ to a node with associated move \move{m}, the move associated with \move{n} is the $i$th son of \move{m} in the arena;
\item if the node \move{r} of odd polarity is the $i$th son of the node \move{n} with associated move \move{m}, the move associated with \move{r} is the $i$th son of \move{m} in the arena.
\end{enumerate}
In this way, we can associate a view with any branch of a closed typed $\lambda\mu$-forest. Nodes of even polarity become Opponent moves. Nodes of odd polarity become Player moves. $\lambda$-edges give the justification pointers. $\mu$-edges give the $\mu$-pointers. Lists of \ovar\pvar-terms give instantiations. Finally we add justification pointers going from each Opponent move to the preceding one.

\begin{defi}[Arena isomorphism]\label{defariso}
An \emph{arena isomorphism} $f$ from $A$ to $B$ is a bijection between the nodes of $A$ and the nodes of $B$ which respects the order, but also the atomic labels up to the first-order labels: the move $\node{m}$ and the move $f(\node{m})$ must have first-order labels of the same length, this induces a mapping of the elements of the first-order label of $\node{m}$ to the elements of the first-order label of $f(\node{m})$; using this mapping, the atomic label of any node $\node{n}$ must be mapped to the atomic label of $f(\node{n})$.
\end{defi}

\begin{lem}[Isomorphic arenas]\label{lemisoar}
If there exists an arena isomorphism between two arenas, they are isomorphic in the category $\Gcat$.
\end{lem}

\proof
The arena isomorphism $f$ induces a strategy
$\{\play{s}\in\pplays{A\imp B} \mid \text{$\play{s}$ $\mu$-rigid}\wedge\forall \play{t}\ppref\play{s}, f(\proj{\play{t}}{A})=\proj{\play{t}}{B}\}$
which is an isomorphism in $\Gcat$ (see~\cite[Proposition~6]{classisos}).
\qed

\begin{thm}[Equivalence completeness]\label{thmcomp}
The game model is equivalence complete: the categories $\Scat$ and $\Gcatatf$ are equivalent.
\end{thm}

\proof
We have already seen in Section~\ref{secsyntcat} that $\CScat$ is equivalent to $\Scat$.
We want to show that the interpretation functor from $\CScat$ to $\Gcatatf$ defines an equivalence of categories.
We have established translations from typed \lmterm s in canonical normal form to typed $\lambda\mu$-forests (and back) and from typed $\lambda\mu$-forests to view functions on the corresponding arena (and back). One can check that all these correspondences are one-to-one.
Moreover the view function associated with a given typed \lmterm{} in canonical normal form by these correspondences is the same as the one obtained through the interpretation given in the beginning of this section (and coming from the control category structure of $\Gcattf$).
This shows the interpretation functor to be full and faithful.

Finally, any arena is isomorphic to the interpretation of a type in canonical form: by Lemma~\ref{lemisoar}, it is enough to prove that any arena is arena isomorphic to the interpretation of a type in canonical form. This is done by induction on the number of nodes of the arena.
\qed

\begin{cor}[Canonical forms are canonical]\label{corcanform}
Two canonical normal forms equal up to $\beta\eta\mu\rho\theta$ are equal.
\end{cor}

\proof
If $M$ and $N$ are two canonical forms such that $M\simeq_{\beta\eta\mu\rho\theta} N$ then $\sem{M}=\sem{N}$ thus $M=N$ by faithful completeness.
\qed

The concrete meaning of Theorem~\ref{thmcomp} is mainly that any total finite strategy on a $1$-arena is the interpretation of a unique closed \lmterm{} in canonical normal form.

This ends the description of our game model.

\section{Related models}

From the game model of first-order classical logic (and the associated completeness result) described in the previous section, we will define complete models for different systems. Some of these derived models are strongly related with (or even equal to) already known ones.

\subsection{Intuitionistic restriction}
\label{secintgames}

Inside the category $\Gcata$, particular strategies (let us call them \emph{$\lambda$-strategies}) are obtained by asking $\mu$-pointers to always have the preceding move as target ($\mu$-rigid strategies are a particular kind of $\lambda$-strategies). Through the completeness results above (Theorem~\ref{thmcomp}), they correspond to \lmterm s in which the $\mu$ and $[\_]$ constructions are always used together in the shape $\mu\alpha[\alpha]$. By $\theta$ equivalence, such a \lmterm{} is equal to a \lterm{} (all the $\mu\alpha[\alpha]$ can be erased). As a consequence, $\lambda$-strategies provide an equivalence complete model of the Church style first-order \lcalc{} and, thus, of first-order intuitionistic logic.

\subsection{\texorpdfstring{\lcalc}{lambda-calculus} over one/many atom(s)}
\label{secmodellcalc}

We consider the simply typed \lcalc{} over one atom. We interpret this atom exactly as $\bot$. In this case there are no labels on the arenas associated with types (neither atomic formulas, nor first-order variables). Such arenas are called \emph{ground arenas} and the full sub-category $\Gcatgr$ of $\Gcat$ is given by the restriction to ground arenas. $\Gcatgr$ is exactly the category of games presented in~\cite{gamesgoi} since plays are using neither instantiations nor $\mu$-pointers. As shown in~\cite{gamesgoi}, it is a fully complete (and even equivalence complete) game model of the simply typed \lcalc{} over one atom.

To take into account multiple atoms, we go back to arenas with atomic labels on nodes. By considering atoms as $0$-ary relation symbols, types are interpreted as arenas with an atomic label of length $1$ associated with each node. 
Simply typed \lterm s are interpreted as $\lambda$-strategies (introduced just above) and $\mu$-pointers are replaced by the condition that the (unique element of the) atomic label of a Player move is always the same as the atomic label of the previous move (exactly in the spirit of \emph{token-reflecting} strategies of~\cite[page~122]{phdmurawski}).
This extends the model of~\cite{gamesgoi} to an equivalence complete model of the general simply typed \lcalc.

Equivalently this gives a complete game model for $\Pi^1$ formulas. Indeed we can distinguish three levels in full completeness results for logics with propositional atomic formulas and second-order quantification:
\begin{enumerate}[(1)]
\item only constant atomic formulas (or just one atomic formula since it can be identified with $\bot$) thus no quantification;
\item many atomic formulas but no quantification (or equivalently $\Pi^1$ formulas since outermost universal quantification has no impact), this is the level corresponding to propositional logic;
\item general quantification over propositional variables (this is much more difficult and will not be addressed here, see Section~\ref{secconcl} for future work and references).
\end{enumerate}

\subsection{Well bracketed HO/N games (propositional logic)}
\label{secunfolding}

In the original works on HO/N games~\cite{ho,gamesn}, nodes in arenas had an additional labelling with \Qn/\Ar{} labels corresponding to a notion of \emph{questions} and \emph{answers}. We are going to compare the information encoded with questions and answers and the one given through $\mu$-pointers.

In this section we consider games (arenas and strategies) without first-order information.
In order to avoid confusion with strategies as given in~\cite{ho}, (innocent) strategies as used in the previous sections will be called \emph{$\mu$-strategies} here.

\begin{defi}[\Qn\Ar-arena]
A \emph{\Qn\Ar-arena} is an arena (without any first-order label) such that if a node has a non-empty atomic label on it, then this label is a singleton and the node is a leaf and is not a root.

Labelled nodes are called \emph{answers} and the others are called \emph{questions}.
\end{defi}

\begin{defi}[Label-rigid strategy]
A play \play{s} on a \Qn\Ar-arena is \emph{rigid} if for any \ppol-prefix $\play{t}\move{m}\move{n}$ of $\play{s}$, \move{m} is an answer if and only if \move{n} is. It is \emph{label-rigid} if moreover the labels of \move{m} and \move{n} are the same.

A strategy on a \Qn\Ar-arena is \emph{rigid} (resp.\ \emph{label-rigid}) if all its plays are rigid (resp.\ label-rigid).
\end{defi}

The notion of rigidity corresponds to having both \emph{forward rigidity} and \emph{backward rigidity} as defined in~\cite{aoirev}. In the case (as here) where answers are leaves and are not roots, forward rigidity is then the same thing as the notion of rigidity introduced in~\cite{anatinnoc}.
Label-rigidity is strongly related with token-reflection in~\cite[page~122]{phdmurawski}.

\bigskip

\Qn\Ar-arenas and label-rigid strategies define a control category~\cite{synsempol} and thus a model of the \lmcalc{} (with or without totality). Let us call it the \emph{\Qn\Ar-game model}.

\bigskip

A rigid view of even length contains either only questions or its only answers are its last two moves (conversely a view satisfying one of these properties is rigid).

\begin{defi}[Well bracketed strategy]
A view \play{s} on a \Qn\Ar-arena is \emph{well bracketed} if any Player answer of \play{s} is justified by the last Opponent question.
A strategy on a \Qn\Ar-arena is \emph{well bracketed} if all its views are well bracketed.
\end{defi}

For a rigid view, being well bracketed means either containing only questions or being of the shape $\newlinplay\play{s}\oq[0]_1\pq[0]_2\oa[2]_1\pa_2$.

\begin{defi}[Folding and unfolding of arenas]
Let $A$ be an arena, the \emph{unfolding} of $A$ is the \Qn\Ar-arena $\unfold{A}$ obtained in the following way: to any node \node{m} of $A$ having atomic label $[X_1,\dots,X_n]$, we add $n$ new sons (denoted $\unfold{\node{m}}_{X_i}$ and labelled with $X_i$, $1\leq i\leq n$) and we erase the atomic label of \node{m}.

Let $A$ be a \Qn\Ar-arena, the \emph{folding} of $A$ is the arena $\fold{A}$ obtained in the following way: we remove all the answers of $A$ and for each of them we put its label in the atomic label of its father.
\end{defi}

\begin{exa}
The arena corresponding to the type $X$ and its unfolding (a \Qn\Ar-arena with answers represented as squares), which is the interpretation of $X$ in the \Qn\Ar-model, are:
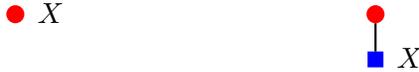
\begin{treeenv}
  \onod{}{X}
\qquad\qquad\qquad\qquad\qquad
  \pstree{\onod{}{}}{\psqnod{}{X}}
\end{treeenv}

The interpretation of $\bot$ is the same in the two models and is its own folding/unfolding.
\end{exa}

\begin{defi}[Folding and unfolding of strategies]
Let $A$ be an arena and \play{s} be a view on $A$, the \emph{unfolding} $\unfold{\play{s}}$ of \play{s} is the set of views given by $\unfold{\eps}=\{\eps\}$ and $\unfold{\play{s}\move{m}\move{n}}=\{\play{s}_0\move{m}\move{n}\}\cup\{\play{s}_0\move{m}\move{n}\unfold{\move{n}}_X\unfold{\move{n}'}_X\mid \text{$X$ in the atomic label of \move{n}}\}$ where $\play{s}_0$ is $\play{s}$ without its $\mu$-pointers, the $\mu$-pointer of $\play{s}\move{m}\move{n}$ starting from the element $X$ of the atomic label of \move{n} goes to the element $X$ of the atomic label of the occurrence of move $\move{n}'$, and $\unfold{\move{n}'}_X$ points to this occurrence of $\move{n}'$.
The unfolding $\unfold{\sigma}$ of a $\mu$-strategy $\sigma$ (in fact of its view function) on $A$ is the union of the unfoldings of its views.

Let $A$ be a \Qn\Ar-arena, $\sigma$ be a total label-rigid strategy on $A$ and \play{s} be a view in $\sigma$ ending with a question (more precisely, not ending with an answer), the \emph{folding} $\fold{\play{s}}$ of \play{s} is the view on $\fold{A}$ given by $\fold{\eps}=\eps$ and $\fold{\play{s}\move{m}\move{n}}=\fold{\play{s}}\move{m}\move{n}$ where the $\mu$-pointers of \move{n} are obtained in the following way: for each atomic label $X$ of \move{n} in $\fold{A}$, there is exactly one corresponding answer $\move{n}_X$ in $A$, we consider the only play of the shape $\play{s}\move{m}\move{n}\move{n}_X\move{n}'_X$ (for some $\move{n}'_X$ which is an answer pointing to some $\move{n}'$) in $\sigma$ and we put a $\mu$-pointer from the label $X$ of \move{n} to the label $X$ of $\move{n}'$.
The folding of $\sigma$ is the set of the foldings of its question-ending views.
\end{defi}

\begin{exa}\label{exunfolding}
The arena corresponding to the type $(((X\imp Y)\imp X)\imp X)$ and its unfolding are:
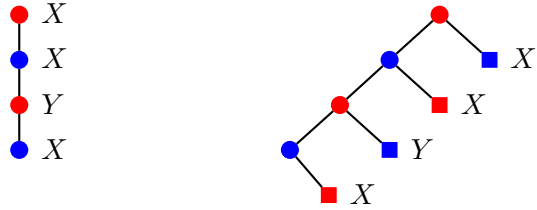
\begin{treeenv}
  \pstree{\onod{}{X}}{\pstree{\pnod{}{X}}{\pstree{\onod{}{Y}}{\pnod{}{X}}}}
\qquad\qquad\qquad
  \pstree{\onod{}{}}{\pstree{\pnod{}{}}{\pstree{\onod{}{}}{\pstree{\pnod{}{}}{\Tn \osqnod{}{X}} \psqnod{}{Y}} \osqnod{}{X}} \psqnod{}{X}}
\end{treeenv}
The following view with its prefixes of even length define a $\mu$-strategy on the starting arena:
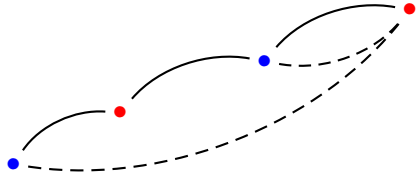
\begin{equation*}
\begin{playenv}{ccccccccccc}{( & ( & X & \imp & Y & ) & \imp & X & ) & \imp & X}
    & & & & & & & & & & \ob[0] \\
    & & & & & & & \pbm{1} \\
    & & & & \ob[2] \\
    & & \pbm[3]{1}
\end{playenv}
\end{equation*}
The maximal views of its unfolding are:
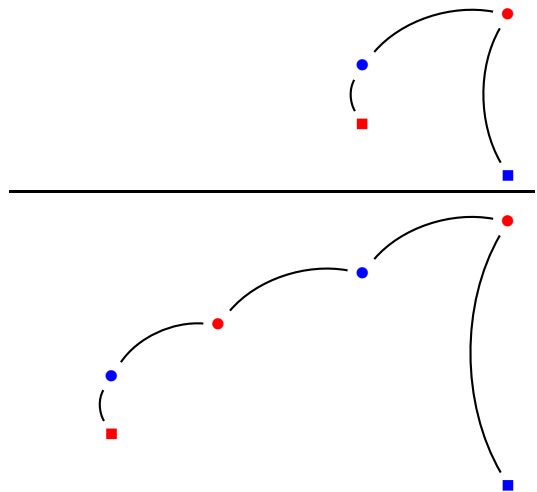
\begin{equation*}
\begin{playenv}{ccccccccccc}{( & ( & X & \imp & Y & ) & \imp & X & ) & \imp & X}
    & & & & & & & & & & \ob[0] \\
    & & & & & & & \pb \\
    & & & & & & & \os[2] \\
    & & & & & & & & & & \ps \\
\nextplay
    & & & & & & & & & & \ob[0] \\
    & & & & & & & \pb \\
    & & & & \ob[2] \\
    & & \pb[3] \\
    & & \os[4] \\
    & & & & & & & & & & \ps
\end{playenv}
\end{equation*}
\end{exa}

\begin{lem}\label{lemuniqmu}
If $\sigma$ is a $\mu$-strategy and \play{s} and \play{t} are two plays in $\sigma$ which differ only on their $\mu$-pointers then $\play{s}=\play{t}$.
\end{lem}

\proof
By induction on the common length of \play{s} and \play{t}, or by Lemma~\ref{lemgrplay}.
\qed

\begin{lem}\label{lemunfoldq}
If $\play{s}\in\unfold{\sigma}$ does not end with an answer, there exists a unique play \play{t} in $\sigma$ such that $\play{s}\in\unfold{\play{t}}$. Moreover \play{s} is obtained from \play{t} by removing its $\mu$-pointers.
\end{lem}

\proof
The existence of \play{t} is given by definition of \unfold{\sigma} and \play{t} is such that removing its $\mu$-pointers gives \play{s}.
If there exist two plays satisfying the required constraints, they must differ only on their $\mu$-pointers (since forgetting them leads to \play{s} in both cases), thus they are equal by Lemma~\ref{lemuniqmu}.
\qed

\begin{thm}[Completeness of unfolding]\label{thmunfolding}
Let $A$ be an arena, $\sigma\mapsto\unfold{\sigma}$ and $\tau\mapsto\fold{\tau}$ define a bijection pair between total $\mu$-strategies on $A$ and total label-rigid strategies on $\unfold{A}$.
\end{thm}

\proof
We are going to prove the following statements:
\begin{enumerate}
\item If $\sigma$ is a $\mu$-strategy on $A$ then $\unfold{\sigma}$ is a label-rigid strategy on $\unfold{A}$ (and if $\sigma$ is total then so is $\unfold{\sigma}$).
\item If $\tau$ is a total label-rigid strategy on $A$ then $\fold{\tau}$ is a total $\mu$-strategy on $\fold{A}$.
\item $\fold{(\unfold{A})}=A$ and $\unfold{(\fold{A})}=A$.
\item If $\sigma$ and $\tau$ are total, $\fold{(\unfold{\sigma})}=\sigma$ and $\unfold{(\fold{\tau})}=\tau$.
\end{enumerate}

\noindent First statement:
  \begin{enumerate}[$\bullet$]
  \item If $\play{s}\move{m}\move{n}\in\unfold{\sigma}$, by construction, either \move{m} is a question and so is \move{n} or \move{m} is an answer and \move{n} is also an answer and has the same label. As a consequence, $\play{s}\move{m}\move{n}$ is label-rigid.
  \item $\unfold{\sigma}$ contains $\eps$, contains only views and is \ppol-prefix closed by definition.
  \item If $\play{s}\move{m}\move{n}\in\unfold{\sigma}$ and $\play{s}\move{m}\move{n}'\in\unfold{\sigma}$, we first consider the case where \move{m} is a question (then both \move{n} and $\move{n}'$ are questions). By Lemma~\ref{lemunfoldq}, there exist two plays $\play{t}\move{m}\move{n}$ and $\play{t}'\move{m}\move{n}'$ in $\sigma$ such that $\play{s}\move{m}\move{n}$ (resp.\ $\play{s}\move{m}\move{n}'$) is obtained from $\play{t}\move{m}\move{n}$ (resp.\ $\play{t}'\move{m}\move{n}'$)  by removing its $\mu$-pointers. By Lemma~\ref{lemuniqmu}, we know that $\play{t}=\play{t}'$ so that, by determinism of $\sigma$, $\play{t}\move{m}\move{n}=\play{t}'\move{m}\move{n}'$  and thus $\play{s}\move{m}\move{n}=\play{s}\move{m}\move{n}'$.

In the case where \move{m} is an answer (then both \move{n} and $\move{n}'$ are answers and these three moves have the same label), then $\play{s}\move{m}\move{n}$ and $\play{s}\move{m}\move{n}'$ belongs to some \unfold{\play{t}} and \unfold{\play{t}'} that can only differ on their $\mu$-pointers thus, by Lemma~\ref{lemuniqmu}, $\play{t}=\play{t}'$. As a consequence \move{n} and $\move{n}'$ are the same move (the unique answer of $\unfold{A}$ corresponding to the label $X$ of the $\mu$-justifier of the label $X$ of the last move of \play{s}) and have the same justifier (this $\mu$-justifier of the last move of \play{s}).
\item We now look at totality. If $\play{s}\in\unfold{\sigma}$ and $\play{s}\move{m}$ is a view on $\unfold{A}$ then \play{s} contains only questions and, by Lemma~\ref{lemunfoldq}, we can find a play \play{t} in $\sigma$ such that \play{s} is obtained from \play{t} by removing its $\mu$-pointers. First, if \move{m} is a question, by totality of $\sigma$, there exists $\play{t}\move{m}\move{n}$ in $\sigma$ and thus $\play{s}\move{m}\move{n}$ (obtained from it by removing the $\mu$-pointers) is in $\unfold{\sigma}$. Second, if \move{m} is an answer, by definition of $\unfold{t}$, there exists some \move{n} such that $\play{s}\move{m}\move{n}\in\unfold{t}\subseteq\unfold{\sigma}$.
\end{enumerate}
 
\noindent Second statement:
\begin{enumerate}[$\bullet$]
\item We first check that $\play{s}\in\fold{\tau}$ is a well defined play. There is exactly one $\mu$-pointer for each formula of each Player move: each such occurrence leads to an answer in $\unfold{A}$ thus corresponds to an Opponent move which has been played in some play of $\tau$ by totality.
\item Then, by definition, $\fold{\tau}$ is a non-empty \ppol-prefix closed set of even length views.
\item If $\play{s}\move{m}\move{n}\in\fold{\tau}$ and $\play{s}\move{m}\move{n}'\in\fold{\tau}$, then $\move{n}=\move{n}'$ and they have the same justification pointer by determinism of $\tau$. Finally their $\mu$-pointers are the same since they are computed in the same way, thus $\play{s}\move{m}\move{n}=\play{s}\move{m}\move{n}'$.
\item $\fold{\tau}$ is total by immediate application of the totality of $\tau$.
\end{enumerate}

\noindent Third statement:
\begin{enumerate}[$\bullet$]
\item The moves of $\fold{(\unfold{A})}$ are the moves of $A$ and their atomic labels are obtained by moving them on a new leaf and then moving them back to their father so that $\fold{(\unfold{A})}=A$.
\item The questions of $\unfold{(\fold{A})}$ are the moves of $\fold{A}$ which are the questions of $A$. The answers of $\unfold{(\fold{A})}$ corresponds to labels of moves of $\fold{A}$ which corresponds to answers of $A$ thus $\unfold{(\fold{A})}=A$.
\end{enumerate}

\noindent Fourth statement: for the two equalities, we consider total strategies and this allows us to prove only one inclusion in each case since totality entails maximality.
\begin{enumerate}[$\bullet$]
\item We consider a play in $\fold{(\unfold{\sigma})}$, and we prove by induction on its length that it belongs to $\sigma$. If it is $\eps$ then it belongs to $\sigma$. Otherwise it is of the shape $\fold{\play{s}\move{m}\move{n}}$ with $\play{s}\move{m}\move{n}\in\unfold{\sigma}$ and \move{n} is a question in $\unfold{A}$. By Lemma~\ref{lemunfoldq}, this entails that we can find a play $\play{t}\in\sigma$ such that $\play{s}\move{m}\move{n}\in\unfold{\play{t}}$, and $\play{s}\move{m}\move{n}$ is obtained from \play{t} by removing all the $\mu$-pointers. $\fold{\play{s}}$ is a \ppol-prefix of $\fold{\play{s}\move{m}\move{n}}$ thus $\fold{\play{s}}\in\fold{(\unfold{\sigma})}$ and, by induction hypothesis, $\fold{\play{s}}\in\sigma$. The only possible difference between \play{t} and $\fold{\play{s}\move{m}\move{n}}$ is about the $\mu$-pointers of \move{n}, but these $\mu$-pointers are built from the elements of \unfold{\play{t}} so that they are exactly the same as the corresponding $\mu$-pointers in \play{t}. Finally $\fold{\play{s}\move{m}\move{n}}=\play{t}\in\sigma$.
\item We prove, by induction on the length of \play{t}, that if $\play{t}\in\fold{\tau}$ then $\unfold{\play{t}}\subseteq\tau$. First, $\unfold{\eps}\subseteq\tau$. Otherwise we have $\play{t}=\play{t}'\move{m}\move{n}$. Let $\play{t}_0\unfold{\move{n}}_X\unfold{\move{n}'}_X$ be an element of \unfold{\play{t}}, we easily see that $\play{t}_0\in\tau$. We look at the $\mu$-pointer of \move{n} corresponding to the label $X$ in $\play{t}'\move{m}\move{n}$: it goes to the label $X$ of the move $\move{n}'$, but since it is an element of $\fold{\tau}$, this means that $\play{t}_0\unfold{\move{n}}_X\unfold{\move{n}'}_X\in\tau$.
\qed
\end{enumerate}

\begin{prop}[Comparing the two models]\label{propcompmodels}
Let $M$ be a \lmterm{} (resp.\ $A$ be a propositional formula) interpreted as $\sigma$ (resp.\ $A_0$) in our game model and as $\tau$ (resp.\ $A_1$) in the \Qn\Ar-game model, we have $\sigma=\fold{\tau}$ (resp.\ $A_1=\fold{A_0}$).
\end{prop}

\proof
The part concerning formulas and arenas is obtained by a simple induction.

Concerning terms and strategies, we prove it by induction on the term $M$ with the interesting cases given by the $\mu\alpha.M$ and $[\alpha]M$ constructions. Since $\sigma$ and $\fold{\tau}$ are total, it is enough to prove $\fold{\tau}\subseteq\sigma$.
\begin{enumerate}[$\bullet$]
\item The interpretation of $\mu\alpha.M$ is almost the same as the interpretation of $M$.
\item For a term of the shape $[\alpha]M$, by induction hypothesis, if $\sigma_0$ (resp.\ $\tau_0$) is the interpretation of $M$ in our model (resp.\ in the \Qn\Ar-model), $\sigma_0=\fold{\tau_0}$.
We go by induction on the length of a play of $\tau$ for proving $\fold{\tau}\subseteq\sigma$.
Assume $\play{s}\move{m}\move{n}\in\tau$ where $\move{n}$ is a question, by induction hypothesis, $\fold{\play{s}}\in\sigma$. There is a unique $\fold{\play{s}}\move{m}\move{n}_1$ in $\sigma$, we want to show $\move{n}$ and $\move{n}_1$ to be the same move with the same pointers in $\fold{\play{s}}\move{m}\move{n}_1$ and in $\fold{\play{s}\move{m}\move{n}}$. Since $\sigma_0=\fold{\tau_0}$, it comes from a look at the contraction strategies on $A\times A\imp A$ ($A_1\times A_2\imp A_0$ with meaningless indexes) in both models. We directly have that $\move{n}$ and $\move{n}_1$ come from the same node in the arena with the same justification pointer. Concerning the $\mu$-pointers, in our model each $\mu$-pointer for a move in $A_0$ coming from a move in $A_1$ is given in a $\mu$-rigid way and goes to the move in $A_1$ (and the same in the other direction). In the \Qn\Ar-model, this corresponds to an answer $\unfold{n}_X$ played by Opponent in $A_0$ and copied in $A_1$ (with a justification pointer going to the last Opponent move in $A_1$ in a well-bracketed way). After folding, one obtains the same $\mu$-pointer.
\qed
\end{enumerate}

\noindent
We now consider Theorem~\ref{thmunfolding} in the particular case where arenas are trees with at most one label on each node and where \Qn\Ar-arenas never have two answers with the same father. This corresponds to the interpretations of types built with propositional variables, $\bot$ and $\imp$.

\begin{prop}[Full completeness of the \Qn\Ar-model]
Let $A$ be a formula built with propositional variables, $\bot$ and $\imp$, and let $\sigma$ be a total finite label-rigid strategy on the \Qn\Ar-arena interpreting $A$,
  \begin{enumerate}[$\bullet$]
  \item $\sigma$ is the interpretation of a \lmterm{} of type $A$
  \item $\sigma$ is the interpretation of a \lterm{} if and only if $\sigma$ is well bracketed
  \end{enumerate}
\end{prop}

\proof
These two results are known and come with direct proofs by induction on the size of the strategy. We give here an alternative proof using our games and the notion of folding.

By Theorem~\ref{thmunfolding}, \fold{\sigma} is a total finite $\mu$-strategy on \fold{A}. By Theorem~\ref{thmcomp}, \fold{\sigma} is the interpretation of a \lmterm{} $M$ of type $A$. By Proposition~\ref{propcompmodels}, the interpretation of $M$ in the \Qn\Ar-game model is $\sigma$.

Finally, $\fold{\sigma}$ is the interpretation of a \lterm{} if and only if the $\mu$-pointers are always going to the previous move. This corresponds in $\sigma$ to the fact that any view is well bracketed.
\qed

There is a purely syntactic counterpart to these semantic results: starting from a \lmterm, it is possible to compute the associated strategy, then to unfold it and by completeness (of our first model as in Section~\ref{secmodellcalc}, not of the \Qn\Ar-game model) to get back a \lterm.
This is a translation of the simply typed \lmcalc{} into the simply typed \lcalc{} already studied in~\cite{lmjsl}. The unfolding of arenas corresponds to a translation of simple types into simple types (see Table~\ref{tablml}).
A typing judgment $\A\vdash M:A\mid\B$ is translated as $\trad{\A},\tradb{\B}\vdash \trad{M}:\trad{A}$ where $\tradb{\B}$ is obtained by transforming any $\alpha:A_1\imp\cdots\imp A_n\imp X$ into $\alpha_1:\trad{A_1},\dots,\alpha_n:\trad{A_n},\alpha:X$, and any $\alpha:A_1\imp\cdots\imp A_n\imp\bot$ into $\alpha_1:\trad{A_1},\dots,\alpha_n:\trad{A_n}$. The translation of \lmterm s is given in Table~\ref{tablml}.
\begin{table}
\paragraph{Types.}
\begin{align*}
\trad{X} &= X\imp\bot \\
\trad{\bot} &= \bot \\
\bigtrad{A\imp B} &= \trad{A}\imp\trad{B}
\end{align*}
\paragraph{Typed terms.}
\begin{align*}
\trad{a} &= a \\
\bigtrad{\lambda a.M} &= \lambda a.\trad{M} \\
\bigtrad{(M) N} &= (\trad{M})\trad{N} \\
\bigtrad{[\alpha]M} &= (\trad{M})\alpha_1\dots\alpha_n\alpha &\text{if $\alpha$ has type $A_1\imp\cdots\imp A_n\imp X$} \\
\bigtrad{[\alpha]M} &= (\trad{M})\alpha_1\dots\alpha_n &\text{if $\alpha$ has type $A_1\imp\cdots\imp A_n\imp\bot$} \\
\bigtrad{\mu\alpha.M} &= \lambda\alpha_1\dots\lambda\alpha_n\lambda\alpha.\trad{M} &\text{if $\alpha$ has type $A_1\imp\cdots\imp A_n\imp X$} \\
\bigtrad{\mu\alpha.M} &= \lambda\alpha_1\dots\lambda\alpha_n.\trad{M} &\text{if $\alpha$ has type $A_1\imp\cdots\imp A_n\imp\bot$}
\end{align*}
\caption{A translation from the \lmcalc{} to the \lcalc}\label{tablml}
\end{table}

The present analysis is developed in the context of propositional logic since the two-moves game model was already known. However following these ideas of folding and unfolding of arenas, one could also define a first-order two-moves game model in correspondence with the one-move one and providing an encoding of the $\mu$-pointers.

\subsection{Forgetting structure}

In the spirit of comparing the various game models for various logical systems presented before, we can define forgetful transformations between them.

This is mainly a digressive section since the remarks mentioned here can easily be given without any use of game semantics. However, we find particularly straightforward to understand them by means of games, following the idea of the forgetful functor $\grfunc$ used in Appendix~\ref{appforgetfunc}.

\paragraph{From Church style first-order to Curry style first-order.}
\label{seccurry}

In type systems dealing with quantification, we usually have the choice between two main presentations: Church style systems and Curry style systems. The first ones are explicitly mentioning the quantification inference rules inside terms (as we do here from the beginning) while the second ones are not modifying the typed term when dealing with such a rule:
\begin{equation*}
\AIC{\A\vdash M:A\mid\B}
\RL{x\notin \A,\B}
\UIC{\A\vdash M:\forall x A\mid\B}
\DP
\qquad\qquad
\AIC{\A\vdash M:\forall x A\mid\B}
\UIC{\A\vdash M:\subst{A}{t}{x}\mid\B}
\DP
\end{equation*}

From these considerations, it is absolutely immediate by starting from a Church style typed \lmterm{} of type $A$ and by erasing the $\Lambda x.\_$ and $\_\foapp{t}$ constructions to get a Curry style typed \lmterm{} of type $A$.

An example is given by:
\begin{align*}
A_1&=\forall x(\forall y(Xx\imp Xy)\imp\bot)\imp\bot\\
M_1&=\lambda f.(f\foapp{x})\Lambda y.\lambda d.\mu\alpha.(f\foapp{y})\Lambda z.\lambda a.\mu\delta[\alpha]a\\
M_2&=\lambda f.(f)\lambda d.\mu\alpha.(f)\lambda a.\mu\delta[\alpha]a
\end{align*}
where $M_1$ is of type $A_1$ in the Church style system and $M_2$ is of type $A_1$ in the Curry style system.

Our game models are useless at this level since we have not considered models of Curry style systems. Let us go to the next step.

\paragraph{From Curry style first-order to classical propositional.}

The term language is the same for a Curry style first-order type system and for a propositional type system. If $M$ is such a \lmterm{} typable with type $A$ in the Curry style first-order system, then $M$ is typable with type $A'$ in the system of simple types (that is in propositional logic) where $A'$ is obtained by erasing all the first-order information in $A$ (\ie{} $\forall x.B\mapsto B$ and $X\vec{t}\mapsto X$).

\begin{prop}[Church to propositional]
By composing these two steps we obtain a \lmterm{} typed in the simple types system from a \lmterm{} typed in the Church style first-order system.
\end{prop}

\proof
If $\sigma$ is a first-order strategy on the first-order arena $A$, by removing all the instantiation information, we obtain a propositional strategy on the propositional arena obtained from $A$ by removing the first-order information in labels (we remove the first-order labels and we apply $X\vec{t}\mapsto X$ in the atomic labels).

Through the correctness and full completeness theorems, these transformations described on strategies and arenas exactly correspond to the syntactic transformations going from Church style typed \lmterm s to simply typed \lmterm s.
\qed

As an example, the \lmterm{} $M_2$ above has type $A_1$ in the Curry style type system but also the type $A_3=((X\imp X)\imp\bot)\imp\bot$ in the system of simple types.

\paragraph{From classical propositional to intuitionistic propositional.}

It is possible to go one step further by erasing all the $\mu\alpha$ and $[\alpha]$ constructs in a simply typed \lmterm{} of type $A$, and by transforming $A$ into $A'$ by mapping all the propositional variables of $A$ to the same atomic formula $\bot$.

\begin{prop}[Classical to intuitionistic]
This transformation gives a simply typed \lterm{} from a simply typed \lmterm.
\end{prop}

\proof
Starting from a propositional strategy $\sigma$ on the arena $A$ and by removing all the $\mu$-pointers, we obtain a strategy on $A'$ in the model with one move for atoms~\cite{gamesgoi} where $A'$ is obtained from $A$ by removing all the labels.

By correctness and full completeness of these models, we derive the syntactic result.
\qed

Going on with our example, while $M_2$ has type $A_3$, the corresponding \lterm{}:
\begin{equation*}
M_4=\lambda f.(f)\lambda d.(f)\lambda a.a
\end{equation*}
has type $A_4=((\bot\imp \bot)\imp\bot)\imp\bot$ in the simply typed \lcalc{} with one atom denoted $\bot$.

The global move from the Church style classical first-order game model to the intuitionistic propositional game model is nothing but the application of the $\grfunc$ functor (see Appendix~\ref{appforgetfunc}).

\subsection{Linear \texorpdfstring{\lmcalc}{lambda-mu-calculus}}

Following the definition of the linear \lcalc, the \emph{linear \lmcalc} is one of the possible presentations of a linearized version of classical logic~\cite[Chapter~14]{phdlaurent}. A closed \lmterm{} is linear if each variable ($\lambda$-variable or $\mu$-variable) has exactly one occurrence (two if we count the occurrence with the binder $\lambda$ or $\mu$).
A typical example is the term $\lambda f.\mu\alpha.(f)\lambda a.[\alpha]a$ of type $((X\imp\bot)\imp\bot)\imp X$.

Since the linear \lmcalc{} is defined as a restriction of the \lmcalc, any model of the \lmcalc{} is a model of the linear \lmcalc. However it is not always easy to define a fully complete sub-model for the linear sub-calculus. Various works on game semantics propose ways of going in this direction~\cite{locus,gamesludics,gameslincps} by mainly asking for each move to be played once in a strategy. We are going to show that it is also possible to use the pointers (the key point being the introduction of $\mu$-pointers in our work).

Concerning $\lambda$-variables (thus $\lambda$-pointers), the very natural definition of a \emph{$\lambda$-linear strategy} is just to ask, in the tree of views, that for each occurrence of an Opponent move \move{m} there is exactly one occurrence of each Player move enabled by \move{m} which is $\lambda$-pointing to \move{m}.

We apply the same kind of idea to $\mu$-pointers: a strategy is \emph{$\mu$-linear} if, in the tree of views, there is exactly one $\mu$-pointer from a Player move going to each atomic label of each Opponent move. A particular case of $\mu$-linear strategy is given by the $\lambda$-strategies of Section~\ref{secintgames}.

\begin{prop}[Linear equivalence completeness]
$\lambda$-linear and $\mu$-linear strategies give an equivalence complete model of the linear \lmcalc.
\end{prop}

\proof
Easy to check by following the proofs of Section~\ref{secinterp}.
\qed

These ideas can easily be extended to notions of $\lambda$-affine and $\mu$-affine strategies. Notice that being linear or affine depends on the type associated with a strategy as shown in the following examples.

\begin{exa}
The simplest proof of the double negation elimination:
\begin{equation*}
\vdash \lambda f.\mu\alpha.(f)\lambda a.[\alpha]a : ((X\imp\bot)\imp\bot)\imp X
\end{equation*}
corresponds to a $\lambda$-linear and $\mu$-linear strategy:
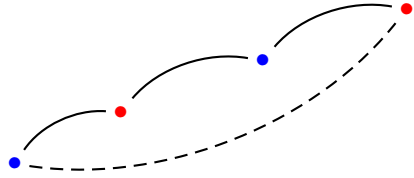
\begin{equation*}
\begin{playenv}{ccccccccccc}{( & ( & X & \imp & \bot & ) & \imp & \bot & ) & \imp & X}
    & & & & & & & & & & \ob[0] \\
    & & & & & & & \pb \\
    & & & & \ob[2] \\
    & & \pbm[3]{1}
\end{playenv}
\end{equation*}
By slightly modifying the type, we have to slightly modify the proof:
\begin{equation*}
\vdash \lambda f.\mu\alpha.(f)\lambda a.\mu\delta[\alpha]a : ((X\imp Y)\imp\bot)\imp X
\end{equation*}
which corresponds to a $\lambda$-linear and $\mu$-affine strategy (which is not $\mu$-linear):
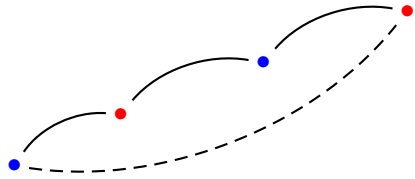
\begin{equation*}
\begin{playenv}{ccccccccccc}{( & ( & X & \imp & Y & ) & \imp & \bot & ) & \imp & X}
    & & & & & & & & & & \ob[0] \\
    & & & & & & & \pb \\
    & & & & \ob[2] \\
    & & \pbm[3]{1}
\end{playenv}
\end{equation*}
The simplest proof of Peirce's law:
\begin{equation*}
\vdash \lambda f.\mu\alpha[\alpha](f)\lambda a.\mu\delta[\alpha]a : ((X\imp Y)\imp X)\imp X
\end{equation*}
corresponds to the $\lambda$-linear strategy (which is not $\mu$-affine) already given in Example~\ref{exunfolding}:
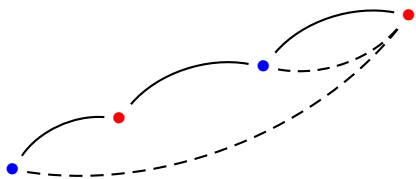
\begin{equation*}
\begin{playenv}{ccccccccccc}{( & ( & X & \imp & Y & ) & \imp & X & ) & \imp & X}
    & & & & & & & & & & \ob[0] \\
    & & & & & & & \pbm{1} \\
    & & & & \ob[2] \\
    & & \pbm[3]{1}
\end{playenv}
\end{equation*}
\end{exa}

\section{Type isomorphisms}
\label{secisos}

It is proved in Appendix~\ref{appisos} that all the equations of Table~\ref{tabtypisos} are syntactic type isomorphisms. We are going to prove that no other isomorphism is syntactically valid, by means of game semantics through the method developed in~\cite{classisos}.

\begin{defi}[Zig-zag play]
  A play \play{s} in the arena $A\imp B$ is a \emph{zig-zag play} if it is $\mu$-rigid (see page~\pageref{defmurigid}) and:
  \begin{enumerate}[$\bullet$]
  \item each Player move following an Opponent move in $A$ (resp.\ $B$) is in $B$ (resp.\ $A$)
  \item each Player move in $A$ following an initial Opponent move in $B$ is $\lambda$-justified by it
  \item the $\lambda$-pointers in $\proj{\play{s}}{A}$ and $\proj{\play{s}}{B}$  are the same
  \end{enumerate}
We denote by $\overline{\play{s}}$ the unique zig-zag play on $B\imp A$ such that $\proj{\overline{\play{s}}}{A}=\proj{\play{s}}{A}$ and $\proj{\overline{\play{s}}}{B}=\proj{\play{s}}{B}$.
\end{defi}

In order to reuse results from~\cite{classisos},
we define the function $\grfunc$ from arenas to ground arenas which erases the labels of its argument. If \play{s} is a play on $A$, we define $\grfunc(\play{s})$ as the play on $\grfunc(A)$ obtained by erasing the instantiations and the $\mu$-pointers. We extend $\grfunc$ to sets of plays by applying it point-wise. The main properties of $\grfunc$ are given in Appendix~\ref{appforgetfunc}.

\begin{lem}[Zig-zag lemma]\label{lemzz}
If $(\sigma,\tau)$ defines an isomorphism between $A$ and $B$ in $\Gcata$, then they contain only zig-zag plays and $\tau=\overline{\sigma}$.
\end{lem}

\proof
Since $(\grfunc(\sigma),\grfunc(\tau))$ is an isomorphism between $\grfunc(A)$ and $\grfunc(B)$ (by Lemma~\ref{lemgrcompid} page~\pageref{lemgrcompid}), we already know that plays in $\grfunc(\sigma)$ and $\grfunc(\tau)$ satisfy all the conditions of zig-zag plays but maybe $\mu$-rigidity (see~\cite[Proof of Theorem~9]{classisos}). We also know that $\grfunc(\sigma)$ and $\grfunc(\tau)$ are total thus $\sigma$ and $\tau$ are total, by Lemma~\ref{lemprestotfin} page~\pageref{lemprestotfin}.

We prove by induction on $k$ that if $\play{s}$ is a play of length $k$ in $\sigma$ then $\play{s}$ is zig-zag, $\overline{\play{s}}$ is in $\tau$ and there is an interaction sequence $\play{u}$ on $A$, $B$ and $A$ such that $\proj{u}{A\imp B}=\play{s}$ and $\proj{u}{B\imp A}=\overline{\play{s}}$.
It is immediate for $k=0$ since $\play{s}=\eps$ and thus $\play{u}=\eps$. Let $\play{s}\move{m}\move{n}$ be a play in $\sigma$ of length $k+2$, by induction hypothesis, we have $\overline{\play{s}}\in\tau$ and an interaction sequence $\play{u}$. Assume $\move{m}$ is in $A$ (the case $\move{m}$ in $B$ is symmetric) with a singleton atomic label (the case of an empty atomic label is simpler). We already know that $\move{n}$ is in $B$. We define $\play{u}'=\play{u}\move{m}_1\move{n}\move{m}'$ where $\move{m}_1$ is a copy of the move $\move{m}$ in the first $A$ and $\move{m}'$ is such that $\proj{\play{u}'}{B\imp A}\in\tau$ (this move exists by totality of $\tau$). We have $\proj{\play{u}'}{A\imp B}=\play{s}\move{m}\move{n}\in\sigma$ thus $\proj{\play{u}'}{A\imp A}\in\id{A}$. This means that in $\proj{\play{u}'}{A\imp A}$ the last move $\move{m}'$ has its $\mu$-pointer going to the previous move $\move{m}_1$ and its instantiation is the same as the instantiation of $\move{m}_1$. The only way to have these properties is that $\move{m}'$ $\mu$-points to $\move{n}$ and $\move{n}$ $\mu$-points to $\move{m}_1$ in $\play{u}'$, and also that if $\vec{x}$ is the $\ovar$-instantiation of $\move{m}_1$, $\vec{y}$ is the $\pvar$-instantiation of $\move{n}$, $\vec{z}$ is the $\ovar$-instantiation of $\move{n}$ and $\vec{w}$ is the $\pvar$-instantiation of $\move{m}'$ in $\play{u}'$ then $\vec{x}=\vec{y}$ and $\vec{z}=\vec{w}$. This proves $\play{s}\move{m}\move{n}$ to be a zig-zag play. Moreover $\proj{\play{u}'}{A\imp A}\in\id{A}$ also entails that $\move{m}'$ and $\move{m}_1$ are ``the same move with the same pointers'' thus $\proj{\play{u}'}{B\imp A}=\overline{\play{s}\move{m}\move{n}}\in\tau$.
\qed

\begin{thm}[Game isomorphisms]\label{thmisos}
Two arenas are isomorphic in $\Gcata$ if and only if they are arena isomorphic (Definition~\ref{defariso} page~\pageref{defariso}).
\end{thm}

\proof
We start with the first direction. By~\cite[Theorem~9]{classisos}, there exists an arena isomorphism $f$ between $\grfunc(A)$ and $\grfunc(B)$. Moreover if $\play{s}\move{m}\move{n}\in\sigma$ (with $\move{m}\in A$) then the node corresponding to $\move{n}$ in $\grfunc(B)$ (thus in $B$) is the image by $f$ of the node corresponding to $\move{n}$ in $\grfunc(A)$ (thus in $A$) and there exists such a play for any move $\move{m}$ of $A$.

We prove by induction on the depth of the node $\node{m}$ in $A$ that $f$ maps the labels of $\node{m}$ to the labels of $f(\node{m})$ turning it into an arena isomorphism between $A$ and $B$.
Let $\play{s}\move{m}\move{n}$ be the view in $\sigma$ dealing with the node $\node{m}$, it is $\mu$-rigid (Lemma~\ref{lemzz}), thus if $X\vec{t}$ is the atomic label of $\node{m}$ and $X'\vec{t'}$ is the atomic label of $\node{n}$, we have $X=X'$ and $\vec{t}\asubst_{\move{m}}=\vec{t'}\asubst_{\move{n}}$. Moreover $\move{m}$ and $\move{n}$ have the same instantiation (thus first-order labels $[x_1,\dots,x_k]$ and $[y_1,\dots,y_k]$ of the same length in their respective arenas).
If $\node{m}$ is a root of $A$, its instantiation is $[\oenum_0,\dots,\oenum_{k-1}]$ and $\asubst_{\move{m}}=\{x_1\mapsto\oenum_0,\dots,x_k\mapsto\oenum_{k-1}\}$. $\node{n}$ is also a root and it has instantiation $[\oenum_0,\dots,\oenum_{k-1}]$ thus $\asubst_{\move{n}}=\{y_1\mapsto\oenum_0,\dots,y_k\mapsto\oenum_{k-1}\}$. From this we can deduce $\vec{t'}=\vec{t}[^{y_1}/_{x_1},\dots,^{y_k}/_{x_k}]$ showing that $f$ respects the atomic label of $\node{m}$ and $\node{n}$. If $\node{m}$ is not a root in $A$ neither is $\node{n}$, and we assume $\node{m}$ is an Opponent move (otherwise we work with $f^{-1}$) so is $\node{n}$. Let $\move{m}_0$ be the justifier of $\move{m}$ in $\play{s}$ (and $\move{n}_0$ be the justifier of $\move{n}$), $\move{m}$ and $\move{n}$ have instantiation $[\oenum_i,\dots,\oenum_{i+k-1}]$ in $\play{s}\move{m}\move{n}$, and $\asubst_{\move{m}}=\asubst_{\move{m}_0}\cup\{x_1\mapsto\oenum_i,\dots,x_k\mapsto\oenum_{i+k-1}\}$ and $\asubst_{\move{n}}=\asubst_{\move{n}_0}\cup\{y_1\mapsto\oenum_i,\dots,y_k\mapsto\oenum_{i+k-1}\}$. From this we obtain that the mapping of first-order labels induced by $f$ maps $\vec{t}$ to $\vec{t'}$.

The second direction is given by Lemma~\ref{lemisoar} page~\pageref{lemisoar}.
\qed

All these results can easily be extended to $\Gcat$ but are not required here. See Section~\ref{secextdisj} for possible applications of this extension.

\begin{cor}[Type isomorphisms]
  Table~\ref{tabtypisos} exactly characterizes the type isomorphisms of the Church style first-order \lmcalc.
\end{cor}

\proof
Let $A$ and $B$ be two isomorphic types. They are both isomorphic to canonical forms $A_0$ and $B_0$ (see Section~\ref{secsyntcat}). By soundness of the game model, the arenas $\sem{A_0}$ and $\sem{B_0}$ interpreting $A_0$ and $B_0$ are isomorphic in $\Gcata$ thus are arena isomorphic (Theorem~\ref{thmisos}).
We prove by induction on the common number of nodes of $\sem{A_0}$ and $\sem{B_0}$ that $A_0$ and $B_0$ are equal up to the equations of Table~\ref{tabtypisos} (so that $A$ and $B$ are also equal up to these equations). Let $f$ be the arena isomorphism between $\sem{A_0}$ and $\sem{B_0}$, $f$ defines a bijection between the roots of the two arenas. If there is more than one root, $A_0$ and $B_0$ are conjunctions and each tree corresponds to one component. We apply the induction hypothesis to the pairs of trees whose roots are related through $f$. Since Table~\ref{tabtypisos} contains the associativity and commutativity of conjunction, one obtains that $A_0$ and $B_0$ are equal up to the equations.
If $\sem{A_0}$ and $\sem{B_0}$ are trees, $A_0$ and $B_0$ are $\imp$-canonical forms. Let $[x_1,\dots,x_k]$ and $[y_1,\dots,y_k]$ be the first-order labels of their roots (they have the same length), their atomic label are either empty or $X\vec{t}$ and $X\vec{t}[^{y_1}/_{x_1},\dots,^{y_k}/_{x_k}]$. Let $\mathcal{F}_A$ and $\mathcal{F}_B$ be the forests under these roots. By induction hypothesis, $\mathcal{F}_A$ and $\mathcal{F}_B[^{x_1}/_{y_1},\dots,^{x_k}/_{y_k}]$ correspond to canonical forms $\bigwedge_{1\leq i\leq n} A_i$ and $\bigwedge_{1\leq j\leq n} B_j$ equal up to the equations. We can conclude since $\forall\vec{x}(A_1\imp\cdots\imp A_n\imp R)$ and $\forall\vec{y}(B_1[^{y_1}/_{x_1},\dots,^{y_k}/_{x_k}]\imp\cdots\imp B_n[^{y_1}/_{x_1},\dots,^{y_k}/_{x_k}]\imp S)$ are equal up to the equations (with $R=S=\bot$ if the roots of $\sem{A_0}$ and $\sem{B_0}$ have an empty atomic label and with $R=X\vec{t}$ and $S=X\vec{t}[^{y_1}/_{x_1},\dots,^{y_k}/_{x_k}]$ otherwise).

Finally, by Appendix~\ref{appisos}, if $A$ and $B$ are equal up to the equations of Table~\ref{tabtypisos}, $A$ and $B$ are isomorphic.
\qed

\section{Krivine's realizability}

Realizability is used in the study of the computational properties of proofs in intuitionistic logic. The setting introduced by J.-L.~Krivine allows for the extension to classical logic and even set theory~\cite{realzf}. In particular, he showed more recently that computational interpretations of classical proofs can be given through a notion of games by means of his realizability interpretation. We are going to explain the close relation between his games and game semantics as presented in this paper.

\subsection{A quick introduction}
\label{secrealintro}

In the setting of Krivine's realizability, terms are studied through their computational behaviour via a notion of execution very similar to what happens in Krivine's abstract machine~\cite{newkam}.

A state is a pair of a Curry style \lmterm{} $M$ (see Section~\ref{seccurry}) and of a stack $\pi$:
\begin{equation*}
  \pi ::= \eps \mid \alpha \mid M.\pi
\end{equation*}
Such a state is written $M\psep\pi$ and the execution is given by a relation $\reduce$ between states:
\begin{align*}
  \lambda a.M\psep N.\pi &\reduce \subst{M}{N}{a}\psep\pi \\
  (M)N\psep\pi &\reduce M\psep N.\pi \\
  \mu\alpha.M\psep\pi &\reduce \subst{M}{\pi}{\alpha}\psep\eps \\
  [\alpha]M\psep\eps &\reduce M\psep\alpha
\end{align*}
where $\subst{M}{M_1\dots M_k.\eps}{\alpha}=\subst{M}{(N)M_1\dots M_k}{[\alpha]N}$ and $\subst{M}{M_1\dots M_k.\beta}{\alpha}=\subst{M}{[\beta](N)M_1\dots M_k}{[\alpha]N}$.

Through the following embedding of states into \lmterm s:
\begin{align*}
  M\psep M_1\dots M_k.\eps &\mapsto (M)M_1\dots M_k \\
  M\psep M_1\dots M_k.\alpha &\mapsto [\alpha](M)M_1\dots M_k
\end{align*}
these execution rules are simulated by the $\beta\mu\rho$ reduction.

\subsection{\texorpdfstring{\U\V\At}{UVA}  provability game}

In this section we give a straightforward adaption of results in~\cite[pages~77--86]{realgeocalslides} to our setting.
We introduce the notion of \emph{\U\V\At{} game}.

A \emph{position} is a triple $(\U,\V,\At)$ where $\U$ and $\V$ are sets of formulas in $\imp$-canonical form and $\At$ is a set of non-constant atomic formulas.

An Opponent move consists in choosing a formula $\forall\vec{x}(A_1\imp\cdots\imp A_n\imp R)$ in $\V$ and first-order terms $\vec{t}$. We then go to the position $(\U\cup\{\subst{A_1}{\vec{t}}{\vec{x}},\dots,\subst{A_n}{\vec{t}}{\vec{x}}\},\V,\At\cup\{\subst{R}{\vec{t}}{\vec{x}}\})$ if $R\neq\bot$ and $(\U\cup\{\subst{A_1}{\vec{t}}{\vec{x}},\dots,\subst{A_n}{\vec{t}}{\vec{x}}\},\V,\At)$ otherwise.

A Player move consists in choosing a formula $\forall\vec{x}(A_1\imp\cdots\imp A_n\imp R)$ in $\U$ and first-order terms $\vec{t}$ such that, if $R\neq\bot$, $\subst{R}{\vec{t}}{\vec{x}}$ is in $\At$. We then go to the position $(\U,\{\subst{A_1}{\vec{t}}{\vec{x}},\dots,\subst{A_n}{\vec{t}}{\vec{x}}\},\At)$.

An \emph{initial position} is a position with $\U=\At=\emptyset$ (in particular the initial position associated with a formula $A$ is $(\emptyset,\{A_1,\dots,A_n\},\emptyset)$ if the canonical form of $A$ is $\bigwedge_{1\leq i\leq n}A_i$). A \emph{final position} is a position with $\V=\emptyset$ (it corresponds to positions where Opponent cannot play).

A \emph{play} is a (possibly empty) sequence of moves in which players alternate and which starts from an initial position by an Opponent move.
If an initial position is given, Player is said to have a \emph{winning strategy} if he is able to choose his moves in such a way that he is always able to play and to eventually reach a final position (meaning that Opponent wins in infinite plays).

The main property of \U\V\At{} games is the following theorem of Krivine:

\begin{thm}
 $A$ is a provable formula if and only if there exists a winning strategy for Player in the associated \U\V\At{} game.
\qed
 \end{thm}

More precisely, if $M$ is a Curry style \lmterm{} of type $A$, $M$ \emph{implements} a winning strategy for Player in the \U\V\At{} game associated with $A$:

With each formula $A$ we associate a $\lambda$-variable $a_A$ and with each non-constant atomic formula $R$ we associate a $\mu$-variable $\alpha_R$. If $M$ has type $A=\forall\vec{x}(A_1\imp\cdots\imp A_n\imp R)$, we choose first-order terms $\vec{t}$ and we look at the execution of $M\psep a_{\subst{A_1}{\vec{t}}{\vec{x}}}\dots a_{\subst{A_n}{\vec{t}}{\vec{x}}}.\alpha_{\subst{R}{\vec{t}}{\vec{x}}}$ (or with $\eps$ instead of $\alpha_{\subst{R}{\vec{t}}{\vec{x}}}$ if $R=\bot$). It is shown that execution will stop in a state $a_{\subst{A_i}{\vec{t}}{\vec{x}}}\psep\pi$ with $\pi=M_1\dots M_k.\alpha$ if $\subst{A_i}{\vec{t}}{\vec{x}}=\forall\vec{y}(B_1\imp\cdots\imp B_k\imp S)$ (or with $\eps$ instead of $\alpha$ if $S=\bot$). There exists a choice of first-order terms $\vec{u}$ such that for any choice of $1\leq j\leq k$ and any choice of the first-order terms $\vec{v}$ with $\subst{B_j}{\vec{u}}{\vec{y}}=\forall\vec{z}(C_1\imp\cdots\imp C_p\imp T)$, the execution of $M_j\psep a_{\subst{C_1}{\vec{v}}{\vec{z}}}\dots a_{\subst{C_p}{\vec{v}}{\vec{z}}}.\alpha_{\subst{T}{\vec{v}}{\vec{z}}}$ will stop in a state $a_D\psep\pi'$ and so on... Whatever choices (for $j$ and $\vec{v}$) we make between the steps of this sequence of runs, it will stop in a state $a\psep\alpha$ (or $a\psep\eps$).

If we interpret execution steps (with the choices of $\vec{u}$) as Player moves and index choices (with the choices of $\vec{v}$) as Opponent moves, this shows that a Curry style \lmterm{} of type $A$ induces a winning strategy for Player in the \U\V\At{} game associated with $A$: each sequence of runs is a play in this game.

\subsection{Relation with game semantics}

We are now going to extend the previous correspondence between execution of terms and the \U\V\At{} provability games to a correspondence between execution and our game model by just a slight modification on the execution sequence. This correspondence is in fact the starting point of the present work on game semantics.

The modifications we have to do concern the use of Church style \lmterm s and the possibility of recovering pointers in games.

The explicit use of first-order terms in the construction of \lmterm s corresponds to the following execution rules:
\begin{align*}
  \Lambda x.M\psep t.\pi &\reduce \subst{M}{t}{x}\psep\pi \\
  M\foapp{t}\psep\pi &\reduce M\psep t.\pi
\end{align*}
where stacks are extended with the construction $t.\pi$.

With each formula $A$ we associate a denumerable set of $\lambda$-variables $(a_A^l)_{l\in\Nat}$ and with each non-constant atomic formula $R$ we associate a denumerable set of $\mu$-variables $(\alpha_R^l)_{l\in\Nat}$. Starting with a Church style \lmterm{} $M$ of type $A=\forall\vec{x}(A_1\imp\cdots\imp A_n\imp R)$ with $\vec{x}$ of length $n'$, we proceed as follows:
\begin{enumerate}[$\bullet$]
\item We start with a state: $M\psep t_1\dots t_{n'}.a_{\subst{A_1}{\vec{t}}{\vec{x}}}^1\dots a_{\subst{A_n}{\vec{t}}{\vec{x}}}^1.\alpha_{\subst{R}{\vec{t}}{\vec{x}}}^1$
.
\item When execution stops (for the $i$th time) in a state: $a^m_B\psep u_1\dots u_{k'}.M_1\dots M_k.\alpha$ with $B=\forall\vec{y}(B_1\imp\cdots\imp B_k\imp S)$ (with $\eps$ instead of $\alpha$ if $S=\bot$), we choose first-order terms $\vec{v}$ and an index $1\leq j\leq k$ and we start a new execution $M_j\psep v_1\dots v_{p'}.a_{\subst{C_1}{\vec{v}}{\vec{z}}}^{i+1}\dots a_{\subst{C_p}{\vec{v}}{\vec{z}}}^{i+1}.\alpha_{\subst{T}{\vec{v}}{\vec{z}}}^{i+1}$ where $\subst{B_j}{\vec{u}}{\vec{y}}=\forall\vec{z}(C_1\imp\cdots\imp C_p\imp T)$ (with $\eps$ instead of $\alpha_{\subst{T}{\vec{v}}{\vec{z}}}^{i+1}$ if $T=\bot$).
\end{enumerate}
An important point in the choice of the $a^i_A$s and $\alpha^i_R$s is that they are always fresh (\ie{} not used yet).

Such an execution sequence can be interpreted as a play (as defined in Section~\ref{secseqmoves}) where Opponent moves are given by the choices ($v_1\dots v_{p'}.a_{\subst{C_1}{\vec{v}}{\vec{z}}}^{i+1}\dots a_{\subst{C_p}{\vec{v}}{\vec{z}}}^{i+1}.\alpha_{\subst{T}{\vec{v}}{\vec{z}}}^{i+1}$ for example) and Player moves are given by the results of executions ($a^m_B\psep u_1\dots u_{k'}.M_1\dots M_k.\alpha$ for example). More precisely we can rebuild a view with its pointers and instantiations from such a sequence. We consider only the case where first-order terms introduced in Opponent moves are fresh variables (by innocence, this is enough to recover the corresponding strategy).

Since we focus here on $\imp$-canonical forms, the corresponding arenas are trees, and the starting state $M\psep x_1\dots x_{n'}.a_{\subst{A_1}{\vec{t}}{\vec{x}}}^1\dots a_{\subst{A_n}{\vec{t}}{\vec{x}}}^1.\alpha_{\subst{R}{\vec{t}}{\vec{x}}}^1$ is interpreted as Opponent playing the root of this arena with instantiation $\vec{x}$ ($a_{\subst{A_1}{\vec{t}}{\vec{x}}}^1\dots a_{\subst{A_n}{\vec{t}}{\vec{x}}}^1$ and $\alpha_{\subst{R}{\vec{t}}{\vec{x}}}^1$ are given as possible targets for future pointers). From that point, a result of execution of the shape $a\psep u_1\dots u_{k'}.M_1\dots M_k.\alpha$ is interpreted as a Player move: its $\lambda$-pointer is going to the Opponent move \move{m} where $a$ has been introduced (and let $i$ be its position in the state corresponding to \move{m}), the move \move{n} itself is the $i$th son of \move{m} in the arena, the instantiation is $\vec{u}$ and the $\mu$-pointer is going to the move where $\alpha$ has been introduced (if we have $\eps$ instead of $\alpha$, there is no $\mu$-pointer). The next starting point of execution $M_j\psep y_1\dots y_{p'}.\pi$ is interpreted as Opponent playing the $j$th son of \move{n} in the arena with a $\lambda$-pointer going to the previous move \move{n} (as required for a view) and instantiations $\vec{y}$.

\begin{prop}
We obtain this way a view on the associated arena if Opponent plays its instantiations according to $(\oenum_i)_{i\in\Nat}$.
\end{prop}

\proof
We build the view $\play{s}$ by induction on the length of the execution using the fact that types are appropriately preserved along the execution (as shown through the embedding into the \lmcalc{} given in Section~\ref{secrealintro}):
\begin{enumerate}[$\bullet$]
\item The starting state of the shape $M\psep \oenum_0\dots \oenum_{n'-1}.a_{\subst{A_1}{\vec{\oenum}}{\vec{x}}}^1\dots a_{\subst{A_n}{\vec{\oenum}}{\vec{x}}}^1.\alpha_{\subst{R}{\vec{\oenum}}{\vec{x}}}^1$ is interpreted as a valid Opponent initial move $\move{m}$ on the arena associated with $A$. $\play{s}$ is just this move.
Moreover, this defines a substitution $\asubst_\move{m}=\{x_1\mapsto\oenum_0,\dots,x_{n'}\mapsto\oenum_{n'-1}\}$ as in the definition of plays.
\item If we arrive at a state $a^q_B\psep u_1\dots u_{k'}.M_1\dots M_k.\alpha^{q'}_S$, let $\move{m}$ be the move of $\play{s}$ corresponding to the state where $a^q_B$ has been introduced and let $i$ be the position of $a^q_B$ in the sequence $\vec{a}$ in this state. Let $\node{n}$ be the node of the arena $A$ which is the $i$th son of the node corresponding to $\move{m}$ in $A$, we extend $\play{s}$ with $\move{n}$ with a $\lambda$-pointer going to $\move{m}$ (this is a correct $\lambda$-pointer). By preservation of typing in the execution, $B$ is of the shape $\forall y_1\dots\forall y_{k'}(B_1\imp\cdots\imp B_k\imp S')$. This means that the node $\node{n}$ in $A$ has a first-order label of length $k'$ and has $k$ sons. As a consequence, $u_1\dots u_{k'}$ is an appropriate instantiation for the move $\move{n}$, and we can define $\asubst_{\move{n}}=\asubst_{\move{m}}\cup\{y_1\mapsto u_1,\dots,y_{k'}\mapsto u_{k'}\}$.
The atomic label of $\node{n}$ in $A$ is $S_0$ with $S_0\asubst_{\move{n}}=S'[^{u_1}/_{y_1},\dots,^{u_{k'}}/_{y_{k'}}]=S=S_1\asubst_{\move{m}'}$ where $S_1$ is the atomic label of the move $\move{m}'$ corresponding to the state where $\alpha^{q'}_S$ has been introduced.
It is then valid to put a $\mu$-pointer from $\move{n}$ to $\move{m}'$ and $\play{s}\move{n}$ is a correct view in $A$.
\item Almost the same with a state $a^q_B\psep u_1\dots u_{k'}.M_1\dots M_k.\eps$ but without $\mu$-pointer.
\item In order to run the execution again, we build a new state:
\begin{equation*}
M_j\psep \oenum_i\dots \oenum_{i+p'-1}.a_{\subst{C_1}{\vec{v}}{\vec{z}}}^l\dots a_{\subst{C_p}{\vec{v}}{\vec{z}}}^l.\alpha_{\subst{T}{\vec{v}}{\vec{z}}}^l
\end{equation*}
(which respects the types). This corresponds to playing the $j$th son $\move{m}$ of the node corresponding to the last move of $\play{s}$ with a $\lambda$-pointer going to it and an instantiation $[\oenum_i\dots \oenum_{i+p'-1}]$ where $\move{m}$ has a first-order label of length $p'$ in $A$. This proves $\play{s}\move{m}$ to be correct view on $A$.
\qed
\end{enumerate}

\section{Extensions and additional comments}
\label{secconcl}

We have chosen to work, on the syntactic side, with a natural deduction system (the \lmcalc). Another possibility would have been to deal with a sequent calculus system (as in~\cite{synsempol} for example). It would not make very important differences. The notion of proofs in canonical form in the sequent calculus is given by cut-free proofs with expanded axioms (introducing only non-constant atomic formulas). In this context, $\mu$-pointers would precisely correspond to these atomic axioms connecting together two dual occurrences of an atomic formula. In~\cite{synsempol}, an encoding was required and $\mu$-pointers were somehow the missing data.

\bigskip

\label{secextdisj}
Another specific choice was to deal with a \lmcalc{} without disjunction or negation. Negation is easily definable through $\neg A=A\imp\bot$ and disjunction can be introduced as in Selinger's calculus~\cite{controlcat} with two new term constructs: $\mu(\alpha,\beta).M$ and $[\alpha,\beta]M$. The associated typing rules are:
\begin{gather*}
\AIC{\A,a:A\vdash M:\bot\mid\B}
\UIC{\A\vdash \ell a.M:\neg A\mid\B}
\DP
\quad\qquad
\AIC{\A\vdash M:\neg A\mid\B}
\AIC{\A\vdash N:A\mid\B}
\BIC{\A\vdash M\mathbin{\bullet}N:\bot\mid\B}
\DP
\\[2ex]
\AIC{\A\vdash M:A\vee B\mid\B,\alpha:A,\beta:B}
\UIC{\A\vdash [\alpha,\beta]M:\bot\mid\B,\alpha:A,\beta:B}
\DP
\quad\qquad
\AIC{\A\vdash M:\bot\mid\B,\alpha:A,\beta:B}
\UIC{\A\vdash \mu(\alpha,\beta).M:A\vee B\mid\B}
\DP
\end{gather*}
All the required material for interpreting these extensions is already given in the game category $\Gcat$ we have described. In particular, the interpretation of disjunction makes $\Gcata$ not big enough and requires us to work with the full control category $\Gcat$ (as done in~\cite{controlcat}).

The explicit treatment of these extensions would require some additional work on the syntactic side, in particular for the notion of canonical form. However no surprise would come from this and the results presented in this paper would extend without any particular problem. We can mention that, concerning type isomorphisms, the following equations would be derived:
\begin{align*}
A\vee(B\vee C) &= (A\vee B)\vee C \\
A\vee B &= B\vee A \\
A\vee\bot &= A \\
A\vee(B\wedge C) &= (A\vee B)\wedge(A\vee C) \\
A\vee\top &= \top \\
A\vee\forall x B &= \forall x (A\vee B) & x\notin A \\
\neg(A\wedge B) &= \neg A\vee\neg B \\
\neg\top &= \bot \\
A\imp B &= \neg A\vee B
\end{align*}

\bigskip

Enumerated data types such as \texttt{Bool} or \texttt{Nat} are usually interpreted in game semantics by \Qn\Ar-arenas with one root (a question) which has as many sons (which are answers) as elements of the data type (possibly an infinity). The traditional approach of game semantics is to build everything from these enumerated data types without any use of atomic formulas. We have shown in Section~\ref{secunfolding} how the label-rigidity constraint makes answers exactly as expressive as $\mu$-pointers. It would be interesting to relax this constraint to deal with systems with both enumerated data types and atomic formulas. Some work has already been done in~\cite{aoirev} to understand the expressive power of various possible restrictions on the use of answers (in particular some are weaker than label-rigidity).
This would help to understand more precisely the possible applications of $\mu$-pointers to the semantics of programming languages in relation with extensions of \PCF{} with control operators (starting from~\cite{controlgames} and~\cite{excepgames}).

The question of introducing enumerated data types in Krivine's realizability as been considered in~\cite{realdatatypes}. It would be nice to also extend the correspondence between games and realizability to data types.

\bigskip

Let us now look at the most natural logical extensions of this work. Concerning first-order logic, it would be important to introduce the \emph{equality predicate} and to be able to deal with given equational axioms and not only with the free first-order language. Once again this could be developed in relation with what happens in Krivine's realizability. What makes such an extension difficult is the requirement of some dynamism in arenas: an arena has to dynamically evolve during a play according to the moves of the players. Concerning equality, it is just a matter of a node being able to disappear (when an equational atomic label $t=u$ of a node becomes true).

This dynamics induced by moves on arenas is at the core of the game interpretations of (propositional) second-order logic (see~\cite{gamesF,phddelataillade} for example). Being able to mix our work with second-order interpretations is the main direction for future work leading to an equivalence complete game model of full second-order logic.
An immediate consequence would be the associated characterization of type isomorphisms. We conjecture the corresponding equational theory to be the union of ours with the second-order one given in~\cite{isoschurch}, together with $\forall x\forall X A=\forall X\forall x A$.

\section*{Acknowledgements}

\noindent I would like to thank Joachim de Lataillade, Russ Harmer, Martin Hyland and Jean-Louis Krivine for important discussions during the development of this work.
I also thank the anonymous referees for their useful suggestions.

\bibliographystyle{alpha}
\bibliography{ll}

\appendix

\section{Some properties of the \texorpdfstring{\lmcalc}{lambda-mu-calculus}}

\subsection{Canonical normal forms}
\label{appcanformterm}

A \lmterm{} $M$ is \emph{simple} if it is in the grammar:
\begin{eqnarray*}
M & ::= & a \mid \lambda a.M \mid (M)M \mid \mu\alpha[\beta]M \mid \Lambda x.M \mid M\foapp{t}
\end{eqnarray*}
We first remark that any typed \lmterm{} without pairs, projections or $\cst$ is equal to a typed simple \lmterm{} up to $\rho$: we choose a particular fresh variable $\xi$ of type $\bot$, we transform any $[\alpha]M$ into $\mu\xi[\alpha]M$ (by $[\alpha]M=_\rho[\xi]\mu\xi[\alpha]M=_\rho\mu\xi[\alpha]M$) and any $\mu\alpha.M$ into $\mu\alpha[\xi]M$ (since $M$ must be of type $\bot$). To see that the result is typed, we add $\xi:\bot$ to the right-hand side typing context.

Our goal is to show the existence of a canonical normal form for any typed simple \lmterm.
We are going to adapt results coming from~\cite[Chapter~5]{phdpy}. We recall the reductions defined in that work in Table~\ref{tabredrules}.
\begin{table}
\begin{equation*}
\begin{array}{rllcl}
(\lambda a.M)N        & \reduct_\beta  & \subst{M}{N}{a} \\
\lambda a.(M)a        & \reduct_\eta   & M               & \qquad & a \notin M \\
(\mu\alpha.M)N        & \reduct_\mu    & \mu\alpha.\subst{M}{[\alpha](L)N}{[\alpha]L} \\
{}[\beta]\mu\alpha.M  & \reduct_\rho   & \subst{M}{\beta}{\alpha} \\
\mu\alpha[\alpha]M    & \reduct_\theta & M               & & \alpha\notin M \\[2ex]
\mu\alpha.M           & \reduct_\nu    & \lambda a.\mu\alpha.\subst{M}{[\alpha](L)a}{[\alpha]L} & & a \notin M 
\end{array}
\end{equation*}
\caption{Reduction rules for the \lmcalc}\label{tabredrules}
\end{table}
Notice that they are all validated by the $\beta\eta\mu\rho\theta$ equational theory. The only interesting case being the $\nu$-reduction:
\begin{equation*}
\mu\alpha.M  =_\eta \lambda a.(\mu\alpha.M)a =_\mu \lambda a.\mu\alpha.\subst{M}{[\alpha](L)a}{[\alpha]L}
\end{equation*}
However, in a typed setting, the $\nu$-reduction cannot always be applied (it requires $\mu\alpha.M$ to be of arrow type), so that we cannot directly apply the results of~\cite{phdpy}.

A simply typed simple \lmterm{} $M$ of type $A=A_1\imp\cdots\imp A_n\imp X$ is in \emph{canonical normal form} if $M=\lambda a_1\dots\lambda a_n\mu\alpha[\beta](b)M_1\dots M_k$ with $a_i$ of type $A_i$, $\alpha$ of type $X$, $\beta$ of type $Y$, $b$ of type $B$ and $M_j$ canonical normal form of type $B_j$ (with $B=B_1\imp\cdots\imp B_k\imp Y$).

\begin{lem}[Canonical normal form (simple types)]\label{lemcannfsimpl}
Let $M$ be a simply typed $\mu$-closed simple \lmterm, there exists a canonical normal form which is $\beta\eta\mu\rho\theta$ equivalent to $M$. More precisely turning $M$ into a canonical normal form only requires $\beta\mu\rho\nu$-reductions and $\eta\theta$-expansions.
\end{lem}

\proof
The $\beta\mu\rho$-reduction is normalizing~\cite{lmjsl}, so that we can concentrate on $\beta\mu\rho$-normal forms.

We first prove, by induction on the size of $A=A_1\imp\cdots\imp A_n\imp X$, that if $a$ is a $\lambda$-variable of type $A$ then it $\eta\theta$-expands into a canonical normal form: by induction hypothesis applied to the $\lambda$-variables $a_i$ of type $A_i$, we obtain the canonical normal forms $M_i$, and we have:
\begin{equation*}
  a \leftarrow^*_{\eta\theta} \lambda a_1\dots\lambda a_n\mu\alpha[\alpha](a)a_1\dots a_n \leftarrow^*_{\eta\theta} \lambda a_1\dots\lambda a_n\mu\alpha[\alpha](a)M_1\dots M_n
\end{equation*}

By induction on the size of $N$, we prove that any $\beta\mu\rho$-normal form $N$ of type $A$ is $\beta\eta\mu\rho\theta$ equivalent to a \emph{pre-canonical form}, that is a \lmterm{} of the shape $\overline{\lambda\mu}(b)M_1\dots M_k$ where $\overline{\lambda\mu}$ is a sequence of $\lambda$s and $\mu[\,]$s ending with a $\mu[\,]$ and containing no pair of consecutive $\mu[\,]$s, $M_j$ is a pre-canonical form of type $B_j$, and moreover $b$ is of type $B=B_1\imp\cdots\imp B_k\imp Y$ (that is each variable has as many arguments as possible).

If $N$ is a $\beta\mu\rho$-normal form (note that it immediately ensures the elimination of consecutive $\mu[\,]$s), under a bunch of $\lambda$s and $\mu[\,]$s, we find a term of the shape $N_0=(b)M_1\dots M_p$ with $b$ of type $B=B_1\imp\cdots\imp B_k\imp Y$ and $p\leq k$. If $p=k$, either $N_0$ is under a $\mu[\,]$ and it is in pre-canonical normal form or $N_0$ is under a $\lambda$ and we apply a $\theta$-expansion to $N_0$. If $p<k$, we replace $N_0$ by $N_1$ (which is equivalent to it) obtained from $\lambda b_{p+1}\dots\lambda b_k\mu\xi[\xi](b)M_1\dots M_p b_{p+1}\dots b_k$ ($\xi$ fresh) by replacing each $b_j$ by its canonical normal form (using the first result above). Finally we replace each $M_i$ by its pre-canonical form (using the induction hypothesis).

Notice that, in a pre-canonical form $M$, if $\alpha$ is of type $A\imp B$ and $[\alpha]N$ is a sub-term of $M$ then $N$ starts with a $\lambda$: otherwise it could only be a $\mu$ (but it would not be $\rho$-normal), or of the shape $(c)\vec{L}$ (but such a sub-term of a pre-canonical form under a $[\,]$ is always of atomic type). This remark allows us to define the following modified $\nu$-reduction on pre-canonical forms:
\begin{equation*}
  \mu\alpha.M   \reduct_\nu  \lambda a.\mu\alpha.\subst{M}{[\alpha]L}{[\alpha]\lambda a.L}
\end{equation*}
if $\alpha$ is of arrow type.

We finally prove the lemma by showing that any pre-canonical form $M$ is equivalent to a canonical normal form. We prove this by induction on the sum of the sizes of the types of the $\mu$-variables in $M$.
If all the types of the $\mu$-variables of $M$ are atomic, then $M$ is in canonical normal form. Otherwise, let $\mu\alpha.N$ be a sub-term of $M$ of arrow type (remember that the \lmterm{} is $\mu$-closed), we apply a $\nu$-reduction, this makes the induction size decrease but we are not sure to obtain directly a pre-canonical form again: we may have to apply some $\rho$-reductions to have a pre-canonical form, but these reductions terminate (the size of the \lmterm{} decreases) and do not make the induction size increase.
\qed

In order to extend this result to the first-order case, we first define an embedding of simple \lmterm s typed in first-order logic (without $\wedge$ or $\top$) into the simply typed \lmcalc. We consider an injective embedding of first-order function symbols, first-order variables and $\lambda$-variables into $\lambda$-variables (we still denote by $f$, $x$ and $a$ the translation of $f$, $x$ and $a$), and an embedding of relation symbols into $0$-ary relation symbols (the image of $X$ will be denoted by $X$). Moreover we choose two particular atomic types $O$ and $F$.
The translation is given in Table~\ref{tabfolprop}.
\begin{table}
\begin{equation*}
\begin{array}{rclcrcl}
\trad{\bot} &=& F & \qquad\qquad & 
\trad{a} &=& a \\
\bigtrad{X\vec{t}} &=& X & & 
\bigtrad{\lambda a.M} &=& \lambda a.\trad{M} \\
\bigtrad{A\imp B} &=& \trad{A}\imp\trad{B} & & 
\bigtrad{(M)N} &=& (\trad{M})\trad{N} \\
\bigtrad{\forall x A} &=& O\imp\trad{A} & &
\bigtrad{\mu\alpha[\beta]M} &=& \mu\alpha[\beta]\trad{M} \\
& & & & 
\bigtrad{\Lambda x.M} &=& \lambda x.\trad{M} \\
\trad{x} &=& x & & 
\bigtrad{M\foapp{t}} &=& (\trad{M})\trad{t} \\
\bigtrad{f t_1\dots t_k} &=& (f)\trad{t_1}\dots\trad{t_k}
\end{array}
\end{equation*}
\caption{A translation from first-order typing to simple types}\label{tabfolprop}
\end{table}

We extend the modified $\nu$-reduction to first-order constructs:
\begin{equation*}
  \mu\alpha.M   \reduct_\nu  \Lambda x.\mu\alpha.\subst{M}{[\alpha]L}{[\alpha]\Lambda x.L}
\end{equation*}
if $\alpha$ is of type $\forall x A$ and all the occurrences of $[\alpha]$ in $M$ are followed by $\Lambda x$.

\begin{lem}[Properties]\label{lempropfotrans}
The translation $\bigtrad{.}$ has the following properties:
\begin{enumerate}[$\bullet$]
\item If $\A\vdash M:A\mid\B$ then $\Lambda,\trad{\A}\vdash\trad{M}:\trad{A}\mid\trad{\B}$, where $\trad{\A}$ contains the translations of the typing declarations in $\A$ and typing declarations $x:O$ for (at least) the free first-order variables of $M$, and $\Lambda$ contains a typing declaration $f:O^k\imp O$ for each function symbol $f$ of arity $k$ occurring in $M$.
\item $\bigtrad{\subst{M}{t}{x}}=\subst{\trad{M}}{\trad{t}}{x}$
\item There is a one-to-one correspondence between the source and the target language for $\beta$-reduction, $\mu$-reduction, $\rho$-reduction, $\nu$-reduction, $\eta$-expansion and $\theta$-expansion. This means that if $\reduct$ is one of these rewriting rules, we have both simulation (if $M\reduct N$ then $\trad{M}\reduct\trad{N}$) and co-simulation (if $\trad{M}\reduct C$, there exists $N$ such that $C=\trad{N}$ and $M\reduct N$).
\end{enumerate}
\end{lem}

\proof\hfill
  \begin{enumerate}[$\bullet$]
  \item We first show that $\Lambda,\trad{\A}\vdash\trad{t}:O\mid\trad{\B}$ if $\trad{\A}$ contains typing declarations $x:O$ for (at least) the free first-order variables of $t$ and $\Lambda$ contains typing declarations for the function symbols occurring in $t$:
    \begin{gather*}
      \ZIC{\Lambda,\trad{\A},x:O\vdash x:O\mid\trad{\B}}
      \DP
      \\[2ex]
      \ZIC{\Lambda,\trad{\A}\vdash f:O^k\imp O\mid\trad{\B}}
      \AIC{\Lambda,\trad{\A}\vdash \trad{t_1}:O\mid\trad{\B}}
      \BIC{\Lambda,\trad{\A}\vdash (f)\trad{t_1}:O^{k-1}\imp O\mid\trad{\B}}
      \PPP
      \UIC{\Lambda,\trad{\A}\vdash (f)\trad{t_1}\dots\trad{t_{k-1}}: O\imp O\mid\trad{\B}}
      \AIC{\Lambda,\trad{\A}\vdash \trad{t_k}:O\mid\trad{\B}}
      \BIC{\Lambda,\trad{\A}\vdash (f)\trad{t_1}\dots\trad{t_k}: O\mid\trad{\B}}
      \DP
    \end{gather*}
  The only two interesting cases of \lmterm s are for first-order constructs:
    \begin{gather*}
      \AIC{\Lambda,\trad{\A},x:O\vdash\trad{M}:\trad{A}\mid\trad{\B}}
      \UIC{\Lambda,\trad{\A}\vdash\lambda x.\trad{M}:O\imp\trad{A}\mid\trad{\B}}
      \DP
      \\[2ex]
      \AIC{\Lambda,\trad{\A}\vdash\trad{M}:O\imp\trad{A}\mid\trad{\B}}
      \AIC{\Lambda,\trad{\A}\vdash \trad{t}:O\mid\trad{\B}}
      \BIC{\Lambda,\trad{\A}\vdash(\trad{M})\trad{t}:\trad{A}\mid\trad{\B}}
      \DP
    \end{gather*}
  \item We first check that $\bigtrad{\subst{u}{t}{x}}=\subst{\trad{u}}{\trad{t}}{x}$ for first-order terms. Then we work by induction on $M$, with only one interesting case:
    \begin{align*}
      \bigtrad{\subst{M\foapp{u}}{t}{x}} &= \bigtrad{\subst{M}{t}{x}\foapp{\subst{u}{t}{x}}} \\
      &= (\trad{\subst{M}{t}{x}})\trad{\subst{u}{t}{x}} \\
      &= (\subst{\trad{M}}{\trad{t}}{x})\subst{\trad{u}}{\trad{t}}{x} \\
      &= \subst{(\trad{M})\trad{u}}{\trad{t}}{x} \\
      &= \subst{\bigtrad{M\foapp{u}}}{\trad{t}}{x}
    \end{align*}
  \item We first prove the simulation results. For redexes without first-order construct, the result is immediate, for the other ones we have:
    \begin{enumerate}[$-$]
    \item $(\Lambda x.M)\foapp{t} \reduct_\beta \subst{M}{t}{x}$ becomes $(\lambda x.\trad{M})\trad{t} \reduct_\beta \subst{\trad{M}}{\trad{t}}{x}$ and we apply the previous result
    \item $M \expand_\eta \Lambda x.M\foapp{x}$ becomes $\trad{M}\expand_\eta \lambda x.(\trad{M})x$
    \item $(\mu\alpha.M)\foapp{t} \reduct_\mu \mu\alpha.\subst{M}{[\alpha]L\foapp{t}}{[\alpha]L}$ becomes $(\mu\alpha.\trad{M})\trad{t} \reduct_\mu \mu\alpha.\subst{\trad{M}}{[\alpha](L)\trad{t}}{[\alpha]L}$ and we conclude from the fact that $[\,]$s are not modified by the translation
    \item $\mu\alpha.M \reduct_\nu \Lambda x.\mu\alpha.\subst{M}{[\alpha]L}{[\alpha]\Lambda x.L}$ becomes $\mu\alpha.\trad{M} \reduct_\nu \lambda x.\mu\alpha.\subst{\trad{M}}{[\alpha]L}{[\alpha]\lambda x.L}$ since each $\Lambda x$ under a $[\alpha]$ becomes a $\lambda x$ and $\forall x A$ becomes $O\imp\trad{A}$.
    \end{enumerate}
  \item We now look at co-simulation for each rewriting rule:
    \begin{enumerate}[$-$]
    \item If $\trad{M}$ contains a $\beta$-redex, either both $\lambda$ and application are already present in $M$ and the result is immediate, or they both come from the corresponding first-order constructs and we can apply the simulation result, or exactly one come from a first-order construct and this would be a typing error in $M$.
    \item If $\trad{M}$ contains a $\mu$-redex, either the application is already present in $M$ and the result is immediate, or it comes from a first-order application and we can apply the simulation result.
    \item If $\trad{M}$ contains a $\rho$-redex, so do $M$, and the result is immediate.
    \item If $\trad{M}$ contains a $\nu$-redex starting with $\mu\alpha$, so do $M$, and we can apply the same reduction in $M$ (with a $\Lambda$ if $\alpha$ is of type $O\imp A$ and with a $\lambda$ otherwise). Finally we conclude with the simulation result.
    \item If $\trad{M}$ $\eta$-expands to $\lambda a.(\trad{M})a$ then $N=\lambda a.(M)a$ and if $\trad{M}$ $\eta$-expands to $\lambda x.(\trad{M})x$ then $N=\Lambda x.M\foapp{x}$. We conclude with the simulation result.
    \item By the simulation result, the case of a $\theta$-expansion is immediate.
\qed
    \end{enumerate}
  \end{enumerate}

\noindent
A simple \lmterm{} $M$ of type $A=\forall x_1\dots\forall x_p (A_1\imp\cdots\imp A_n\imp R)$ is in \emph{quasi canonical normal form} if $M=\Lambda x_1\dots\Lambda x_p\lambda a_1\dots\lambda a_n\mu\alpha[\beta](b\foapp{t_1}\dots\foapp{t_q})M_1\dots M_k$ with $a_i$ of type $A_i$, $\alpha$ of type $R$, $\beta$ of type $S[^{t_1}/_{y_1},\dots,^{t_q}/_{y_q}]$, $b$ of type $B$ and $M_j$ canonical normal form of type $B_j[^{t_1}/_{y_1},\dots,^{t_q}/_{y_q}]$ (with $B=\forall y_1\dots\forall y_q (B_1\imp\cdots\imp B_k\imp S)$).
A canonical form is obtained from a quasi canonical normal form by removing the $\mu$s and $[\,]$s acting on $\mu$-variables of type $\bot$.

\begin{prop}[Canonical normal form]
Let $M$ be a typed $\mu$-closed simple \lmterm, there exists a canonical normal form which is $\beta\eta\mu\rho\theta$ equivalent to $M$.
\end{prop}

\proof
We first show how to obtain quasi canonical normal forms. We translate $M$ into $\trad{M}$ and we apply Lemma~\ref{lemcannfsimpl} to get a simply typed canonical normal form $C$. By Lemma~\ref{lempropfotrans} (since going from $\trad{M}$ to $C$ requires only $\beta\mu\rho\nu$-reduction and $\eta\theta$-expansion)  there exists a \lmterm{} $N$ equivalent to $M$ and such that $\trad{N}=C$. Going from $C$ to $N$ transforms some $\lambda$s into $\Lambda$s and some applications into first-order applications, but using the type we can see that we do not have choices: the $\Lambda$s are all arriving before the $\lambda$s (for each bunch of $\lambda$s of $C$) and the same for applications. As a consequence $N$ is a quasi canonical normal form.

Finally, if the quasi canonical normal form $M$ contains some $\mu\alpha[\beta]$ with $\alpha$ of type $\bot$ we apply $\mu\alpha[\beta]N=_\rho[\alpha]\mu\alpha[\beta]N=_\rho [\beta]N$, and if $M$ contains some $[\alpha]N$ with $\alpha$ of type $\bot$, we erase the $[\alpha]$ by $\rho$.
\qed

\subsection{Isomorphisms}
\label{appisos}

The notion of isomorphism is very standard in algebra and in category theory. There is a natural corresponding notion in extensions of the \lcalc~\cite{inversenf}. We consider such an extension of the \lcalc{} endowed with the equational theory on terms generated by the reduction rules (containing the $\beta\eta$ equality). The term $M$ is an isomorphism if there exists a term $N$ such that $\lambda x.(M)(N)x=\lambda x.x$ and $\lambda y.(N)(M)y=\lambda y.y$ (we say that $M$ and $N$ give an isomorphism pair).

In a typed setting (\ie{} if the calculus comes with a type system extending the simply typed \lcalc), we can consider isomorphisms between types~\cite{isotypes}. The types $A$ and $B$ are isomorphic if there exist two terms $M$ and $N$ such that $\vdash M:A\imp B$, $\vdash N:B\imp A$ and $M$ and $N$ give an isomorphism pair. The main question about type isomorphisms is usually to find the equational theory characterizing them in a given calculus.

Proving that a given equation on types is a valid isomorphism only requires us to exhibit a pair of typed terms (of appropriate types) and to prove that their compositions in both directions are equal to the identity up to the appropriate equational theory on terms.

This is what we are looking at here with the Church style first-order \lmcalc{} with terms equal up to the $\beta\eta\mu\rho\theta$ equality.

\begin{table}
\begin{align*}
\lambda a.\pair{\pair{\projl{a}}{\projl{\projr{a}}}}{\projr{\projr{a}}} &:A\wedge(B\wedge C) \imp (A\wedge B)\wedge C \\
\lambda a.\pair{\projl{\projl{a}}}{\pair{\projr{\projl{a}}}{\projr{a}}} &:(A\wedge B)\wedge C \imp A\wedge (B\wedge C) \\[2ex]
\lambda a.\projl{a} &:A\wedge\top \imp A \\
\lambda a.\pair{a}{\cst} &:A \imp A\wedge\top \\[2ex]
\lambda a.\projr{a} &:\top\wedge A \imp A \\
\lambda a.\pair{\cst}{a} &:A \imp \top\wedge A \\[2ex]
\lambda a.\lambda b.\lambda c.(a)\pair{b}{c} &:((A\wedge B)\imp C) \imp A \imp B \imp C \\
\lambda a.\lambda b.(a)\projl{b}\projr{b} &: (A\imp B\imp C)\imp (A\wedge B)\imp C \\[2ex]
\lambda a.(a)\cst &:(\top\imp A) \imp A \\
\lambda a.\lambda d.a &:A\imp \top\imp A \\[2ex]
\lambda a.\pair{\lambda b.\projl{(a)b}}{\lambda b.\projr{(a)b}} &:(A\imp(B\wedge C))\imp (A\imp B)\wedge(A \imp C) \\
\lambda a.\lambda b.\pair{(\projl{a})b}{(\projr{a})b} &:((A\imp B)\wedge(A \imp C))\imp A\imp(B\wedge C)  \\[2ex]
\lambda a.\cst &:(A\imp\top) \imp \top \\
\lambda a.\lambda d.a &:\top\imp A\imp\top \\[2ex]
\lambda a.\pair{\Lambda x.\projl{a\foapp{x}}}{\Lambda x.\projr{a\foapp{x}}} &:\forall x(A\wedge B) \imp \forall x A\wedge\forall x B \\
\lambda a.\Lambda x.\pair{(\projl{a})\foapp{x}}{(\projr{a})\foapp{x}} &:(\forall x A\wedge\forall x B) \imp \forall x(A\wedge B) \\[2ex]
\lambda a.\cst &:\forall x\top \imp \top \\
\lambda a.\Lambda x.a &:\top\imp \forall x\top \\[2ex]
\lambda a.\Lambda x.\lambda b.((a)b)\foapp{x} &: (A\imp\forall x B) \imp \forall x(A\imp B) \\
\lambda a.\lambda b.\Lambda x.(a\foapp{x})b &:\forall x(A\imp B) \imp A\imp\forall x B \\[2ex]
\lambda a.\pair{\projr{a}}{\projl{a}} &:A\wedge B \imp B\wedge A \\
\lambda a.\pair{\projr{a}}{\projl{a}} &:B\wedge A \imp A\wedge B \\[2ex]
\lambda a.\Lambda y.\Lambda x.a\foapp{x}\foapp{y} &:\forall x\forall y A \imp \forall y\forall x A \\
\lambda a.\Lambda x.\Lambda y.a\foapp{y}\foapp{x} &:\forall y\forall x A \imp \forall x\forall y A
\end{align*}
\caption{Isomorphisms}\label{tabtermisos}
\end{table}

\begin{prop}[Syntactic isomorphisms]\label{propsynisos}
For each equation of Table~\ref{tabtypisos}, the Table~\ref{tabtermisos} gives such a pair of terms validating the equation.
\end{prop}

\proof
This proof is left to the reader. We just give one example (for the equation $\forall x(A\wedge B) = \forall x A\wedge\forall x B$) of the kind of computation we have to apply:
\begin{align*}
\lambda c.&
(\lambda a.\pair{\Lambda x.\projl{a\foapp{x}}}{\Lambda x.\projr{a\foapp{x}}})
(\lambda a.\Lambda x.\pair{(\projl{a})\foapp{x}}{(\projr{a})\foapp{x}})
c \\
&=_\beta 
\lambda c.
(\lambda a.\pair{\Lambda x.\projl{a\foapp{x}}}{\Lambda x.\projr{a\foapp{x}}})
\Lambda x.\pair{(\projl{c})\foapp{x}}{(\projr{c})\foapp{x}} \\
&=_\beta 
\lambda c.
\pair{\Lambda x.\projl{(\Lambda x.\pair{(\projl{c})\foapp{x}}{(\projr{c})\foapp{x}})\foapp{x}}}{\Lambda x.\projr{(\Lambda x.\pair{(\projl{c})\foapp{x}}{(\projr{c})\foapp{x}})\foapp{x}}} \\
&=_\beta 
\lambda c.
\pair{\Lambda x.\projl{\pair{(\projl{c})\foapp{x}}{(\projr{c})\foapp{x}}}}{\Lambda x.\projr{(\Lambda x.\pair{(\projl{c})\foapp{x}}{(\projr{c})\foapp{x}})\foapp{x}}} \\
&=_\beta 
\lambda c.
\pair{\Lambda x.\projl{\pair{(\projl{c})\foapp{x}}{(\projr{c})\foapp{x}}}}{\Lambda x.\projr{\pair{(\projl{c})\foapp{x}}{(\projr{c})\foapp{x}}}} \\
&=_\beta 
\lambda c.
\pair{\Lambda x.(\projl{c})\foapp{x}}{\Lambda x.\projr{\pair{(\projl{c})\foapp{x}}{(\projr{c})\foapp{x}}}} \\
&=_\beta 
\lambda c.
\pair{\Lambda x.(\projl{c})\foapp{x}}{\Lambda x.(\projr{c})\foapp{x}} \\
&=_\eta 
\lambda c.
\pair{\projl{c}}{\Lambda x.(\projr{c})\foapp{x}} \\
&=_\eta 
\lambda c.
\pair{\projl{c}}{\projr{c}} \\
&=_\eta 
\lambda c.
c
\end{align*}
\qed

\noindent
The difficult point about type isomorphisms is always the opposite direction: showing that all the equations have been found. This is the purpose of Section~\ref{secisos}.

\section{Categorical properties of games}
\label{appcat}
\label{appforgetfunc}

Let $A$ be an arena without any label (such an arena is called a \emph{ground arena}), strategies on $A$ are exactly those defined in~\cite{gamesgoi,synsempol}. In particular ground arenas and strategies allow us to define a category $\Gcatgr$ (this is the category also used in Section~\ref{secmodellcalc}).

We define the function $\grfunc$ from arenas to ground arenas which erases the labels of its argument. If \play{s} is a play on $A$, we define $\grfunc(\play{s})$ as the play on $\grfunc(A)$ obtained by erasing the instantiations and the $\mu$-pointers (and the same with interaction sequences). We extend $\grfunc$ to sets of plays by applying it point-wise.

\begin{lem}\label{lemgrplay}
Let $\sigma:A$ be a strategy, if $\play{s}\move{m}\in\sigma$ and $\play{t}\move{n}\in\sigma$ with $\grfunc(\play{s})=\grfunc(\play{t})$ then there exists an \ovar-renaming $\oren$ such that $\play{s}\move{m}=\ren{\play{t}\move{n}}{\oren}$.
\end{lem}

\proof
We prove it by induction on the common length of $\play{s}\move{m}$ and $\play{t}\move{n}$.
If $\grfunc(\move{m}_1\move{m}_2)=\grfunc(\move{n}_1\move{n}_2)$ then $\move{m}_1$ and $\move{n}_1$ are Opponent moves corresponding to the same initial move in $A$, thus they can only differ on their \ovar-instantiations $[x_1,\dots,x_k]$ and $[y_1,\dots,y_k]$ (which must have the same length $k$). We consider $\oren$ such that $\oren(y_1)=x_1$,~...,~$\oren(y_k)=x_k$. We have $\move{m}_1=\ren{\move{n}_1}{\oren}$, $\move{m}_1\move{m}_2\in\sigma$, $\move{n}_1\move{n}_2\in\sigma$ thus $\ren{\move{n}_1\move{n}_2}{\oren}\in\sigma$ (by uniformity), and finally $\move{m}_1\move{m}_2=\ren{\move{n}_1\move{n}_2}{\oren}$ by determinism.
In a similar way, if $\grfunc(\play{s}\move{m}_1\move{m}_2\move{m}_3)=\grfunc(\play{t}\move{n}_1\move{n}_2\move{n}_3)$ then $\grfunc(\play{s})=\grfunc(\play{t})$ and by induction hypothesis $\play{s}\move{m}_1=\ren{\play{t}\move{n}_1}{\oren}$. Moreover $\move{m}_2$ and $\move{n}_2$ are Opponent moves corresponding to the same move in $A$, they have the same justification pointer and thus they can only differ on their \ovar-instantiations $[x_1,\dots,x_k]$ and $[y_1,\dots,y_k]$. We modify $\oren$ into $\oren'$ in such a way that $\oren'(y_1)=x_1$,~...,~$\oren'(y_k)=x_k$ and $\oren$ and $\oren'$ agree on the \ovar-variables appearing in $\play{t}\move{n}_1$. We have $\play{s}\move{m}_1\move{m}_2=\ren{\play{t}\move{n}_1\move{n}_2}{\oren'}$, $\play{s}\move{m}_1\move{m}_2\move{m}_3\in\sigma$, $\play{t}\move{n}_1\move{n}_2\move{n}_3\in\sigma$ thus $\ren{\play{t}\move{n}_1\move{n}_2\move{n}_3}{\oren'}\in\sigma$ (by uniformity), and finally $\play{s}\move{m}_1\move{m}_2\move{m}_3=\ren{\play{t}\move{n}_1\move{n}_2\move{n}_3}{\oren'}$ by determinism.
\qed

\begin{lem}[Preservation of strategies]\label{lemgrstrat}
If $\sigma$ is a strategy on $A$, then $\grfunc(\sigma)$ is a strategy on $\grfunc(A)$.
\end{lem}

\proof
It is immediate that $\grfunc(\sigma)$ is a non-empty set of even length plays which is closed under even length prefixes.

If $\play{s}\move{m}=\grfunc(\play{t}_1\move{m}')$ and $\play{s}\move{n}=\grfunc(\play{t}_2\move{n}')$ with $\play{t}_1\move{m}'\in\sigma$ and $\play{t}_2\move{n}'\in\sigma$. By Lemma~\ref{lemgrplay}, $\play{t}_1\move{m}'=\ren{\play{t}_2\move{n}'}{\oren}$ for some \ovar-renaming $\oren$. As a consequence $\play{s}\move{m}=\play{s}\move{n}$.

There is nothing to say about uniformity since moves in $\grfunc(\sigma)$ have no instantiation.
\qed

\begin{lem}[Preservation of composition and identity]\label{lemgrcompid}
Let $\sigma$ and $\tau$ be strategies on $A\imp B$ and $B\imp C$, $\grfunc(\sigma\fol\tau)=\grfunc(\sigma)\fol\grfunc(\tau)$ and $\grfunc(\id{A})=\id{\grfunc(A)}$.
\end{lem}

\proof
We first prove the inclusion $\grfunc(\sigma\fol\tau)\subseteq\grfunc(\sigma)\fol\grfunc(\tau)$. Let \play{s} be a play in $\grfunc(\sigma\fol\tau)$ coming from the interaction sequence \play{u}. $\grfunc(\play{u})$ is an interaction sequence on $\grfunc(A)$, $\grfunc(B)$ and $\grfunc(C)$ with $\grfunc(\proj{\play{u}}{A\imp B})=\proj{\grfunc(\play{u})}{A\imp B}$, $\grfunc(\proj{\play{u}}{B\imp C})=\proj{\grfunc(\play{u})}{B\imp C}$, and $\grfunc(\proj{\play{u}}{A\imp C})=\proj{\grfunc(\play{u})}{A\imp C}$. Since $\grfunc(\play{s})=\proj{\grfunc(\play{u})}{A\imp C}$, we conclude $\grfunc(\play{s})\in\grfunc(\sigma)\fol\grfunc(\tau)$.

Concerning $\grfunc(\sigma)\fol\grfunc(\tau)\subseteq\grfunc(\sigma\fol\tau)$, we consider an interaction sequence $\play{v}$ on $\grfunc(A)$, $\grfunc(B)$ and $\grfunc(C)$ with $\proj{\play{v}}{\grfunc(A)\imp \grfunc(B)}=\grfunc(\play{s})$ ($\play{s}\in\sigma$), $\proj{\play{v}}{\grfunc(B)\imp \grfunc(C)}=\grfunc(\play{t})$ ($\play{t}\in\tau$). By applying an \ovar-renaming if required, we can assume that the \ovar-variables appearing in \play{s} and \play{t} are different.
We build from \play{v} an interaction sequence \play{u} on $A$, $B$ and $C$ in the following way: if \move{m} is a move in $C$, we add on it the $\mu$-pointers and instantiations coming from \play{t}, if \move{m} is a move in $A$, we add on it the $\mu$-pointers and instantiations coming from \play{s}, if \move{m} is a move in $B$, we add on it the $\mu$-pointers and instantiations coming from both \play{s} and \play{t}. We can check that $\proj{\play{u}}{A\imp C}\in\pplays{A\imp C}$ so that $\proj{\play{u}}{A\imp C}\in\sigma\fol\tau$ (the only point to verify is the condition relating $\mu$-pointers and atomic labels, but it comes easily).
This gives $\proj{\play{v}}{\grfunc(A)\imp \grfunc(C)}=\grfunc(\proj{\play{u}}{A\imp C})\in\grfunc(\sigma\fol\tau)$.

The case of the identity is immediate by definition of $\id{A}$.
\qed

\begin{lem}[Zipping]\label{lemzipping}
Let $\sigma:A\imp B$ and $\tau:B\imp C$ be two strategies, let \play{u} and \play{v} be two interaction sequences on $A$, $B$ and $C$ such that $\proj{\play{u}}{A\imp B}\in\sigma$, $\proj{\play{v}}{A\imp B}\in\sigma$, $\proj{\play{u}}{B\imp C}\in\tau$ and $\proj{\play{v}}{B\imp C}\in\tau$, the first move which differs between \play{u} and \play{v} is an Opponent move in $A\imp C$.
\end{lem}

\proof
This is the very natural extension (with $\mu$-pointers and instantiations) of the usual result for HO/N games proved by induction on the length of the maximal common prefix of \play{u} and \play{v}, using the determinism of $\sigma$ and $\tau$ (see~\cite{classisos} for example).
\qed

\begin{lem}
The identity is a strategy, and the composition of two strategies is a strategy.
\end{lem}

\proof
$\id{A}$ is a non-empty set of even length plays which is closed under even length prefixes. It is clearly uniform. It is deterministic: if $\play{s}\move{m}\in\id{A}$ and $\play{s}\move{n}\in\id{A}$, the node in $A$ underlying \move{m} and \move{n} is the same and the justification pointers are the same since $\grfunc(\id{A})$ is a strategy (Lemma~\ref{lemgrcompid}), and the $\mu$-pointers and the instantiations are the same by definition.

Let $\sigma:A\imp B$ and $\tau:B\imp C$ be two strategies, $\sigma\fol\tau$ is a non-empty set of even length plays which is closed under even length prefixes.

If $\play{s}\move{m}\in\sigma\fol\tau$ and $\play{s}\move{n}\in\sigma\fol\tau$, let \play{u} be an interaction sequence corresponding to $\play{s}\move{m}$ and \play{v} be an interaction sequence corresponding to $\play{s}\move{n}$, by Lemma~\ref{lemzipping}, one of \play{u} and \play{v} must be a prefix of the other. As a consequence $\play{s}\move{m}=\play{s}\move{n}$ since they have the same length.

Let \play{s} be a play in $\sigma\fol\tau$, \play{u} be a corresponding interaction sequence and $\oren$ be an \ovar-renaming, we define $\oren'$ as an \ovar-renaming which coincides with $\oren$ on the \ovar-instantiations of \play{u} in $A$ and $C$ (so that $\ren{\play{s}}{\oren}=\ren{\play{s}}{\oren'}$) and which maps \ovar-instantiations of \play{u} in $B$ to fresh \ovar-variables (neither appearing in \play{u} nor in the image of the \ovar-instantiations of \play{u} by $\oren$). $\ren{\play{u}}{\oren'}$ is an interaction sequence on $A$, $B$ and $C$ such that $\proj{\ren{\play{u}}{\oren'}}{A\imp B}=\ren{\proj{\play{u}}{A\imp B}}{\oren'}\in\sigma$, $\proj{\ren{\play{u}}{\oren'}}{B\imp C}=\ren{\proj{\play{u}}{B\imp C}}{\oren'}\in\tau$, and $\proj{\ren{\play{u}}{\oren'}}{A\imp C}=\ren{\play{s}}{\oren'}=\ren{\play{s}}{\oren}$, thus $\ren{\play{s}}{\oren}\in\sigma\fol\tau$.
\qed

To turn arenas and strategies into a category, we still have to show the composition to be associative and the identity to be neutral for composition.

\begin{prop}[Category of strategies]
Arenas and strategies define a category.
\end{prop}

\proof
Let $\sigma:A\imp B$, $\tau:B\imp C$ and $\rho:C\imp D$ be three strategies, by Lemma~\ref{lemgrcompid}, we know that $\grfunc((\sigma\fol\tau)\fol\rho)=\grfunc(\sigma\fol(\tau\fol\rho))$. This means that we only have to look at $\mu$-pointers and instantiations to show that $(\sigma\fol\tau)\fol\rho=\sigma\fol(\tau\fol\rho)$. We just sketch the arguments (a more precise proof would go through a zipping lemma~\cite[Lemma~3.2.3]{phdharmer}).

If $\play{s}\move{m}\in(\sigma\fol\tau)\fol\rho$ with $\move{m}\in A$ equipped with a $\mu$-pointer to a move in $D$, it is obtained by following $\mu$-pointers through $B$ in an interaction sequence coming from the composition $\sigma\fol\tau$ until arriving in $C$ and then by following $\mu$-pointers in $C$ in an interaction sequence coming from the composition $(\sigma\fol\tau)\fol\rho$ until arriving in $D$. In $\sigma\fol(\tau\fol\rho)$, this $\mu$-pointer is obtained by building a $\mu$-pointer $p$ from $B$ to $D$ by following the $\mu$-pointers through $C$ in an interaction sequence coming from the composition $\tau\fol\rho$ and then by following $\mu$-pointers from $A$ to the source of $p$ through $B$ in an interaction sequence coming from the composition $\sigma\fol(\tau\fol\rho)$. This means that in both case we build a path going from $A$ to $D$ through moves in $B$ and $C$ in the same way, and thus we obtain the same $\mu$-pointer.

Concerning instantiations, they are built by applying an \ovar-substitution $\osubst_1$ corresponding to pairs of \ovar-instantiations and \pvar-instantiations in $B$ and an \ovar-substitution $\osubst_2$ corresponding to pairs of \ovar-instantiations and \pvar-instantiations in $C$. By disjointness of the \ovar-instantiations in $B$ and $C$, applying first $\osubst_1$ and then $\osubst_2$ or the converse leads to the same instantiations in $(\sigma\fol\tau)\fol\rho$ and $\sigma\fol(\tau\fol\rho)$.

The neutrality of the identity with respect to composition is easy and left to the reader.
\qed

Lemmas~\ref{lemgrstrat} and~\ref{lemgrcompid} show that $\grfunc$ defines a functor from the category of arenas and strategies to the full sub-category of ground arenas and strategies.

We are now able to prove Proposition~\ref{propcat} (page~\pageref{propcat}).

\begin{lem}\label{lempvvc}
Let $\sigma$ be a view function, $\pview{\vc{\sigma}}=\sigma$.
\qed
\end{lem}

\begin{lem}\label{lemvcpv}
Let $\sigma$ be an innocent strategy, $\vc{\pview{\sigma}}=\sigma$.
\end{lem}

\proof
By definition, there exists a view function $\tau$ such that $\sigma=\vc{\tau}$ thus:
\begin{align*}
  \vc{\pview{\sigma}} &= \vc{\pview{\vc{\tau}}} \\
                      &= \vc{\tau} & \text{by Lemma~\ref{lempvvc}} \\
                      &= \sigma
\end{align*}
\qed

\begin{lem}[Composition of view closures]\label{lemcompvc}
Let $\sigma$ be a view function on $A\imp B$ and $\tau$ be a view function on $B\imp C$, there exists a view function $\rho$ on $A\imp C$ such that $\vc{\sigma}\fol\vc{\tau}=\vc{\rho}$.
\end{lem}

\proof
In the usual setting of games for extensions of the simply typed \lcalc, there is a well known direct characterization of innocent strategies, and they are known to compose (see~\cite{phdmccusker,phdharmer} for example). This tells us that $\vc{\sigma}\fol\vc{\tau}=\vc{\rho_0}$ with $\rho_0=\ppreview{\vc{\sigma}\fol\vc{\tau}}$ (since our definition of pre-view is the usual definition of view and our definition of view closure is the usual one).

What remains to be checked are our constraints on first-order instantiations which are not present in the traditional setting. More precisely, we define $\rho=\pview{\vc{\sigma}\fol\vc{\tau}}$ and we have to show that $\rho$ is a view function such that $\vc{\rho}=\vc{\rho_0}$.
$\rho$ is a non-empty set of even length views closed under even length prefixes. Moreover $\rho\subseteq\rho_0$ since $\rho_0$ is closed under $\ovar$-renaming (by uniformity of $\vc{\sigma}\fol\vc{\tau}$). We deduce that $\rho$ is a view function and that $\vc{\rho}\subseteq\vc{\rho_0}$. Finally $\vc{\rho_0}\subseteq\vc{\rho}$ since, for any play $\play{s}$, $\pview{\play{s}}\in\rho_0$ implies $\pview{\play{s}}\in\rho$.
\qed

\begin{prop}
Arenas and view functions give a category $\Gcat$.
\end{prop}

\proof
The identity view function is a view function on $A\imp A$ (whose view closure is the identity strategy). If $\sigma$ is a view function on $A\imp B$ and $\tau$ is a view function on $B\imp C$, the composition of $\sigma$ and $\tau$ is $\pview{\vc{\sigma}\fol\vc{\tau}}$ which is $\pview{\vc{\rho}}=\rho$ (for some view function $\rho$, by Lemmas~\ref{lemcompvc} and~\ref{lempvvc}) thus it is a view function.

The identity view function is neutral for composition (we give only one side):
\begin{align*}
  \sigma\fol\pview{\id{}} &= \pview{\vc{\sigma}\fol\vc{\pview{\id{}}}} \\
                          &= \pview{\vc{\sigma\fol\id{}}} & \text{by Lemma~\ref{lemvcpv}} \\ 
                          &= \pview{\vc{\sigma}} \\
                          &= \sigma & \text{by Lemma~\ref{lempvvc}}
\end{align*}
The composition of view functions is associative:
\begin{align*}
\sigma\fol(\tau\fol\rho) &= \pview{\vc{\sigma}\fol\vc{\pview{\vc{\tau}\fol\vc{\rho}}}} \\
&=  \pview{\vc{\sigma}\fol(\vc{\tau}\fol\vc{\rho})} & \text{by Lemmas~\ref{lemcompvc} and~\ref{lemvcpv}} \\
&=  \pview{(\vc{\sigma}\fol\vc{\tau})\fol\vc{\rho}} \\
&=  \pview{\vc{\pview{\vc{\sigma}\fol\vc{\tau}}}\fol\vc{\rho}} & \text{by Lemmas~\ref{lemcompvc} and~\ref{lemvcpv}} \\
&= (\sigma\fol\tau)\fol\rho
\end{align*}
\qed

\noindent
$\grfunc$ turns view functions into view functions, and thus defines a functor from $\Gcat$ to $\Gcatgr$.

Additional structure of this category is given by Theorem~\ref{thmcontrolcat} (page~\pageref{thmcontrolcat}). We will prove it after a few lemmas (Theorem~\ref{thmcontrolcatapp}).

\begin{lem}[Composition of $\mu$-rigid strategies]\label{lemcompmurigid}
Let $\sigma:A\imp B$ and $\tau:B\imp C$ be two $\mu$-rigid strategies, $\sigma\fol\tau$ is $\mu$-rigid.
\end{lem}

\proof
We prove (by induction on its length) that any interaction sequence \play{u} on $A$, $B$ and $C$, such that $\proj{\play{u}}{A\imp B}\in\sigma$, $\proj{\play{u}}{B\imp C}\in\tau$, has its $\mu$-pointers given in a $\mu$-rigid way. If \play{u} is empty, the result is immediate, if the last move corresponds to a Player move of $\sigma$, the result comes from the $\mu$-rigidity of $\sigma$ and the same with $\tau$. As a consequence, the $\mu$-pointers in $\proj{\play{u}}{A\imp C}$ respect the $\mu$-rigidity.

In a similar way, between two moves \move{m} and \move{n} (\move{n} Player move in $A\imp C$) in $A$ or $C$ in \play{u}, the \ovar-substitution induced by \play{u} identifies each \ovar-instantiation with the previous one in moves in $B$, and we deduce that the \pvar-instantiation of \move{n} is the same as the \ovar-instantiation of \move{m}.
\qed

\begin{lem}\label{lemgrmurigid}
Let $\sigma$ and $\tau$ be two $\mu$-rigid strategies on $A$, if $\grfunc(\sigma)=\grfunc(\tau)$ then $\sigma=\tau$.
\end{lem}

\proof
By symmetry, it is enough to show $\sigma\subseteq\tau$. We prove by induction on the length of \play{s} that $\play{s}\in\sigma$ entails $\play{s}\in\tau$. The case of $\eps$ is immediate. If $\play{s}\move{m}\move{n}\in\sigma$, there exists $\play{t}\move{m}'\move{n}'\in\tau$ such that $\grfunc(\play{s}\move{m}\move{n})=\grfunc(\play{t}\move{m}'\move{n}')$. By induction hypothesis, $\play{s}\in\tau$, thus by Lemma~\ref{lemgrplay} there exists an \ovar-renaming $\oren$ such that $\play{s}=\ren{\play{t}}{\oren}$. We consider an \ovar-renaming $\oren'$ such that $\play{s}=\ren{\play{t}}{\oren'}$ and $\oren'$ maps each \ovar-variable in the \ovar-instantiation of $\move{m}'$ to the corresponding \ovar-variable in the \ovar-instantiation of \move{m}. We look at $\play{s}\move{m}\move{n}''=\ren{(\play{t}\move{m}'\move{n}')}{\oren'}\in\tau$. We want to prove $\play{s}\move{m}\move{n}=\play{s}\move{m}\move{n}''$. We have $\grfunc(\play{s}\move{m}\move{n})=\grfunc(\play{s}\move{m}\move{n}'')$, thus \move{n} and $\move{n}''$ correspond to the same node in $A$ and have the same justification pointer. Moreover they are $\mu$-rigid thus their $\mu$-pointers and instantiations are obtained in the same way.
\qed

These two lemmas show that $\grfunc$, restricted to the sub-category of $\Gcat$ given by $\mu$-rigid strategies, is faithful.

\begin{thm}[Control category of games]\label{thmcontrolcatapp}
The category $\Gcat$ of arenas and view functions is a control category.
\end{thm}

\proof
Since $\Gcatgr$ is a control category~\cite{synsempol} and $\grfunc$ preserves the various constructions on arenas and strategies ($\grfunc(A+B)=\grfunc(A)+\grfunc(B)$, $\grfunc(\sigma+\tau)=\grfunc(\sigma)+\grfunc(\tau)$, $\grfunc(A\times B)=\grfunc(A)\times\grfunc(B)$, $\grfunc(\sigma\times\tau)=\grfunc(\sigma)\times\grfunc(\tau)$,~...) as well as basic morphisms (such as associativity and commutativity of the constructions, which are $\mu$-rigid), any commutative diagram required in the definition of a control category and concerning only $\mu$-rigid strategies commutes in $\Gcat$ (by faithfulness of $\grfunc$ on $\mu$-rigid strategies).

The other properties are about the monoid structure (with respect to the pre-monoidal product) defined on each object and about cartesian closedness. They are not difficult to check and left to the reader.
\qed

\begin{lem}\label{lemprestotfin}
$\grfunc$ reflects totality and finiteness: $\grfunc(\sigma)$ is total if and only if $\sigma$ is total, $\grfunc(\sigma)$ is finite if and only if $\sigma$ is finite.
\qed
\end{lem}

\begin{prop}[Composition of total finite strategies]
  The composition of two total finite strategies is a total finite strategy.
\end{prop}

\proof
Let $\sigma$ and $\tau$ be two total finite strategies, $\sigma\fol\tau$ is total finite iff $\grfunc(\sigma\fol\tau)$ is total finite (by Lemma~\ref{lemprestotfin}) iff $\grfunc(\sigma)\fol\grfunc(\tau)$ is total finite. By the full completeness result of~\cite{gamesgoi}, $\grfunc(\sigma)$ and $\grfunc(\tau)$ are the interpretations of two simply typed \lterm s $M$ and $N$ and $\grfunc(\sigma)\fol\grfunc(\tau)$ is the interpretation of $\lambda x.(M)(N)x$ thus a total finite strategy.
\qed

It would be possible to extend this categorical analysis of our game model by introducing a notion of first-order control hyperdoctrines (in the spirit of control hyperdoctrines~\cite{isoschurch}), and by proving our games to give such a first-order control hyperdoctrine.
We do not think it would help a lot in the present work.

\end{document}